\documentclass[a4paper]{easychair}

\usepackage{graphicx}
\usepackage{sectsty}
\usepackage{amssymb}
\usepackage{amsmath}
\usepackage{tikz-cd}

\usepackage{pslatex}
\usepackage{paralist}
\usepackage{mathtools}
\usepackage{color}
\usepackage{stmaryrd}
\usepackage{todonotes}
\usepackage[inline]{enumitem}

\newtheorem{theorem}{Theorem}

\newtheorem{definition}{Definition}

\newtheorem{corollary}{Corollary}

\newtheorem{example}{Example}

\usepackage[createShortEnv,conf={end,link to proof,restate}]{proof-at-the-end}

\renewcommand{\paragraph}[1]{\vspace*{.5\baselineskip}\noindent{\emph{#1}}}

\newcommand{\emptyseq}{\varepsilon}
\newcommand{\nat}{\mathbb{N}}

\newcommand{\arity}{\#}
\newcommand{\arityof}[1]{{\arity{#1}}}

\newcommand{\lenof}[1]{\mathrm{len}({#1})}
\newcommand{\cardof}[1]{\mathrm{card}({#1})}

\newcommand{\width}[1]{\mathrm{wd}({#1})}

\newcommand{\vars}{\mathcal{V}}
\newcommand{\Vars}{\mathcal{X}}

\newcommand{\isdef}{\stackrel{\scalebox{0.5}{$\mathsf{def}$}}{=}}
\newcommand{\iffdef}{\stackrel{\scriptscriptstyle{\mathsf{def}}}{\iff}}
\newcommand{\interv}[2]{[{#1},{#2}]}
\newcommand{\tuple}[1]{\langle {#1} \rangle}

\newcommand{\set}[1]{\{ {#1} \}}

\newcommand{\pow}[1]{\mathrm{pow}({#1})}

\newcommand{\dom}[1]{\mathrm{dom}({#1})}
\newcommand{\img}[1]{\mathrm{img}({#1})}

\newcommand{\alphabet}{\mathbb{A}}

\newcommand{\auto}[2]{\mathcal{A}_{
    {#1}
    \ifthenelse{\equal{#2}{}}{}{,{#2}}
}}

\newcommand{\autsat}[2]{\mathcal{A}^{\scriptscriptstyle\mathsf{sat}}_{
    {#1}
    \ifthenelse{\equal{#2}{}}{}{,{#2}}
}}

\newcommand{\autcut}[2]{\mathcal{A}^{\scriptscriptstyle\mathsf{cut}}_{
    {#1}
    \ifthenelse{\equal{#2}{}}{}{,{#2}}
}}

\newcommand{\autcst}[2]{\mathcal{A}^{\scriptscriptstyle\mathsf{cst}}_{
    {#1}
    \ifthenelse{\equal{#2}{}}{}{,{#2}}
  }
}

\makeatletter
\newcommand*{\da@rightarrow}{\mathchar"0\hexnumber@\symAMSa 4B }
\newcommand*{\da@leftarrow}{\mathchar"0\hexnumber@\symAMSa 4C }
\newcommand*{\xdashrightarrow}[2][]{%
  \mathrel{%
    \mathpalette{\da@xarrow{#1}{#2}{}\da@rightarrow{\,}{}}{}%
  }%
}
\newcommand{\xdashleftarrow}[2][]{%
  \mathrel{%
    \mathpalette{\da@xarrow{#1}{#2}\da@leftarrow{}{}{\,}}{}%
  }%
}
\newcommand*{\da@xarrow}[7]{%
  \sbox0{$\ifx#7\scriptstyle\scriptscriptstyle\else\scriptstyle\fi#5#1#6\m@th$}%
  \sbox2{$\ifx#7\scriptstyle\scriptscriptstyle\else\scriptstyle\fi#5#2#6\m@th$}%
  \sbox4{$#7\dabar@\m@th$}%
  \dimen@=\wd0 %
  \ifdim\wd2 >\dimen@
    \dimen@=\wd2 %
  \fi
  \count@=2 %
  \def\da@bars{\dabar@\dabar@}%
  \@whiledim\count@\wd4<\dimen@\do{%
    \advance\count@\@ne
    \expandafter\def\expandafter\da@bars\expandafter{%
      \da@bars
      \dabar@
    }%
  }%
  \mathrel{#3}%
  \mathrel{%
    \mathop{\da@bars}\limits
    \ifx\\#1\\%
    \else
      _{\copy0}%
    \fi
    \ifx\\#2\\%
    \else
      ^{\copy2}%
    \fi
  }%
  \mathrel{#4}%
}
\makeatother

\newcommand{\store}{\mathfrak{s}}

\newcommand{\cardgeconstr}[1]{\mathsf{card}({#1})}
\newcommand{\cardconstr}[3]{\mathsf{card}_{{#2},{#3}}({#1})}

\newcommand{\fv}[1]{\mathrm{fv}({#1})}

\newcommand{\cmso}{$\mathsf{CMSO}$}

\newcommand{\mso}{\textsf{MSO}}

\newcommand{\Models}{\models}
\newcommand{\edgrel}[1]{\mathsf{edg}_{#1}}

\newcommand{\seplog}{\textsf{SL}}

\newcommand{\gl}{\textsf{GL}}

\newcommand{\sintfusion}[2]{\widetilde{\mathtt{IF}}({#1}\ifthenelse{\equal{#2}{}}{}{,{#2}})}

\newcommand{\lang}[1]{\mathcal{L}({#1})}
\newcommand{\langof}[2]{\mathcal{L}_{#1}({#2})}
\newcommand{\langoflimited}[3]{\mathcal{L}_{#1 \downarrow #2}({#3})}

\newcommand{\domrestrict}[2]{{#1}\!\!\downharpoonleft_{\scriptscriptstyle{#2}}}

\newcounter{index}

\newcommand{\graph}{G}
\newcommand{\graphsof}[1]{{\mathcal{G}^{\scriptscriptstyle{#1}}}}
\newcommand{\graphs}{\graphsof{}}

\newcommand{\vertices}{V}
\newcommand{\vertof}[1]{\vertices_{\scriptscriptstyle{#1}}}
\newcommand{\edgeof}[1]{\edges_{\scriptscriptstyle{#1}}}
\newcommand{\labels}{\lambda}
\newcommand{\labof}[1]{\labels_{\scriptscriptstyle{#1}}}
\newcommand{\edgerel}{\upsilon}
\newcommand{\edgerelof}[1]{\edgerel_{\scriptscriptstyle{#1}}}

\newcommand{\sources}{\xi}
\newcommand{\sourceof}[1]{\sources_{\scriptscriptstyle{#1}}}
\newcommand{\ksubst}[5]{{#1}[{#2}({#3}),\ldots,{#4}({#5})]}
\newcommand{\ksubsttwo}[5]{{#1}[{#2}({#3}),{#4}({#5})]}
\newcommand{\ksubstthree}[7]{{#1}[{#2}({#3}),\ldots,{#4}({#5}),{#6}({#7})]}
\newcommand{\ksubstfour}[9]{{#1}[{#2}({#3}),\ldots,{#4}({#5}),{#6}({#7}),\ldots,{#8}({#9})]}

\newcommand{\singlegraph}{\overline{\mathbf{1}}}

\newcommand{\grammar}{\Gamma}

\newcommand{\node}{\mathsf{node}}

\newcommand{\hr}{\textsf{HR}}
\newcommand{\ac}{\textsf{AC}}
\newcommand{\rules}{\mathcal{R}}
\newcommand{\nonterm}{\mathcal{U}}

\newcommand{\edges}{{E}}

\newcommand{\tree}{T}

\newcommand{\twof}[1]{\mathrm{tw}({#1})}
\newcommand{\etwof}[1]{\mathrm{etw}({#1})}

\newcommand{\pop}{\parallel}

\newcommand{\tedge}[2]{\overline{\mathbf{#1}}^{#2}}
\newcommand{\cgr}{\cong}
\newcommand{\expon}{q}
\newcommand*{\ruleA}{A}
\newcommand*{\ruleAPred}{\underline{A}}
\newcommand*{\ruleASet}{\mathcal{A}}
\newcommand*{\ruleB}{B}

\newcommand*{\ruleBSet}{\mathcal{B}}
\newcommand*{\ruleC}{C}
\newcommand*{\ruleCPred}{\underline{C}}
\newcommand*{\ruleCSet}{\mathcal{C}}

 \newcommand*{\stem}{\mathsf{stem}}
\newcommand*{\lasso}{\mathsf{lasso}}
\newcommand*{\rootSymb}{\mathsf{rt}}
\newcommand*{\rootsSymb}{\mathsf{fut}}
\newcommand*{\rootof}[1]{\rootSymb({#1})}
\newcommand*{\rootsof}[1]{\rootsSymb({#1})}
\newcommand*{\nrootsSymb}{\mathsf{nrts}}
\newcommand*{\nrootsof}[1]{\nrootsSymb({#1})}
\newcommand*{\langu}{\mathcal{L}}
\newcommand*{\iroot}{z}
\newcommand*{\internalSymb}{\mathsf{intern}}
\newcommand*{\internal}[1]{\internalSymb({#1})}
\newcommand*{\iroots}{\pi}
\newcommand*{\type}{m}
\newcommand*{\classSymb}{\mathsf{cl}}
\newcommand*{\classof}[1]{\classSymb({#1})} \newcommand*{\rmsources}[1]{\widehat{#1}}
\newcommand*{\atindex}{\mathsf{at}}
\newcommand*{\head}{\mathsf{hd}}
\newcommand*{\rootVar}{r}
\newcommand*{\position}{\mathsf{pos}}

\newcommand*{\rpos}{\mathsf{rpos}}
\newcommand*{\cpos}{\mathsf{cpos}}
\newcommand*{\graphofrule}{\mathsf{graph}}
\newcommand*{\vertedge}{\mathsf{vtx}} 
\newcommand*{\equalformula}{\mathsf{equal}} 

\usepackage{mathdots}

\newcommand{\edge}{\epsilon}

\title{Tree-Verifiable Graph Grammars}

\author{
Mark Chimes\inst{1}
\and
Radu Iosif\inst{2}
\and
Florian Zuleger\inst{1}
}

\institute{
   Technische Universtität Wien, Vienna, Austria\\
   \email{\{mark.chimes,florian.zuleger\}@tuwien.ac.at}
\and
   Université Grenoble Alpes, Grenoble, France\\
   \email{Radu.Iosif@univ-grenoble-alpes.fr}
 }

\authorrunning{Chimes, Iosif, Zuleger}
\titlerunning{Tree-Verifiable Grammars}

\begin{document}

\maketitle	


\begin{abstract}
Hyperedge-Replacement grammars (\hr) have been introduced by Courcelle in order to extend the notion of context-free sets from words and
trees to graphs of bounded tree-width. While for words and trees the
syntactic restrictions that guarantee that the associated languages of
words resp. trees are regular - and hence, \mso-definable - are known,
the situation is far more complicated for graphs. Here, Courcelle
proposed the notion of regular graph grammars, a syntactic restriction
of \hr\ grammars that guarantees the definability of the associated
languages of graphs in Counting Monadic Second Order Logic (\cmso).
However, these grammars are not complete in the sense that not every
\cmso-definable set of graphs of bounded tree-width can be generated
by a regular graph grammar. In this paper, we introduce a new
syntactic restriction of \hr\ grammars, called tree-verifiable graph
grammars, and a new notion of bounded tree-width, called embeddable
bounded tree-width, where the later restricts the trees of a
tree-decomposition to be a subgraph of the analyzed graph. The main
property of tree-verifiable graph grammars is that their associated
languages are \cmso-definable and that the have bounded embeddable
tree-width. We show further that they strictly generalize the regular
graph grammars of Courcelle. Finally, we establish a completeness
result, showing that every language of graphs that is \cmso-definable
and of bounded embeddable tree-width can be generated by a
tree-verifiable graph grammar.
\end{abstract}

\section{Introduction}
The notion of \emph{regular} word and tree languages is one of the pillars of the theory of formal languages because many equivalent representations of regular languages are known.
In particular, we have that Monadic Second Order Logic (\mso) over words resp. trees defines exactly the class of regular word resp. tree languages.
On the other hand, regular word resp. tree languages are a subset of the context-free word resp. tree languages, and regular word resp. tree grammars can be seen as syntactic restrictions of context-free grammars that ensure the regularity of the generated languages.
Regular languages enjoy many desirable properties, in particular, the following problems are known to be decidable:
\begin{inparaenum}[(1)]
\item \emph{emptiness}: is the language empty?
\item \emph{membership}: does a given word resp. tree belong to the language?   %
\item \emph{inclusion}: given two languages, is the first
  included in the second?
\end{inparaenum}
Because of the importance of regular languages and their decidability properties, it would be highly desirable to lift the connection between regular grammars and \mso\ to graphs, which are important for many branches of computing, such as e.g.,
static analysis \cite{10.1145/292540.292552}, databases and knowledge
representation \cite{AbitebouldBunemanSuciu00}, concurrency
\cite{DBLP:conf/birthday/2008montanari} and aspects of machine
learning, such as neural networks~\cite{4700287}.

In this paper, we study graph languages that lie at the intersection
of~\begin{inparaenum}[(i)]
\item \hr-\emph{context-free} and
\item \cmso-\emph{definable}
\end{inparaenum} classes. The sets of graphs in this class are,
at the same time~\begin{inparaenum}[(i)]
\item least solutions of a system of recursive equations using
  operations that substitute a hyper-edge of a graph by another graph,
  and
\item models of a formula that uses (quantification over) variables
  denoting combined sets of vertices and edges and relation symbols
  (from a fixed finite signature) describing the incidence relation of
  a graph. In particular, \cmso\ is the extension of \mso\ with atomic
  propositions that express constraints on the cardinality of a set,
  modulo a constant.
\end{inparaenum}
This intersection class has nice algorithmic properties, because
emptiness, membership and inclusion are all decidable. In particular,
the decidability of inclusion follows from a celebrated theorem proved
by Courcelle, stating that the satisfiability of a \cmso\ formula is
decidable over graphs of bounded tree-width, a parameter that,
intuitively speaking, measures the difference between a graph and a
tree encoding of that graph~\cite{journals/iandc/Courcelle90}.
We recall that the \mso-definability of a word language defined by a context-free grammar is, in general, undecidable~\cite{Greibach68}.
Hence, as for words and trees, we will put syntactic restrictions on \hr-grammars in order to ensure \cmso-definability.

An important precursor of our work is the class of \emph{regular
graph grammars} introduced by
Courcelle~\cite{journals/tcs/Courcelle91a} as a graph counterpart to the left- or right-recursive word grammars, which define regular, and, hence, \mso-definable~\cite{Buechi90} word languages. 
Regular graph grammars are a subclass of \hr-grammars.
In order to state some intuition on regular graph grammars, we first recall the \hr-grammars.
We note that \hr-grammars generalize context-free word and tree grammars to graphs of bounded tree-width.
\hr-grammars describe the construction of graphs, using zero or
more distinguished vertices, called \emph{sources}, labeled by numbers
$1, \ldots, n$. 
A \emph{substitution operation} replaces a
\emph{nonterminal edge} attached to vertices $v_1, \ldots, v_n$ in a
graph $H$ by a disjoint graph $K$ with sources $1, \ldots, n$, such
that the $i$-th source of $K$ is glued with $v_i$, for each
$i=1,\ldots,n$ and this is the only overlapping between the two
graphs. 
A \emph{parallel composition} is a special substitution that
glues the $i$-th sources of two graphs, for $i=1,\ldots,n$.
The rules of a regular graph grammar use two disjoint sets of
nonterminals, called \emph{recursive} and
\emph{non-recursive}. Intuitively, a recursive nonterminal can be
composed in parallel with any number of occurrences of the same
non-recursive terminal in a derivation, but a substitution may only generate non-recursive terminals. The rules of a regular graph grammar are defined by two
syntactic conditions: \begin{inparaenum}[(i)]
\item each two vertices of a graph $H$ in a substitution operation are
  connected by a path that crosses only terminal edges and internal
  (i.e., non-source) vertices, except for endpoints, and
\item the parallel composition involves at most one recursive
  nonterminal.
\end{inparaenum}
The graphs generated by regular graph grammars have the property that
each parse tree from the derivation of a graph lies embedded in the
structure of the graph and can be extracted by an \mso-definable
relation between the relational structures that represent
graphs. Then, for each set of graphs defined by the grammar there
exists an \mso\ formula that defines it. This formula is obtained
from~\begin{inparaenum}[(i)]
\item the \mso\ definition of the set of parse trees, the existence of
  which is guaranteed by the alternation between recursive and
  non-recursive nonterminals in the derivation, and
\item the definition (i.e., tuple of \mso\ formul{\ae}) of the
  relation that extracts the parse tree from the graph.
\end{inparaenum}
A natural question that arises is whether each set of graphs that is both \mso-definable and \hr-context-free can be defined by a regular graph grammar.

\paragraph{Our contributions} In this paper, we give this question a negative answer and suggest an alternative class that lies at the intersection of \cmso-definable and \hr-context-free classes, by introducing \emph{tree-verifiable grammars}. 
We demonstrate that tree-verifiable grammars strictly generalize the regular graph grammars of Courcelle, and establish a characterization result:
We show that a language of graphs is \cmso-definable
and of bounded embeddable tree-width, a notion we introduce below, if and only if this language can be generated by a tree-verifiable graph grammar.

The syntactic condition that defines tree-verifiable grammars is that each rule maps a nonterminal $u$ to a
graph operation, which is either~\begin{inparaenum}[(A)]
\item\label{it1:intro} a substitution described by a graph with a
  single terminal hyper-edge attached to an internal \emph{root} node
  and several \emph{future roots} that are distributed among the
  nonterminal edges (in this case $u$ is a
  non-recursive nonterminal),
\item\label{it2:intro} a parallel composition of $u$ with several
  occurrences of the same non-recursive nonterminal, or
\item\label{it3:intro} a parallel composition of non-recursive
  nonterminals ($u$ is a recursive nonterminal in the
  (\ref{it2:intro}) and (\ref{it3:intro}) cases).
\end{inparaenum}

We define the \emph{embeddable tree-width} of a graph to be the
minimal width of a tree decomposition whose backbone is isomorphic to
some spanning tree of the graph. The embeddable tree-width of a graph
is, in general, greater than its tree-width, i.e., there are graphs
for which the optimal tree decomposition is not isomorphic to a
spanning tree. 
We show that this stronger notion allows us to precisely capture that class of graphs generated by tree-verifiable grammars: 
We establish that a language of graphs is \cmso-definable
and of bounded embeddable tree-width if and only if this language can be generated by a tree-verifiable graph grammar.

In the special case of trees (i.e., graphs obtained by restricting the parallel composition to graphs with one root source) tree-verifiable grammars define precisely the \cmso-definable unranked sets of trees.
Moreover, if all the repetition counts from the rules
(\ref{it2:intro}) equal one, tree-verifiable grammars are the same as the \mso-definable unranked sets of trees. 
This settles positively another question relative to the definition of associative-commutative (\ac) tree grammars of Courcelle~\cite[Definition 3.7]{journals/tcs/Courcelle91a}, namely whether \ac\ grammars define all the \mso-definable sets of unranked trees.

\paragraph{Related work}
The intersection of the \cmso-definable and \hr-context-free sets of
graphs has been investigated in the seminal paper of
Courcelle~\cite{journals/tcs/Courcelle91a} that introduced regular
graph grammars~\cite[\S5]{journals/tcs/Courcelle91a} and proves that
the set of series-parallel (i.e., graphs with sources $1$ and $2$ that
can be either serialized or composed in parallel) belongs to this
class graphs~\cite[\S6]{journals/tcs/Courcelle91a}. More recently,
Doumane refined this result by developping a language of regular
expressions that describes the class of \cmso-definable graphs that
can be built using at most two
sources~\cite{DBLP:conf/icalp/Doumane22}. An interesting question that
remains open, for the time being, is whether the sets of graphs
defined by these regular expressions have bounded embeddable
tree-width. It is known that their optimal tree decompositions can be
extracted from the structure of the graphs in a canonical fashion, but
these canonical decompositions do not correspond to spanning trees.

Substructural logics~\cite{Paoli2002} are first-order logics extended
with a non-idempotent \emph{separating conjunction} operator that
decomposes logical structures. A distinguished substructural logic is
Separation Logic (\seplog) \cite{Ishtiaq00bias,Reynolds02}, that is
interpreted over finite partial functions of fixed arity, called
\emph{heaps}. With heaps one can model certain graphs of bounded
degree, i.e., where the number of edges attached to a vertex is bounded by a constant.
Another model of composition, used in Graph Logic (\gl) of
Cardelli et al \cite{Cardelli2002Spatial} or the Separation Logic of
Relations~\cite{10.1007/978-3-540-27864-1_26}, splits graphs into
subgraphs with disjoint sets of edges. These logics are more general
than \seplog, because they describe classes of graphs of unbounded
degree (e.g., stars and unranked sets of trees). Used in combination
with recursive definitions, the separating conjunction becomes a
powerful means to describe shape-related graph properties (lists,
trees, stars, etc.).

One of the first fragments of \seplog\ with recursive definitions
having a decidable entailment problem used an ad-hoc translation into
equivalent \mso\ formul{\ae}, together with a syntactic guarantee of
tree-width boundedness~\cite{DBLP:conf/cade/IosifRS13}. 
In subsequent work, a decision procedure, not based on \mso, for the entailment problem in this \seplog\ fragment was proposed and implemented ~\cite{conf/tacas/KatelaanMZ19,conf/lpar/KatelaanMNZ18} and the complexity of this entailment problem was established~\cite{conf/lpar/EchenimIP20,journals/tocl/MathejaPZ23}.
The ideas used in the definition of the \seplog\ fragment motivated the definition of tree-verifiable grammars in this paper.
In particular, considering substitution
operations described by graphs with a single edge that connects the
root to the future roots is similar to the \emph{connectivity}
condition from~\cite{DBLP:conf/cade/IosifRS13}. Moreover, distributing
the future roots among the nonterminals guarantees that each future
root will eventually become the root of a subgraph, which corresponds
to the \emph{establishment} condition
of~\cite{DBLP:conf/cade/IosifRS13}. These conditions guarantee that
the \seplog-defined heaps of~\cite{DBLP:conf/cade/IosifRS13} have
bounded embeddable tree-width.
We note that the fragment of~\cite{DBLP:conf/cade/IosifRS13} is restricted to degree-bounded heaps and that~\cite{DBLP:conf/cade/IosifRS13} did not attempt to achieve any completeness resp. characterization result that corresponds to the proposed syntactic restriction.

\section{Preliminaries}
The set of natural numbers is denoted by $\nat$. Given $i,j\in\nat$,
we write $\interv{i}{j}$ for the set $\set{i,i+1,\ldots,j}$, assumed
to be empty if $i>j$.  The cardinality of a finite set $A$ is denoted
by $\cardof{A}$.  For a set $A$, we denote by $\pow{A}$ its powerset,
$A^0 \isdef \set{\epsilon}$, $A^{i+1} \isdef A^i \times A$, for all
$i \geq 0$, $A^* \isdef \bigcup_{i\geq0} A^i$ and $A^+ \isdef
\bigcup_{i\geq1} A^i$, where $\times$ is the Cartesian product and
$\emptyseq$ denotes the empty sequence. Intuitively, $A^*$
(resp. $A^+$) denotes the set of possibly empty (resp. nonempty)
sequences of elements from $A$.  The length of a sequence $\alpha \in
A^*$ is denoted by $\lenof{\alpha}$ and its $i$-th element by
$\alpha_i$, for $i \in \interv{1}{\lenof{\alpha}}$. For a function $f:
A \rightarrow B$, we denote its \emph{domain} by $\dom{f} = A$ and its
image by $\img{f} \isdef \set{b \in B \mid f(a)=b \text{ for some }
a \in A}$. The domain-restriction of a function $f$ to a set $C$ is
denoted as $\domrestrict{f}{C} \isdef \set{ (a,f(a)) \mid a \in C}$.



\paragraph{Graphs}
in this paper are multigraphs consisting of edges of arity one or
more. Let $\alphabet$ be a finite alphabet of edge labels, ranged
over by $a$ and equipped with arities $\arityof{a}\geq 1$.

\begin{definition}\label{def:c-graph}
A \emph{concrete graph (c-graph) of type $n\geq0$} is a tuple $\graph
= \tuple{\vertof{\graph}, \edgeof{\graph}, \labof{\graph},
  \edgerelof{\graph}, \sourceof{\graph}}$, where: \begin{compactitem}
\item $\vertof{\graph}$ is a finite set of \emph{vertices},
\item $\edgeof{\graph}$ is a finite set of \emph{edges}, such that
  $\vertof{\graph}\cap\edgeof{\graph}=\emptyset$,
\item $\labof{\graph} : \edgeof{\graph} \rightarrow \alphabet$ is a
  mapping that defines the labels of the edges,
\item $\edgerelof{\graph} : \edgeof{\graph} \rightarrow
  \vertof{\graph}^+$ is a mapping that associates each edge a nonempty
  sequence of vertices attached to the edge, such that
  $\arityof{(\labof{\graph}(e))} = \lenof{\edgerelof{\graph}(e)}$, for
  each $e \in \edgeof{\graph}$,
\item $\sourceof{\graph} : \interv{1}{n} \rightarrow \vertof{\graph}$
  is an injective mapping that designates $n$ distinct
  \emph{sources}. The vertex $\sourceof{\graph}(i)$ is the
  \emph{$i$-th source} of $\graph$.
  A vertex $v \in \vertof{\graph}$ is \emph{internal} if $v \not\in \img{\sourceof{\graph}}$.
  By $\internal{\graph} = \vertof{\graph} \setminus \{\sourceof{\graph}(i) \mid i \in [1,n]\}$ we denote the internal nodes of $\graph$.
  For simplicity, we omit the sources for graphs of type $0$.
\end{compactitem}
We denote by $\graphsof{\alphabet}{}$ the set of c-graphs with edge
labels from $\alphabet$.
\end{definition}
\noindent For example, Fig. \ref{fig:graphs} (a,b) shows several
c-graphs with (numbered) sources depicted in black and internal
vertices in white. A \emph{path} from $v_1$ to $v_{k+1}$ in a c-graph
$\graph$ is a sequence of tuples:
\[\tuple{v_1,i_1,e_1,v_2,j_1}, \tuple{v_2,i_2,e_2,v_3,j_2},
\ldots, \tuple{v_k,i_k,e_k,v_{k+1},j_k}\]
where $v_1,\ldots,v_{k+1} \in \vertof{\graph}$, $e_1, \ldots,
e_k \in \edgeof{\graph}$ and $v_\ell$ and $v_{\ell+1}$ are the
$i_\ell$-th and $j_\ell$-th vertices attached to $e_\ell$,
respectively, for all $\ell \in \interv{1}{k}$.
A set $C \subseteq \vertof{\graph}$ is \emph{connected} in $\graph$
iff for any $u,v\in C$ there exists a path from $u$ to $v$ in $\graph$
containing only vertices from $C$.

Two c-graphs are \emph{disjoint} iff their sets of vertices and edges
are disjoint, respectively.  Two c-graphs are \emph{isomorphic} iff
there exists a bijection between their vertices and edges that
preserves the edge labels, the sequences of vertices attached to edges
and the sources. A \emph{graph} is the equivalence class of a c-graph
for isomorphism.  For any graph $\graph$ we denote by
$\rmsources{\graph}$ the graph of type $0$ that is obtained from
$\graph$ by removing its sources. A language $\langu$ of graphs is
some set of graphs.  We say $\langu$ is of type $n$ if all graphs
$\graph \in \langu$ are of type $n$. By $\rmsources{\langu}
= \{\rmsources{\graph} \mid \graph \in \langu \}$ we denote the
language of type $0$ that is obtained by removing the sources of the
graphs of $\langu$.

\paragraph{Graph Operations and Grammars.}
If $\graph$ is a c-graph, $e \in \edgeof{\graph}$ is an edge such that
$\arityof{\labof{\graph}}(e)=m$ and $H$ is a c-graph of type $m$
disjoint from $\graph$, then the \emph{substitution} $\graph[e/H]$
deletes the edge $e$ from $\graph$, adds $H$ to $\graph$ and joins the
$i$-th vertex from $\edgerelof{\graph}(e)$ with the $i$-th source of
$H$, for all $i \in \interv{1}{m}$.  The sources and type of
$\graph[e/H]$ are the same as for $\graph$. Fig. \ref{fig:graphs} (a)
shows an example of substitution, formally defined below:

\begin{figure}[t!]
  \centerline{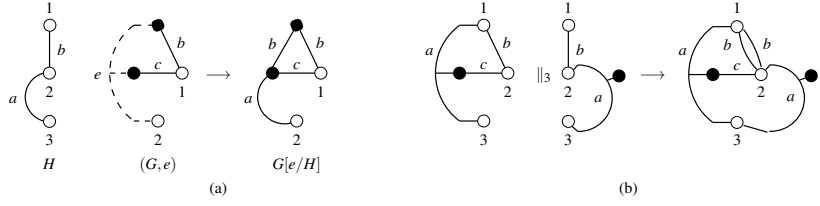}

  \vspace*{-.5\baselineskip}
  \caption{Graph operations: substitution (a) parallel composition (b)}
  \label{fig:graphs}
  \vspace*{-.5\baselineskip}
\end{figure}

\begin{definition}[Substitution]
  Let $\graph$ be a c-graph of type $n$, $e \in \edgeof{\graph}$ is an
  edge such that $\edgerelof{\graph}(e) = \tuple{v_1, \ldots, v_m}$
  and $H$ is a c-graph of type $m$ disjoint from $\graph$.  Let
  $\sim \ \subseteq (\vertof{\graph} \cup \vertof{H})^2$ be the least
  equivalence relation such that $v_i \sim \sourceof{H}(i)$ for all
  $i \in [1,m]$.  We denote by $[u]_\sim$ the $\sim$-equivalence class
  of $u \in \vertof{\graph} \cup \vertof{H}$.  Then $\graph'
  = \graph[e/H]$ is defined by setting: \begin{compactitem}
        \item $\vertof{\graph'} = \{[u]_\sim \mid u \in \vertof{\graph} \cup \vertof{H} \}$,
        \item $\edgeof{\graph'} = \edgeof{\graph} \setminus \{e\} \cup \edgeof{H}$,
        \item $\labof{\graph'}\isdef \domrestrict{(\labof{\graph})} {(\edgeof{\graph} \setminus \set{e})} \cup~ \labof{H}$,
        \item
        $\edgerelof{\graph'}(e') \isdef \tuple{[u_1]_\sim, \ldots,
        [u_k]_\sim}$, for all $e' \in \edgeof{\graph} \setminus \{e\}$
        with $\edgerelof{\graph}(e') = \tuple{u_1, \ldots, u_k}$,
        resp. all $e' \in \edgeof{H}$ with $\edgerelof{H}(e')
        = \tuple{u_1, \ldots, u_k}$,
        \item
          $\sourceof{\graph'}(i) \isdef
          [\sourceof{\graph}(i)]_\sim$, for all $i \in \interv{1}{n}$.
    \end{compactitem}
    The substitution $\graph[e/H]$ is lifted from c-graphs to graphs
    (one can always chose c-graphs that are isomorphic to $\graph$ and $H$ and are disjoint).
    Then, the graph $\graph[e/H]$ is defined as the isomorphism class of the c-graph that
    results from the substitution involving these c-graphs.
\end{definition}

The graph operation $(\graph, e_1, \ldots, e_k)$, where $e_1, \ldots,
e_k \in \edgeof{\graph}$ are pairwise distinct edges, takes pairwise
distinct c-graphs $H_1, \ldots, H_k$ of types
$\arityof{\labof{\graph}(e_1)}, \ldots, \arityof{\labof{\graph}(e_k)}$,
respectively, and returns the c-graph $\graph[e_1/H_1, \ldots,
e_k/H_k]$, where the substitutions can be done in any order.  Again
this operation can be lifted from concrete graphs to graphs.

We now introduce an important special case of a graph operation:
The \emph{parallel composition} $\parallel_n$ denotes the operation
$(\graph, e_1, e_2)$, where $\graph$ has exactly $n$ vertices, all of
which are sources, and $e_1$ and $e_2$ are the only edges of $\graph$,
both of arity $n$ and attached to all $n$ sources.
Intuitively, $\graph_1 \parallel_n \graph_2$ denotes the graphs that can be obtained from two graphs $\graph_1$ and
$\graph_2$ of type $n$ by merging their sources. Fig. \ref{fig:graphs}
(b) shows an example of parallel composition.  Note that the
composition operation $\parallel_n$ is both commutative and
associative.

\begin{figure}[t!]
  \centerline{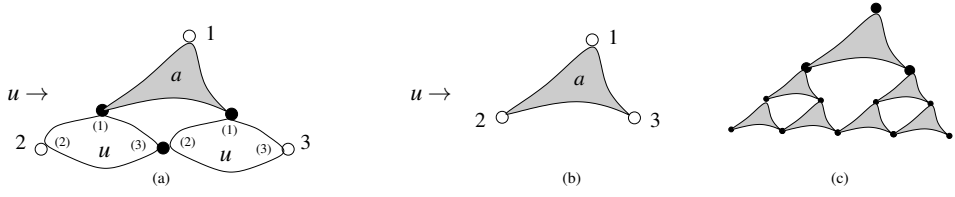}

  \vspace*{-.5\baselineskip}
  \caption{Graph Grammar for Trees with Linked Leaves}
  \label{fig:tll}
  \vspace*{-.5\baselineskip}
\end{figure}

A \emph{graph grammar} is a pair $\grammar = (\nonterm,\rules)$, where
$\nonterm$ is a set of \emph{nonterminals} distinct from $\alphabet$,
ranged over $u,v,w$ with associated arities $\arityof{u}\geq1$, and
$\rules$ is a finite set of rules of one of the following forms, with
$u,v,w\in \nonterm$:
\begin{compactenum}
\item $u \rightarrow v_1 \parallel_n \ldots \parallel_n v_k$, where
  $u,v_1,\ldots,v_k\in\nonterm$ are nonterminals of arity $n$,
\item $u \rightarrow \ksubst{\graph}{e_1}{v_1}{e_k}{v_k}$, where
  $u,v_1,\ldots,v_k\in\nonterm$ are nonterminals and
  $\ksubst{\graph}{e_1}{v_1}{e_k}{v_k}$ denotes the graph operation
  $(\graph,e_1,\ldots,e_k)$, for some c-graph $\graph$ of type
  $\arityof{u}$, such that $\img{\labof{\graph}} \subseteq \alphabet
  \cup \set{v_1,\ldots,v_k}$ and $\labof{\graph}(e) = v_i \iff e =
  e_i$, for all $e \in \edgeof{\graph}$; if $\labof{\graph}(e) \in
  \nonterm$ (resp. $\labof{\graph}(e) \in \alphabet$), we say that $e$
  is a nonterminal (resp. terminal) edge of $\graph$,
\item $\rightarrow u$, for some nonterminal $u \in \nonterm$, is
  called an \emph{axiom}.
\end{compactenum}
We note that in rules of shape 1., we explicitly allow $k=0$.
With this we denote the rule $u \rightarrow \singlegraph_n$, where $\singlegraph_n$ denotes the graph of type $n$ that consists of $n$ vertices (all of which are sources), and that has no edges.
The nonterminal on the left-hand side of $\rightarrow$ in a non-axiom
rule is called the \emph{head} of that rule. The \emph{language of
  $\grammar$ associated with nonterminal $u$} is the set
$\langof{\grammar}{u}$ of graphs corresponding to $u$ in the unique
least solution\footnote{By Tarski's fixpoint theorem, since the graph
  operations are monotonic (w.r.t. to inclusion of sets) and
  continuous (w.r.t. infinite unions of sets), the system of recursive
  equations defined by the grammar has a unique least solution.} of
the system of recursive equations defined by the grammar rules, with
nonterminals interpreted as sets, operations lifted to sets and
$\rightarrow$ interpreted as the right-to-left set inclusion
$\supseteq$. Note that each $\langof{\grammar}{u}$ is a language of
type $\arityof{u}$. The \emph{language of $\grammar$}, denoted by
$\lang{\grammar}$, is the union of the sets
$\rmsources{\langof{\grammar}{u}}$ for all axioms $\rightarrow u$ of
$\grammar$. Note that $\lang{\grammar}$ is a language of type $0$.  A
set $\langu$ of graphs is \emph{hyperedge-replacement} (\hr) iff
$\langu=\lang{\grammar}$ for some graph grammar $\grammar$.

\begin{example}\label{ex:tll}
Fig. \ref{fig:tll} shows a graph grammar with a single nonterminal
$u$, of arity $3$, and two rules. The first rule (a) is of the form $u
\rightarrow \ksubsttwo{\graph}{e_1}{u}{e_2}{u}$, where $\graph$ is a
c-graph with a (grey-filled) terminal edge labeled by $a$, where
$\arityof{a}=3$, and two nonterminal (unfilled) edges $e_1$ and $e_2$,
both labeled by $u$.  The order of the vertices attached to these
edges is (1) middle, (2) left and (3) right. The second rule (b) is of
the form $u \rightarrow \graph$, where $\graph$ consists of a single
terminal edge labeled by $a$. The language of this grammar consists of
all graphs of the form depicted in Fig. \ref{fig:tll} (c).
\end{example}

\paragraph{Monadic Second Order Logic with Counting}
(\cmso) is the set of formul{\ae} written in the syntax from
Fig. \ref{fig:logics} (a) using a set $\vars = \set{x,y,\ldots}$ of
\emph{first-order variables}, a set $\Vars = \set{X,Y,\ldots}$ of
\emph{second-order variables} and relation symbols $\edgrel{a}$ of
arity $\arityof{a}+1$, for all $a \in \alphabet$. By \mso\ we denote
the subset of \cmso\ formul{\ae} consisting of formul{\ae} that do not
contain atomic subformul{\ae} of the form $\cardconstr{X}{q}{p}$, also
called \emph{cardinality constraints}.  A variable is \emph{free} if
it occurs outside the scope of any quantifier. The set of free
variables of a formula $\psi$ is denoted by $\fv{\psi}$.  A
\emph{sentence} is a formula without free variables.

\begin{figure}[t!]
  \begin{center}
    \[\psi := x=y \mid \edgrel{a}(x_1, \ldots, x_{\arityof{a}+1}) \mid \cardconstr{X}{q}{p} \mid X(x) \mid
    \neg\psi \mid \psi \wedge \psi \mid \exists x ~.~ \psi \mid \exists X
    ~.~ \psi\]

    \vspace*{-2mm}(a) \cmso\ Syntax

    \vspace*{-4mm}
    \[\begin{array}{rclcl}
    \graph & \Models^\store & x=y & \iff & \store(x)=\store(y) \\
    \graph & \Models^\store & \edgrel{a}(x_1, \ldots, x_{\arityof{a}+1}) & \iff &
    \store(x_1) \in \edgeof{\graph},~ \labof{\graph}(\store(x_1))=a \text{ and }
    \edgerelof{\graph}(\store(x_1))=\tuple{\store(x_2), \ldots, \store(x_{\arityof{a}+1})} \\
    \graph & \Models^\store & \cardconstr{X}{q}{p} & \iff & \cardof{\store(X)} = kq+p \text{ for some } k\in\nat \\
    \graph & \Models^\store & X(x) & \iff & \store(x) \in \store(X) \\
    \graph & \Models^\store & \neg \phi & \iff & \graph \not\Models^\store \phi\\
    \graph & \Models^\store & \phi_1 \land \phi_2 & \iff & \graph \Models^\store \phi_1 \text{ and }  \graph \Models^\store \phi_2 \\
    \graph & \Models^\store & \exists x ~.~ \psi & \iff & \graph \Models^{\store[x\leftarrow u]} \psi
    \text{, for an element } u \in \vertof{\graph} \cup \edgeof{\graph} \\
    \graph & \Models^\store & \exists X ~.~ \psi & \iff & \graph \Models^{\store[X\leftarrow U]} \psi
    \text{, for a set } U \subseteq \vertof{\graph} \cup \edgeof{\graph}
    \end{array}\]

    \vspace*{-2mm}(b) \cmso\ Semantics
  \end{center}
  \vspace*{-\baselineskip}
  \caption{Monadic Second Order Logic with Counting (\cmso)}
  \label{fig:logics}
  \vspace*{-\baselineskip}
\end{figure}

We define the semantics of \cmso, for graphs of type $0$, by a
satisfaction relation $\graph \Models^\store \psi$ between graphs and
formul{\ae}, where the valuation $\store : \vars \cup \Vars \rightarrow
\vertof{\graph}\cup\edgeof{\graph} \cup
\pow{\vertof{\graph}\cup\edgeof{\graph}}$ maps each variable $x \in
\vars$ to a vertex or and edge, i.e.,
$\store(x)\in\vertof{\graph}\cup\edgeof{\graph}$, and each variable $X
\in \Vars$ to a finite subset of vertices and edges, i.e.,
$\store(X)\subseteq\vertof{\graph}\cup\edgeof{\graph}$. The
satisfaction relation is defined inductively on the structure of
formul{\ae} in Fig. \ref{fig:logics} (b).
If $\psi$ is a sentence, the satisfaction relation does not depend on the variable
mapping and we write $\graph \Models \psi$ instead of $\graph
\Models^\store \psi$.
A language $\langu$ of graphs of type $0$ is
\emph{\cmso-definable} iff there exists a \cmso\ sentence $\psi$ such
that $\graphs = \set{\graph \mid \graph \Models \psi}$.

We define the formula
$e \in \edges \iffdef \bigvee_{a \in \alphabet} \exists x_1, \ldots, x_{\arityof{a}}. \ \edgrel{a}(e, x_1, \ldots, x_{\arityof{a}})$, which denotes that $e$ is an edge, and the formula $v \in \vertices \iffdef \bigvee_{a \in \alphabet} \exists e, x_1, \ldots, x_{\arityof{a}}. \ \edgrel{a}(e, x_1, \ldots, x_{\arityof{a}}) \wedge \bigvee_{1 \le i \le \arityof{a}} x_i = v$, which denotes that $v$ is a vertex.
As usual we define $\phi \rightarrow \psi \iffdef \neg \phi \lor \psi$ and $\phi \leftrightarrow \psi \iffdef \phi \rightarrow \psi \wedge \psi \rightarrow \phi$.
For $\Delta = \edges,\vertices$, we further define $\forall x \in \Delta  ~.~ \psi(x) \iffdef \forall x  ~.~ x \in \Delta \rightarrow \psi(x)$, $\exists x \in \Delta  ~.~ \psi(x) \iffdef \exists x  ~.~ x \in \Delta \wedge \psi(x)$, $\exists! x \in \Delta ~.~ \psi(x) \iffdef
\exists x \in \Delta ~.~ \psi(x) \wedge \forall y \in \Delta ~.~ \psi(y)
\rightarrow y=x$, and $\exists X \subseteq \Delta ~.~ \psi \iffdef \exists X
~.~ \left( \forall x ~.~  X(x) \rightarrow x \in \Delta \right) \wedge \psi$.

\paragraph{Recognizable Sets of Graphs.}
A congruence relation on graphs is an equivalence relation $\cgr$ such
that~\begin{inparaenum}[(1)]
\item any two equivalent graphs are of the same type, and
\item for every graph $H$, for every edge $e$ of $H$, for every graph
  $\graph$ of type $\arityof{(\labof{H}(e))}$, and every $\graph' \cgr
  \graph$, one has $H[e/\graph] \cgr H[e/\graph']$.
\end{inparaenum}
Such a congruence is called \emph{locally-finite} iff it has finitely
many equivalence classes of each type (by the definition of
congruences, each equivalence class consists of graphs of the same
type). A set of graphs $\langu$ is \emph{recognizable} iff there
exists a locally-finite congruence $\cgr$ such that $\graph \cgr
\graph'$ only if $\graph \in \langu$ iff $\graph' \in \langu$.

\paragraph{Trees.}
We introduce trees as a subset of the set of graphs of type 1, whose
only source denotes the root. We consider the following restricted
graph operations. Let $\singlegraph$ be the graph that consists of a
singleton vertex, which is also its only source, and that has no edges.
For an edge label $a \in \alphabet$ and $1 \le i \le \arityof{a}$
denote by $\tedge{a}{i}$ the operation
$(\graph,e_1,\ldots,e_{i-1},e_{i+1},\ldots,e_{\arityof{a}})$, where
$\graph$ is a c-graph that consists of a singleton edge $e$ labelled
by $a$ attached to pairwise distinct vertices
$v_1,\ldots,v_{\arityof{a}}$, such that $\sourceof{\graph}(1)=v_i$ and
the other vertices $v_1, \ldots, v_{i-1}, v_{i+1}, \ldots,
v_{\arityof{a}}$ are attached to distinct edges $e_1, \ldots, e_{i-1},
e_{i+1}, \ldots, e_{\arityof{a}}$ labeled with unary symbols,
respectively. Intuitively,
$\tedge{a}{i}(\graph_1,\ldots,\graph_{i-1},\graph_{i+1},
\ldots,\graph_{\arityof{a}})$ is obtained by taking the disjoint union
of the graphs $\graph_1,\ldots,\graph_{i-1},\graph_{i+1},
\ldots,\graph_{\arityof{a}}$ and adding a new edge labelled by $a$
with sequence of vertices $(\sourceof{\graph_1}(1),\ldots,
\sourceof{\graph_{i-1}}(1),v,\sourceof{\graph_{i+1}}(1),\ldots,
\sourceof{\graph_{\arityof{a}}}(1))$, where $v$ is a new vertex that
becomes the only source of the resulting graph. This operation is
depicted in Fig. \ref{fig:trees} (a). The parallel composition
operation for trees is $\parallel_1$, depicted in Fig. \ref{fig:trees}
(b).

The set of trees over $\alphabet$ is defined as the least set closed
under the operations $\singlegraph$, $\parallel_1$ and $\tedge{a}{i}$,
for all $a \in \alphabet$ and $1 \le i \le \arityof{a}$. We retrieve
the standard terminology for trees from this definition. The vertices
of a tree $\tree$ are called \emph{nodes}. For an edge $e \in
\edgeof{\tree}$ introduced by an operation $\tedge{a}{i}$, we say that
$\edgerelof{\tree}(i)$ is the \emph{parent} of $\edgerelof{\tree}(1),
\ldots, \edgerelof{\tree}(i-1), \edgerelof{\tree}(i+1), \ldots,
\edgerelof{\tree}(\arityof{a})$ and the latter nodes are the
\emph{children} of the former. Note that the trees we consider in this
paper are \emph{unordered}, i.e., the relative order of the edges
attached to a node is not important, and \emph{unranked}, i.e., there
is no bound on the number of children of a node.

\begin{figure}[t!]
  \centerline{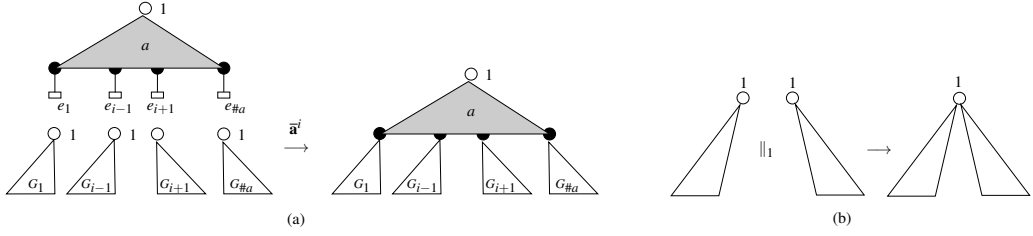}

  \vspace*{-.5\baselineskip}
  \caption{Tree operations: substitution (a) parallel composition (b)}
  \label{fig:trees}
  \vspace*{-.5\baselineskip}
\end{figure}




\section{Tree Grammars}
\newcommand{\phiofgrammar}[1]{\phi_{\langof{\grammar}{#1}}}
\newcommand{\phiofgrammarlimited}[2]{\phi_{\langoflimited{\grammar}{#1}{#2}}}
\newcommand{\psiofgrammar}[1]{\psi_{\langof{\grammar}{#1}}}
\newcommand{\psiofgrammarlimited}[2]{\psi_{\langoflimited{\grammar}{#1}{#2}}}
\newcommand{\deriv}{D}
\newcommand{\ruleAShape}{\ref{it1:regular-tree-grammar}}
\newcommand{\ruleBShape}{\ref{it2:regular-tree-grammar}}
\newcommand{\ruleCShape}{\ref{it3:regular-tree-grammar}}

\newcommand\dtwidth{0.9}
\newcommand\dtcolor{green!20!white}
\newcommand\dtcolorlight{green!5!white}

\usetikzlibrary{shapes}

This section is concerned with unranked trees, whose nodes may have unbounded numbers of
children, the order of which is not important.
We are interested in \hr\ sets of trees, generated by \emph{tree grammars}, which are graph grammars, whose rules use only tree operations.
In general, \hr\ sets of trees are not \cmso-definable.
For instance, the grammar with rules $u \rightarrow \tedge{a}{1}
\parallel_1 \tedge{b}{1} \parallel_1 u$ and $u \rightarrow
\singlegraph$, where $a$ and $b$ are binary edge labels, defines the
set of trees of height one, whose root has the same number of $a$- and
$b$-labeled edges. Since \cmso\ cannot describe sets of equal
cardinality (see, e.g.,~\cite[Proposition
  3.9]{journals/iandc/Courcelle90}), this set is not \cmso-definable.

This motivates the quest for trees grammars defined by easy-to-check
syntactic restrictions, that guarantee the \cmso-definability of their
languages. The following definition introduces such a class of tree
grammars. In the rest of this section, we prove that the languages of
the grammars that conform with the following definition coincide with
the \cmso-definable tree languages.

\begin{definition}
\label{def:regular-tree-grammar}
  A tree grammar $\grammar = (\nonterm,\rules)$ is \emph{regular} iff
  there exists a set $\mathcal{W} \subseteq \nonterm$ of nonterminals
  such that the rules in $\rules$ are of one of the following forms,
  either:~\begin{compactenum}[A.]
  \item\label{it1:regular-tree-grammar} $w \rightarrow
    \tedge{a}{i}(u_1,\ldots,u_{\arityof{a}-1})$, for some $a \in
    \alphabet$ and $1 \le i \le \arityof{a}$, where $w\in\mathcal{W}$
    and $u_1,\ldots,u_{\arityof{a}-1} \in \nonterm$,
  \item\label{it2:regular-tree-grammar} $u \rightarrow u
    \parallel_1 w^\expon$, for some $\expon \in \nat$, where $u \in \nonterm\setminus\mathcal{W}$ and $w \in \mathcal{W}$,
  \item\label{it3:regular-tree-grammar} $u \rightarrow w_1 \parallel_1
    \cdots \parallel_1 w_k$, where $u \in
    \nonterm\setminus\mathcal{W}$ and $w_1,\ldots,w_k\in\mathcal{W}$;
    this rule becomes $u \rightarrow \singlegraph$, in case $k=0$.
  \end{compactenum}
  We further require that the axioms are of the form $\rightarrow u$,
  for some $u \in \nonterm \setminus \mathcal{W}$.
\end{definition}

We call a set $\langu$ of trees \emph{regular} iff there is a regular
tree-grammar $\grammar$ and such that $\langu=\lang{\grammar}$.  The
shape of regular tree-grammars as given in
Definition~\ref{def:regular-tree-grammar} is inspired by the AC-tree
grammars introduced by
Courcelle~\cite[Section~3]{journals/tcs/Courcelle91a}. However,
Courcelle's definition of AC-grammars considers $\expon = 1$ in all
rules of the form rules~\ref{it2:regular-tree-grammar} in
Definition~\ref{def:regular-tree-grammar}. In this section, we show
that the languages of these grammars coincide with the \mso-definable
tree languages. Hence, the AC-grammars are not complete wrt
\cmso-definablility, because there are tree languages that are
\cmso-definable but not \mso-definable, see e.g.,~\cite[Corollary
  6.6]{journals/iandc/Courcelle90}.

First, we prove that the \cmso-definable tree-languages are exactly
the regular tree languages. For this we will make use of the following
lemma:

\begin{lemmaE}\label{lemma:singleB}
  Given a regular tree-grammar $\grammar = (\nonterm,\rules)$ with a
  set $\mathcal{W}$ as in Definition \ref{def:regular-tree-grammar},
  we can build a regular tree-grammar $\grammar' =
  (\nonterm',\rules')$ with $\lang{\grammar} = \lang{\grammar'}$, such
  that, for each pair of non-terminals $u \in
  \nonterm\setminus\mathcal{W}$ and $w \in \mathcal{W}$, there is at
  most one rule $u \rightarrow u \parallel_1 w^\expon$ of shape
  \ruleBShape\ in $\rules'$ .
\end{lemmaE}
\begin{proofE}
  Let $u$ and $w$ be fixed nonterminals, with $u \in \nonterm
  \setminus \mathcal{W}$ and $w \in \mathcal{W}$ and $u \rightarrow u
  \parallel_1 w^{q_1}$, $\ldots$, $u \rightarrow u \parallel_1
  w^{q_k}$ be the set of rules of shape \ruleBShape\ of
  $\rules$, for the given nonterminals. Let $d\geq1$ be the gcd of the
  integers $q_1, \ldots, q_k \geq 1$. By Schur's theorem, there exists
  a computable integer $n\geq1$ such that for each integer $x\geq n$
  there exist $x_1,\ldots,x_k \in \nat$ such that $d x = q_1 x_1 +
  \ldots + q_k x_k$. We further observe that the set $M \isdef \set{q_1 x_1 + \ldots
    + q_k x_k \mid x_1,\ldots,x_k \in \nat} \cap \interv{1}{dn-1}$ is
  computable. Let $u', u''$ be fresh nonterminals, not in $\nonterm$. We
  define $\nonterm'\isdef\nonterm\cup\set{u',u''}$.
  We now obtain the set of rules $\rules'$ from $\rules$ as follows:
  For every rule $w \rightarrow \tedge{a}{i}(u_1,\ldots,u_{\arityof{a}-1})$ with some $u_j = u$ we replace this rule by two copies, one where we replace $u_j$ with $u'$ and one where we replace $u_j$ with $u''$.
  We keep all the rules $w \rightarrow \tedge{a}{i}(u_1,\ldots,u_{\arityof{a}-1})$ with $u_j \neq u$ for all $j$.
  We remove the rules $u \rightarrow u \parallel_1 w^{q_1}, \ldots, u
  \rightarrow u \parallel_1 w^{q_k}$ and add the rule $u'' \rightarrow u'' \parallel_1 w^\expon$.
  Further, we replace every rule $u \rightarrow w_1 \parallel_1
    \cdots \parallel_1 w_k$ with the rules $u' \rightarrow w_1 \parallel_1
    \cdots \parallel_1 w_k \parallel_1 w^m$ for all $m \in M$ and the rule $u'' \rightarrow w_1 \parallel_1 \cdots \parallel_1 w_k \parallel_1 w^{dn}$.
  Then $\grammar' =(\nonterm',\rules')$ is a regular tree grammar, for
  the set $\mathcal{W}\subseteq\nonterm'$, where
  $u',u''\not\in\mathcal{W}$. Moreover, it is immediate to show that each
  derivation of $\grammar$ can be simulated by a derivation of
  $\grammar'$ with the same outcome, and vice-versa.
  As we can iteratively perform the above transformation for every pair of
non-terminals $u \in \nonterm\setminus\mathcal{W}$ and $w \in
\mathcal{W}$, the claim follows.
\end{proofE}

The first direction of the equivalence between regular and
\cmso-definable tree languages is given by the following theorem:

\begin{theoremE}\label{thm:regular-tree-grammar-definability}
  Let $\langu$ be a regular tree language.  Then, $\langu$ is
  \cmso-definable.
\end{theoremE}
\begin{proofE}
  Let $\grammar = (\nonterm,\rules)$ be a regular tree grammar such
  that $\lang{\grammar} = \langu$.  Let $\mathcal{W} \subseteq
  \nonterm$ be the set that partitions $\nonterm$ as required by the
  definition of a regular tree-grammar.  By Lemma~\ref{lemma:singleB},
  we can assume that for each pair of non-terminals $u, w \in
  \nonterm$, there is at most one rule $u \rightarrow u \parallel_1
  w^\expon$ of shape \ruleBShape\ in $\rules$.  We now build a
  \cmso-sentence $\phi$ such that $\langu = \set{\graph \mid \graph
    \Models \phi}$.

  Let $\ruleASet, \ruleBSet, \ruleCSet$ be the set of rules of the
  form shape~\ref{it1:regular-tree-grammar},
  \ref{it2:regular-tree-grammar}, \ref{it3:regular-tree-grammar} in
  $\rules$, respectively, where by $n_\ruleASet = \cardof{\ruleASet}$
  and $n_\ruleCSet = \cardof{\ruleCSet}$ we denote the number of rules
  of the respective form. For each of the rules $\ruleA \in \ruleASet,
  \ruleC \in \ruleCSet$ we will use a set variable $\ruleAPred,
  \ruleCPred$.  Further, we will use a first-order variable $\rootVar$
  for the root of the tree.  Below we state a formula $\psi$ with free
  variables $\rootVar$ and $\ruleAPred_1 \ldots \ruleAPred_{n_A},
  \ruleCPred_1 \ldots \ruleCPred_{n_C}$ such that $\graph
  \Models^\store \psi$ implies that $\graph$ is a tree with root
  $\rootVar$ whose edges can be partitioned into the sets
  $\ruleAPred_1 \ldots \ruleAPred_{n_A}$ and whose vertices can be
  partitioned into sets $\ruleCPred_1 \ldots \ruleCPred_{n_C}$ such
  that the rules denoted by these sets together with some applications
  of rules $\ruleB \in \ruleBSet$ constitute a derivation of $\graph$
  (note that this derivation does not necessarily need to start with
  an axiom).  Then, we define the formula:
  $$\phi \iffdef \exists \rootVar, \ruleAPred_1, \ldots,
  \ruleAPred_{n_A} , \ruleCPred_1, \ldots, \ruleCPred_{n_C}. \left(
  \psi \wedge \bigvee_{\ruleC \in \ruleCSet, \ruleC \text{ has head }
    u, \text{ axiom } \rightarrow u \in \grammar } \ruleCPred(r)
  \right).$$ As we will require that every edge is labelled by some
  rule $\ruleA \in \ruleASet$ of shape $w \rightarrow
  \tedge{a}{i}(u_1,\ldots,u_{\arityof{a}-1})$, we will call $i$ the
  \emph{root position} of this rule, and say that a vertex $v$ is
  \emph{at the root position} of some edge $\edge$, if we have
  $\ruleAPred(\edge)$ for some $\ruleA = w \rightarrow
  \tedge{a}{i}(u_1,\ldots,u_{\arityof{a}-1}) \in \ruleASet$ and
  $\edgrel{a}(\edge,v_1,\ldots,v_{i-1},v,v_{i+1},\ldots,v_{\arityof{a}})$
  for some vertices
  $v_1,\ldots,v_{i-1},v_{i+1},\ldots,v_{\arityof{a}}$.
  The formula $\psi$ is the conjunction of the sentences
  below:~\begin{compactitem}
  \item $\phi_1 \iffdef \forall e \in \edges. \bigvee_{\ruleA \in
    \ruleASet} \ruleAPred(e) \land \bigwedge_{\ruleA_1 \neq \ruleA_2 \in
    \ruleASet} \neg (\ruleAPred_1(e) \land \ruleAPred_2(e))$. This
    formula states that every edge of the tree is labelled by exactly
    one rule of shape~\ref{it1:regular-tree-grammar}.
  \item $\phi_2 \iffdef \forall v \in \vertices. \bigvee_{\ruleC \in
  \ruleCSet} \ruleCPred(v) \land \bigwedge_{\ruleC_1 \neq \ruleC_2 \in
  \ruleCSet} \neg (\ruleCPred_1(v) \land \ruleCPred_2(v))$.  This
  formula states that every vertex of the tree is labelled by exactly
  one rule of shape~\ref{it3:regular-tree-grammar}.
\item $\phi_3$ is the conjunction of the formulae $\phi_{3a}$,
  $\phi_{3b}$, $\phi_{3c}$, and $\phi_{3d}$, and states that the graph
  is a tree with root $r$, where:~\begin{compactitem}
  \item $\phi_{3a} \iffdef \bigwedge_{a \in \alphabet} \forall e \in
    \edges. \forall x_1 \ldots x_{\arityof{a}} \in
    \vertices. \ \edgrel{a}(e,x_1,\ldots,x_{\arityof{a}}) \rightarrow
    \bigwedge_{1 \le i<j \le \arityof{a}} x_i \neq x_j$ states that
    for every edge the vertices incident to that edge are pairwise
    different.
  \item The formula
  \begin{multline*}\phi_{3b} \iffdef  \forall e \in \edges. \\
 \bigwedge_{\ruleA = w
    \rightarrow \tedge{a}{i}(u_1,\ldots,u_{\arityof{a}-1}) \in
    \ruleASet} \left( \ruleAPred(e) \rightarrow \forall x_1 \ldots
    x_{\arityof{a}} \in \vertices.
    \edgrel{a}(e,x_1,\ldots,x_{\arityof{a}}) \rightarrow \bigwedge_{1
      \leq t \leq \arityof{a}, t \neq i} x_t \neq r \right)\end{multline*}
     states that the root $r$ is only incident to some edge $\edge$, if $r$ is
    at the root position of $\edge$.
  \item The formula \begin{multline*}
    \phi_{3c} \iffdef \forall v \in \vertices. v \neq r \rightarrow \exists! e \in \edges.\\
     \bigvee_{\ruleA = w \rightarrow \tedge{a}{i}(u_1,\ldots,u_{\arityof{a}-1}) \in \ruleASet} \left( \ruleAPred(e) \land \exists x_1 \ldots x_{\arityof{a}} \in \vertices.
    \edgrel{a}(e,x_1,\ldots,x_{\arityof{a}}) \land \bigvee_{1 \leq t \leq \arityof{a}, t \neq i} x_t = v \right )
  \end{multline*}
    states that for every vertex $v \neq r$ there is exactly one edge $\edge$ such that $v$ is incident to $\edge$ and $v$ is not at the root position of $\edge$.
%
  \item $\phi_{3d}$ states that every vertex in the graph is reachable
    from the root (as this formula is standard we omit it for
    brevity).
  \end{compactitem}
\item $\phi_4 \iffdef \forall e \in \edges \;.\; \bigwedge_{\ruleA \in
  \ruleASet} \left( \ruleAPred(e) \rightarrow \psi^\ruleA_4(e)
  \right)$ states that every edge $\edge$ labelled by some rule
  $\ruleA = w \rightarrow
  \tedge{a}{i}(u_1,\ldots,u_{i-1},u_{i+1},\ldots,u_{\arityof{a}}) \in
  \ruleASet$ needs to satisfy the formula
	\[
	\psi^\ruleA_4(e) \iffdef
	\forall v \in \vertices \; . \;
	\bigwedge_{1 \leq t \leq \arityof{a}, t \neq i}
	\left(
	\atindex_t(e,v)
	\rightarrow
	\bigvee_{\ruleC \in \ruleASet, \ \ruleC \text{ has head } u_t}
	\ruleCPred(v)
	\right)
	,
	\]
        which states that the vertices attached to $\edge$ that are
        not at position $i$ are labelled by some rule whose head
        agrees with the required non-terminal, where the auxiliary
        formula $\atindex_t(e,v)$ denotes that $v$ is at the $t$-th
        position of edge $\edge$, i.e.,
        \[
        \atindex_t(e,v)
        \iffdef
        \bigvee_{a \in \alphabet}
        \exists x_1, \ldots, x_{t-1}, x_{t + 1}, \ldots, x_{\arityof{a}} \in \vertices.
        \edgrel{a}(e,x_1 \ldots x_{t-1}, \;v,\;
        x_{t + 1}, \ldots, x_{\arityof{a}}).
        \]
      \item $\phi_5 \iffdef \forall v \in \vertices.
        \bigwedge_{\ruleC \in \ruleCSet} \ruleCPred(v) \rightarrow
        \left(\bigwedge_{w \in \mathcal{W}} \psi^C_w(v)\right)$ states
        that each vertex labelled by some rule $\ruleC = u \rightarrow
        w_1 \parallel_1 \ldots \parallel_1 w_k \in \ruleCSet$
        satisfies the formula $\psi^C_w$ for every $w \in
        \mathcal{W}$.  This formula is obtained as follows: Let us
        consider some $w \in \mathcal{W}$.  Let $p$ be the number of
        times $w$ appears among the $w_1 \parallel_1 \ldots
        \parallel_1 w_k$.  Let $\expon$ be the number such that there
        is a rule $u \rightarrow u \parallel_1 w^\expon \in \ruleBSet$
        (we recall that there is at most one rule of this shape), and
        let $\expon = 0$, otherwise.  We now define the formula
	\[
	\psi^C_w(v) \iffdef
	\exists X \subseteq \edges \;.\;
	\cardconstr{X}{q}{p}
	\wedge
	\forall e \;.\;
	e \in X
	\leftrightarrow
	\bigvee_{\ruleA = w \rightarrow \tedge{a}{i}(u_1,\ldots,u_{\arityof{a}-1}) \in \ruleASet}
	\left(
	\ruleAPred(e)
	\wedge
	\atindex_{i}(e,v)
	\right),
	\]
        which states that there are exactly $k\expon+p$ edges $\edge$,
        for some $k\in\nat$, such that $v$ is the root of 
        $\edge$, i.e., such that $\edge$ is labelled by some
        rule $w \rightarrow \tedge{a}{i}(u_1,\ldots,u_{\arityof{a}-1})
        \in \ruleASet$ and $v$ is incident to $\edge$ at position $i$.
\item $\phi_6 \iffdef \forall e \in \edges. \bigwedge_{\ruleA = w
  \rightarrow \tedge{a}{i}(u_1,\ldots,u_{\arityof{a}-1}) \in
  \ruleASet} \ruleAPred(e) \rightarrow \exists x_1 \ldots
  x_{\arityof{a}} \in \vertices. \ \edgrel{a}(e,x_1,\ldots
  x_{\arityof{a}})$ states that if edge $\edge$ is labelled by some
  rule $\ruleA = w \rightarrow
  \tedge{a}{i}(u_1,\ldots,u_{\arityof{a}-1}) \in \ruleASet$, then
  $\edge$ is labelled $a$.
\end{compactitem}

We now prove that $\langu = \set{\graph \mid \graph \Models \phi}$.
We begin by proving $\langu \subseteq \set{\graph \mid \graph \Models
  \phi}$.
For this it is sufficient to prove for every $u \in
\nonterm \setminus \mathcal{W}$ that $\graph \in \langof{\grammar}{u}$
implies $\graph \Models \phi(u)$, where we introduce the formula
$\phi(u) \iffdef \exists \rootVar, \ruleAPred_1, \ldots,
\ruleAPred_{n_A} , \ruleCPred_1, \ldots, \ruleCPred_{n_C}. \left( \psi
\wedge \bigvee_{\ruleC \in \ruleCSet, \ruleC \text{ has head } u}
\ruleCPred(\rootVar) \right)$.  We proceed by induction on the height
of the tree $\graph$.

We consider some $\graph \in \langof{\grammar}{u}$ for some $u \in
\nonterm \setminus \mathcal{W}$.  Then, the derivation of $\graph$
starts with zero or more applications of rules $\ruleB \in \ruleBSet$
of form $u \rightarrow u \parallel_1 w^\expon$.  At some point there
is an application of a rule $\ruleC \in \ruleCSet$ of form $u
\rightarrow w_1 \parallel_1 \ldots \parallel_1
w_n$.
Then,
indexing by $m$, for each non-terminal $w_m$, the derivation continues
with an application of some rule $\ruleA \in \ruleASet$ of form $w_m
\rightarrow \tedge{a_m}{i_m}(u^m_1,\ldots,u^m_{i_m - 1} , u^m_{i_m +
  1}, \ldots u^m_{n_m})$.  Let us now assume that for each $u^m_j$ we
have derived a tree $\graph^m_j \in \langof{\grammar}{u^m_j}$.  Then,
we obtain
\begin{align*}
&\graph =
\tedge{a_1}{i_1}(\graph^1_1,\ldots,\graph^1_{i_1 - 1}, \graph^1_{i_1 + 1}, \ldots \graph^1_{n_1}) \parallel_1
\cdots
\parallel_1
\tedge{a_m}{i_m}(\graph^m_1,\ldots,\graph^m_{i_m - 1}, \graph^m_{i_m + 1}, \ldots \graph^m_{n_m}) \parallel_1
\cdots
\\
&\qquad\qquad
\parallel_1
\tedge{a_k}{i_k}(\graph^k_1,\ldots,\graph^k_{i_k - 1}, \graph^k_{i_k + 1}, \ldots \graph^k_{n_k}).
\end{align*}
Let $e_m$ be the edge of $\graph$ that is generated by the operation $\tedge{a_m}{i_m}(\graph^m_1,\ldots,\graph^m_{i_m - 1}, \graph^m_{i_m + 1}, \ldots \graph^m_{n_m})$, i.e., the edge that connects the roots of the graphs $\graph^m_j$.
We now note that each of the subtrees $\graph^m_j$ is of smaller height than $\graph$.
Hence, we can apply the induction assumption and obtain that $\graph^m_j \Models \phi(u^m_j)$.
In particular, each of the $\phi(u^m_j)$ begins with the quantifier prefix
$\exists r^{m,j},\ruleAPred_1^{m,j}, \ldots, \ruleAPred_{n_A}^{m,j},
\ruleCPred_1^{m,j}, \ldots, \ruleCPred_{n_C}^{m,j}$,
meaning that we can choose valuations for these variables that witness the satisfaction of the formula $\psi$.
We will below use these valuations in order to assign values to $\rootVar, \ruleAPred_1, \ldots, \ruleAPred_{n_A},
\ruleCPred_1, \ldots, \ruleCPred_{n_C}$ and argue these values witness
$\graph \Models \phi(u)$.
Let $\rootVar$ be the root of $\graph$.
For each edge $e_m$, let $\ruleA^m \in \ruleASet$ be the rule that was used to derive this edge.
Further, let $\ruleC \in \ruleCSet$ be the rule that was used in the beginning of the derivation of $\graph$ (after some applications of rules $\ruleB \in \ruleBSet$).
We now set $\ruleAPred_t = \bigcup_{m,j} \ruleAPred^{m,j}_t
\cup \bigcup_{m} \{e_m \mid \ruleA^m = \ruleA_t\}$ for all $1 \le t \le n_A$, $\ruleCPred_t = \bigcup_{m,j} \ruleCPred^{m,j}_t$ for all $1 \le t \le n_C$ with $\ruleC_t \neq \ruleC$, and $\ruleCPred_t = \bigcup_{m,j} \ruleCPred^{m,j}_t \cup \{\rootVar\}$, otherwise.

We now verify that $\graph \Models \phi(u)$:
\begin{compactitem}
\item $\phi_1$:
Each edge $e$ is either part of some subgraph
$\graph^m_j$, or is equal to some edge $e_m$.
The way we defined the sets $\ruleAPred_t$ still partitions all the edges in $\graph$.
\item $\phi_2$: Each vertex is either part of some subgraph or equals the root $\rootVar$.
     The way we defined the sets $\ruleCPred_t$ above still partitions all the vertices in $\graph$.
\item $\phi_3$: Since $\grammar$ is a tree grammar, $\graph$ is a tree.
\item $\phi_4$: Clearly, this property holds for all edges of  $\graph^m_j$.
    We have that the root $\rootVar^m_j$ of every $\graph^m_j$ is labeled by some rule with head $u^m_j$.
    Hence, the property holds for all edges $e_m$.
\item $\phi_5$: Clearly, this property holds for all vertices of
  $\graph^m_j$.  We now consider the root $\rootVar$.  We recall that
  the root is labelled by some rule $\ruleC = u \rightarrow w_1
  \parallel_1 \ldots \parallel_1 w_k \in \ruleCSet$.  Let $w \in
  \nonterm \setminus \mathcal{W}$ be some non-terminal.  Let $p$ be
  the number of times $w$ appears among the $w_1 \parallel_1 \ldots
  \parallel_1 w_k$.  Let $\expon$ be the number such that there is a
  rule $u \rightarrow u \parallel_1 w^\expon \in \ruleBSet$ (we recall
  that there is at most one rule of this shape), and let $\expon = 0$,
  otherwise.  It is now easy to verify that there are at least $p$
  edges, which are labelled by rules with head $w$, and the number of
  these edges equals $p$ modulo $q$.  As this holds for every $w \in
  \nonterm \setminus \mathcal{W}$, the property holds for the root
  $\rootVar$.
\item $\phi_6$: Clearly, this property holds for all edges of  $\graph^m_j$.
    The property further holds for the edges $e_m$.
\end{compactitem}

We now prove that $\langu \supseteq \set{\graph \mid \graph \Models \phi}$.
For this it is sufficient to prove that $\graph \Models \phi(u)$ implies that $\graph \in \langof{\grammar}{u}$ for all $u \in \nonterm \setminus \mathcal{W}$.
We now consider some graph $\graph \Models \phi(u)$ for some non-terminal $u \in \nonterm \setminus \mathcal{W}$.
We note that $\graph \Models \phi(u)$ implies that $\graph$ is a tree.
This allows us to proceed by induction on the height of the tree.
Let $\rootVar$ be the root of $\graph$, and let $e_1, \ldots, e_k$ be the edges to which $\rootVar$ is incident.
Let $\ruleA^m = w^m \rightarrow \tedge{a_m}{i_m}(u_1^m,\ldots,u_{\arityof{a}-1}^m) \in \ruleASet$ be the rule such that $\ruleAPred^m(e_m)$ and let $\edgerelof{\graph}(e_m) = \rootVar^m_1 \ldots \rootVar^m_{n_m}$ be the incident vertices, for $1 \le m \le k$.
We note that $\graph \Models \phi(u)$ implies that $\rootVar = \rootVar^m_{i^m}$ and $a_m = \labof{\graph}(e_m)$ for all $1 \le m \le k$.
For each vertex $\rootVar^m_j \neq \rootVar$, define $\graph^m_j$ as the subtree starting with root $\rootVar^m_j$ (note that such a subtree can be defined because  there is exactly one edge, namely $e_m$, to which $\rootVar^m_j$ is incident and where $\rootVar^m_j$ is not at the root position of this edge).
Thus, we have
\begin{align*}
&\graph =
\tedge{a_1}{i_1}(\graph^1_1,\ldots,\graph^1_{i_1 - 1}, \graph^1_{i_1 + 1}, \ldots \graph^1_{n_1}) \parallel_1
\cdots
\parallel_1
\tedge{a_m}{i_m}(\graph^m_1,\ldots,\graph^m_{i_m - 1}, \graph^m_{i_m + 1}, \ldots \graph^m_{n_m}) \parallel_1
\cdots
\\
&\qquad\qquad
\parallel_1
\tedge{a_k}{i_k}(\graph^k_1,\ldots,\graph^k_{i_k - 1}, \graph^k_{i_k + 1}, \ldots \graph^k_{n_k}) \quad (*).
\end{align*}
Because of $\graph \Models \phi(u)$, there are sets $\ruleAPred_1,
\ldots, \ruleAPred_{n_A}, \ruleCPred_1, \ldots, \ruleCPred_{n_C}$ that
witness the satisfaction of $\psi$.  We now choose the sets
$\ruleAPred_1^{m,j}, \ldots, \ruleAPred_{n_A}^{m,j},
\ruleCPred_1^{m,j}, \ldots, \ruleCPred_{n_C}^{m,j}$ as the restriction
of the above sets to the vertices and edges of graph $\graph^m_j$.  It
is now routine to verify that $\graph^m_j \Models \phi(u^m_j)$.
Hence, we can apply the induction assumption and obtain that
$\graph^m_j \in \langof{\grammar}{u^m_j}$.  Let $\ruleC = u
\rightarrow w_1 \parallel_1 \ldots \parallel_1 w_n \in \ruleCSet$ be
the rule such that $\ruleCPred(r)$.
Let $w \in \nonterm \setminus
\mathcal{W}$ be some non-terminal.  Let $p$ be the number of times $w$
appears among the $w_1 \parallel_1 \ldots \parallel_1 w_n$.  Let
$\expon$ be the number such that there is a rule $u \rightarrow u
\parallel_1 w^\expon \in \ruleBSet$ (we recall that there is at most
one rule of this shape), and let $\expon = 0$, otherwise.  Because of
$\graph \Models \phi(u)$ there are at least $p$ edges among the
$e_1,\ldots,e_k$, which are labelled by rules with head $w$, and the
number of these edges equals $p$ modulo $q$. As this holds for every
$w \in \nonterm \setminus \mathcal{W}$, we can apply zero or more
times rules $\ruleB \in \ruleBSet$, followed by a single application
of rule $\ruleC = u \rightarrow w_1 \parallel_1 \ldots \parallel_1 w_n
\in \ruleCSet$, in order to obtain a partial derivation that can be
summarized as $u \rightarrow w_1 \parallel_1 \cdots \parallel_1 w_k$,
where $w_m$ is the head of the rule $\ruleA^m \in \ruleASet$.  Then,
we continue the derivation by applying the rule $\ruleA^m$ for every
$m \in [1,k]$ in order to obtain the partial derivation
\begin{align*}
&u \rightarrow
\tedge{a_1}{i_1}(u^1_1,\ldots,u^1_{i_1 - 1}, u^1_{i_1 + 1}, \ldots u^1_{n_1}) \parallel_1
\cdots
\parallel_1
\tedge{a_m}{i_m}(u^m_1,\ldots,u^m_{i_m - 1}, u^m_{i_m + 1}, \ldots u^m_{n_m}) \parallel_1
\cdots
\\
&\qquad\qquad
\parallel_1
\tedge{a_k}{i_k}(u^k_1,\ldots,u^k_{i_k - 1}, u^k_{i_k + 1}, \ldots u^k_{n_k}).
\end{align*}
This partial derivation together with (*) and $\graph^m_j \in \langof{\grammar}{u^m_j}$ then establishes $\graph \in \langof{\grammar}{u}$.
\end{proofE}

In order to establish the completeness of regular tree-grammars, we
use the already known equivalence between \cmso-definable and
recognizable languages of unranked trees~\cite[Theorem
  5.3]{journals/iandc/Courcelle90}:

\begin{theorem}[\cite{journals/iandc/Courcelle90}]\label{thm:rec-equals-cmso-def-trees}
Let $\langu$ be a language of trees.
Then, $\langu$ is \cmso-definable iff $\langu$ is recognizable.
\end{theorem}

The dual direction of the equivalence between regular tree languages
and \cmso-definable tree languages is given by the following result:

\begin{theoremE}
\label{thm:regular-tree-grammar-completeness}
Let $\langu$ be a \cmso-definable tree language. Then, $\langu$ is a
regular tree language.
\end{theoremE}
\begin{proofE}
By Theorem~\ref{thm:rec-equals-cmso-def-trees}, $\langu$ is recognizable.
Hence, there is a locally-finite congruence relation $\cgr$ that recognizes $\langu$.
Let $A$ be the finite set of equivalence classes of graphs of type 1 (as all trees are graphs of type 1, these are the only relevant equivalence classes).
We now observe that $\cgr$ induces operations
$\parallel_1$ and $\tedge{a}{i}$ on $A$.
In particular, we note that $A$ and $\parallel_1$ form a finite commutative monoid.
Let $n = \cardof{A}$ be the cardinality of $A$.
Because $A$ is finite, we have for every $a \in A$ that there are some indices $0 \le d < n$ and $1 \le \expon \le n$ such that $a^d = a^{d+\expon}$ (*), where $a^\expon$ denotes the $\expon$-fold composition of $a$ with itself wrt the operation $\parallel_1$.
We set $\stem(a) = d$ and $\lasso(a) = \expon$ for the least such $d$ and $\expon$ and note that they are unique.
We now create non-terminals $u_a$ and $w_a$ for all elements $a \in A$, and define a grammar $\grammar$ by the following rules:
\begin{compactenum}[A.]
\item\label{it1:tree-complete} $w_b \rightarrow
    \tedge{a}{i}(u_{a_1},\ldots,u_{a_{\arityof{a}-1}})$ iff
    $b = \tedge{a}{i}(a_1, \ldots, a_{\arityof{a}-1})$
\item\label{it2:tree-complete} $u_b \rightarrow u_b \parallel_1 w_a^\expon$ iff $b = b \parallel_1 a^\expon$ for some $1 \le \expon \le n$
\item\label{it3:tree-complete} $u_b \rightarrow w_{a_1}^{p_1} \parallel_1 \ldots \parallel_1 w_{a_n}^{p_n}$ iff $b = a_1^{p_1} \parallel_1 \cdots \parallel_1 a_n^{p_n}$ for some $0 \le p_i < 2n$
\end{compactenum}
We finally add axioms $\rightarrow u_a$ for all $a \in A$ such that there is some tree $\graph$ with $\graph \in a \subseteq \langu$.

We now prove that $\langu = \lang{\grammar}$:
We first verify for all non-terminals $u_a, w_a \in \nonterm$ that $\graph \in \langof{\grammar}{u_a}$ resp. $\graph \in \langof{\grammar}{w_a}$ implies that $\graph \in a$;
this by done by a straightforward induction on the length of the derivation.
In particular, this implies that $\langu \supseteq \lang{\grammar}$.
We now prove that $\graph \in a$ implies $\graph \in \langof{\grammar}{u_a}$ and $\graph \in \langof{\grammar}{w_a}$ in case of $\graph = \tedge{a}{i}(\graph_1,\ldots,\graph_{i-1},\graph_{i+1}, \ldots,\graph_{\arityof{a}})$ for some trees $\graph_1,\ldots,\graph_{i-1},\graph_{i+1}, \ldots,\graph_{\arityof{a}}$.
We prove the claim by induction on the size of $\graph$.

We first assume that $\graph =
\tedge{a}{i}(\graph_1,\ldots,\graph_{i-1},\graph_{i+1},
\ldots,\graph_{\arityof{a}})$ for some trees
$\graph_1,\ldots,\graph_{i-1},\graph_{i+1},
\ldots,\graph_{\arityof{a}}$.  Let $w_j \in A$ be the equivalence
class such that $\graph_j \in w_j$.  Then, we have that $a =
\tedge{a}{i}(w_1,\ldots,w_{i-1},w_{i+1}, \ldots,w_{\arityof{a}})$.  By
induction assumption, we have that $\graph_j \in
\langof{\grammar}{w_j}$.  Now, we can apply a rule of
shape~\ref{it1:tree-complete} and deduce that $\graph \in
\langof{\grammar}{w_a}$ and a rule of shape~\ref{it3:tree-complete},
that is the rule $u_a \rightarrow w_a$, in order to deduce that
$\graph \in \langof{\grammar}{u_a}$. Note that the rule $u_a
\rightarrow w_a$ exists as a special case of rules of shape~\ref{it3:tree-complete}.

We now assume that $\graph = \graph_1 \parallel_1 \graph_2 \parallel_1 \cdots$ for some sequence of trees $\graph_j$ such that for all $j$ we have $\graph_j = \tedge{a_j}{i_j}(\graph_1^j,\ldots,\graph_{i-1}^j, \graph_{i+1}^j, \ldots,\graph_{\arityof{a_j}}^j)$ for some trees $\graph_1^j,\ldots,\graph_{i-1}^j,\graph_{i+1}^j, \ldots,\graph_{\arityof{a_j}}^j$.
Let $a_i \in A$ be the equivalence class such that $\graph_i \in a_i$.
Note that we have that $a = a_1 \parallel_1 a_2 \parallel_1 \cdots $.
Now, because $\parallel_1$ is associative and commutative, we can represent this product as
$a = a_1^{l_1} \parallel_1 \cdots \parallel_1 a_n^{l_n}$ for some enumeration $a_1, \ldots, a_n$ of the equivalence classes of $A$ and $l_i \in\nat$.
By (*), we have $a = a_1^{p_1} \parallel_1 \cdots \parallel_1 a_n^{p_n}$ for some $0 \le p_i < 2n$ such that $l_i = \lasso(a_i)k_i+p_i$ for some $k_i \in\nat$, and $a = a \parallel_1 a_i^{\lasso(a_i)}$ for every $a_i$ with $l_i \ge 2n$.
Now we obtain that $\graph \in \langof{\grammar}{u_a}$, using that $\graph_j \in \langof{\grammar}{w_j}$ by induction assumption and then applying rules of shape~\ref{it2:tree-complete} as often as needed followed by a single application of a rule of shape~\ref{it3:tree-complete}.
\end{proofE}

In the rest of this section, we characterize the class of
\mso-definable tree languages by tree grammars. We show that the
\mso-definable tree language are exactly the languages produced by the
tree grammars having $\expon=1$ in each rule of the form
(\ref{it2:regular-tree-grammar}).

\begin{theoremE}
\label{thm:regular-tree-grammar-mso-definability}
Let $\langu$ be a tree language such that $\langu=\lang{\grammar}$ for
some regular tree grammar $\grammar$ such that $\expon = 1$ for all
rules of the form (\ref{it2:regular-tree-grammar}).  Then, $\langu$ is
\mso-definable.
\end{theoremE}
\begin{proofE}
The proof agrees with the proof of
Theorem~\ref{thm:regular-tree-grammar-definability}, except for a
small modification.  Because of $\expon = 1$ for all rules of
shape~\ref{it2:regular-tree-grammar}, we are able to replace the
constraints $\cardconstr{X}{q}{p}$ in the proof of
Theorem~\ref{thm:regular-tree-grammar-definability}.  We recognize
that the formula $\cardconstr{X}{q}{p}$ is equivalent to the
\mso-formula $\cardgeconstr{X} \ge p$ for $q = 1$, and
$\cardgeconstr{X} \ge p \wedge \neg (\cardgeconstr{X} \ge p+1)$ for $q
= 0$, where $\cardgeconstr{X} \ge k \iffdef \exists
x_1,\ldots,x_k.\ \bigwedge_{1 \le i \le k} x_i \in X \wedge
\bigwedge_{1 \le i < j \le k} x_i \neq x_j$.
\end{proofE}

\begin{theoremE}
\label{thm:regular-tree-grammar-mso-completeness}
Let $\langu$ be a \mso-definable tree language. Then, there exists
some regular tree grammar $\grammar$ with $\langu=\lang{\grammar}$
such that $\expon = 1$ for all rules of the form
(\ref{it2:regular-tree-grammar}).
\end{theoremE}
\begin{proofE}
The proof agrees with the proof of Theorem~\ref{thm:regular-tree-grammar-completeness}, except for one additional argument:
Again we use that $\langu$ is recognizable by Theorem~\ref{thm:rec-equals-cmso-def-trees}.
Thus, there is a locally-finite congruence relation $\cgr$ that recognizes $\langu$.
Let $A$ be the finite set of equivalence classes of type 1.
We recall that $\cgr$ induces an operation
$\parallel_1$ on $A$.
In particular, we note that $A$ and $\parallel_1$ form a finite commutative monoid.
We recall that a monoid $A$ is called \emph{aperiodic} iff $\lasso(a) = 1$ for all elements $a \in A$.
We now use that a set of trees is \mso-definable iff the monoid $(A,\parallel_1)$ is aperiodic, see Theorem~5.10 from Bojanczyk~\cite{journals/corr/abs-2008-11635}.
We conclude by observing that the rules resulting from the proof of Theorem~\ref{thm:regular-tree-grammar-completeness} have the required form.
\end{proofE}

\section{Tree Verifiable Graph-Grammars}
In this section, we introduce a syntactic restriction on \hr\ graph grammars, that ensures the \cmso-definability of the generated graph languages.
The restriction is given in terms of easy-to-check syntactic conditions on the rules of the
\hr\ grammar.
Like in the case of regular tree grammars, we partition the set of nonterminals in two alternating subsets.
The crucial restriction is then on the shape of the graphs $\graph$ in the rules of the form $w \leftarrow \ksubst{\graph}{e_1}{v_1}{e_k}{v_k}$, where we require $\graph$ to consist of a single terminal edge~$e$, which needs to contain at least some source and some vertex of each $e_i$.
Some additional conditions ensures that these edges and vertices give rise to a spanning tree of the generated graph that witnesses the bounded embeddable tree-width of the graph (see Definition~\ref{def:embeddable-tree-decomposition} and Proposition~\ref{prop:tree-verifiable-embeddable} below).
The spanning tree property also underlies our proof of \cmso-definability, which motivates the term \emph{tree-verifiable}.
As we shall argue next (\S\ref{sec:simulation}), tree-verifiable grammars define an expressive class of graph languages, as they strictly subsume the regular graph grammars of Courcelle~\cite[\S5]{journals/tcs/Courcelle91a}.

\begin{definition}
\label{def:tree-verifiable-grammar}
  A graph grammar $\grammar = (\nonterm,\rules)$ is
  \emph{tree-verifiable} iff there exist two functions:~\begin{compactitem}
  \item $\rootSymb : \nonterm \rightarrow \nat$ mapping
    each non-terminal $u$ to an index
    $\rootof{u}\in\interv{1}{\arityof{u}}$, called the \emph{root} of
    $u$,
  \item $\rootsSymb : \nonterm \rightarrow \pow{\nat}$ mapping each
    non-terminal $u$ to a subset $\rootsof{u} \subseteq
    \interv{1}{\arityof{u}} \setminus \set{\rootof{u}}$, called the
    \emph{future roots} of $u$,
  \end{compactitem}
   and a set $\mathcal{W} \subseteq \nonterm$ of nonterminals, such
   that the rules in $\rules$ are of one of the following forms,
   either:
  \begin{compactenum}[A.]
  \item\label{it1:tree-verifiable-graph-grammar} $w \rightarrow
    \ksubst{\graph}{e_1}{u_1}{e_k}{u_k}$, for some $w\in\mathcal{W}$
    and $u_1,\ldots,u_k \in \nonterm \setminus \mathcal{W}$, where
    $\graph$ has exactly one terminal edge~$e$, the vertices
    $\set{\sourceof{\graph}(\rootof{w})}\cup\set{\edgerelof{\graph}(e_i)_{\rootof{u_i}}}_{i\in\interv{1}{k}}$ are pairwise distinct and attached to $e$, and
    \[\internal{\graph} \cup \{ \sourceof{\graph}(i) \mid i \in \rootsof{w} \}= \biguplus_{i=1..k} \{\edgerelof{\graph}(e_i)_j \mid j \in \{\rootof{u_i}\} \cup \rootsof{u_i} \} \]
    i.e., the internal nodes of $\graph$ plus the future roots associated to $w$ can be
    partitioned into the roots and future roots associated to all the non-terminals
    $u_i$.
    %
  \item\label{it2:tree-verifiable-graph-grammar} $u \rightarrow u
    \parallel_n w^\expon$, for some $\expon \in \nat$, where $u \in
    \nonterm\setminus\mathcal{W}$, $w \in \mathcal{W}$ with
    $\arityof{u}=\arityof{w}=n$, such that $\rootof{w} = \rootof{u}$
    and $\rootsof{w} = \emptyset$, i.e., the roots of $u$ and $w$
    agree and, moreover, $w$ has no future roots,
  \item\label{it3:tree-verifiable-graph-grammar} $u \rightarrow w_1
    \parallel_n \cdots \parallel_n w_k$, where $u \in
    \nonterm\setminus\mathcal{W}$, $w_1,\ldots,w_k\in\mathcal{W}$ with
    $\arityof{u}=\arityof{w_i}=n$, such that $\rootof{u} =
    \rootof{w_i}$ and $\rootsof{u} = \biguplus_{i=1..k}
    \rootsof{w_i}$, i.e., the roots associated to the non-terminals of
    the right-hand side agree with the root of the left-hand side and
    that future roots of the left-hand side are partitioned into the
    future roots of the non-terminals of the right-hand side.
  \end{compactenum}
  Furthermore, the axioms are of the shape $\rightarrow u$ for some $u
  \in \nonterm \setminus \mathcal{W}$ with $[1,\arityof{u}] =
  \{\rootof{u}\} \cup \rootsof{u}$, i.e., for every start non-terminal
  $u$ every source must be a root or future root.
\end{definition}
A set of graphs $\langu$ is said to be \emph{tree-verifiable} iff
there is a tree-verifiable graph-grammar $\grammar$ such that
$\langu=\lang{\grammar}$. The following example states a  tree-verifiable grammar for the trees with linked leaves from Example~\ref{ex:tll} plus an additional binary edge from the bottom-right node of the tree to the root of the tree (see the left of Fig.~\ref{fig:etd} for the depiction of such a graph)\footnote{We note that we chose to add the binary edge as the easiest way of setting up a tree-verifiable grammar (with axioms) that meets the requirements of Def.~\ref{def:tree-verifiable-grammar}.
However, one can also state a slightly more complicated tree-verifiable grammar that exactly produces the trees with linked leaves from Example~\ref{ex:tll}.}:

\begin{figure}[t!]
  \centerline{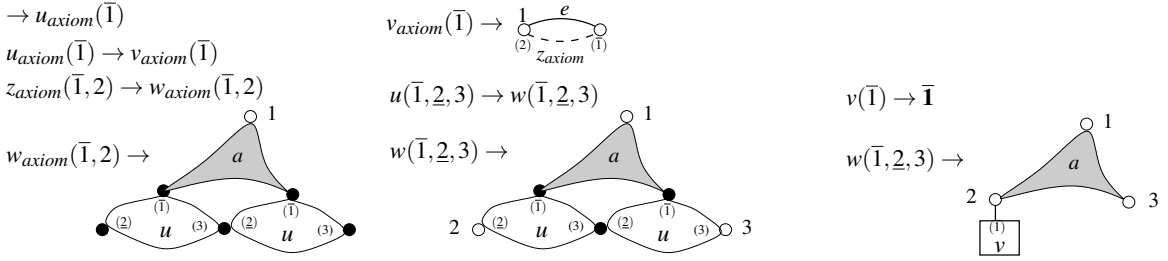}

  \vspace*{-.5\baselineskip}
  \caption{Tree-Verifiable Grammar for Trees with Linked Leaves}
  \label{fig:tvtll}
  \vspace*{-.5\baselineskip}
\end{figure}

\begin{example}\label{ex:tree-verifiable-tll}
  Let us consider the tree-verifiable grammar from
  Fig.~\ref{fig:tvtll}, with the set of non-terminals
  $\nonterm=\set{u_{\mathit{axiom}},v_{\mathit{axiom}},z_{\mathit{axiom}},w_{\mathit{axiom}},u,v,w}$ and
  distinguished non-terminals
  $\mathcal{W}=\set{v_{\mathit{axiom}},w_{\mathit{axiom}},w}$, with arities, roots
  and future roots, as follows:~\begin{compactitem}
    \item
      $\arityof{u_{\mathit{axiom}}}=\arityof{v_{\mathit{axiom}}}=1$,
      $\rootof{u_{\mathit{axiom}}}=\rootof{v_{\mathit{axiom}}}=1$,
      $\rootsof{u_{\mathit{axiom}}}=\rootsof{v_{\mathit{axiom}}}=\emptyset$,
    \item
      $\arityof{z_{\mathit{axiom}}}=\arityof{w_{\mathit{axiom}}}=2$,
      $\rootof{z_{\mathit{axiom}}}=\rootof{w_{\mathit{axiom}}}=1$,
      $\rootsof{z_{\mathit{axiom}}}=\rootsof{w_{\mathit{axiom}}}=\emptyset$,
    \item $\arityof{u}=\arityof{w}=3$, $\rootof{u}=\rootof{w}=1$,
      $\rootsof{u}=\rootsof{w}=\set{2}$,
    \item $\arityof{v}=1$, $\rootof{v}=1$, $\rootsof{v}=\emptyset$.
  \end{compactitem}
  We write the indices $1, \ldots, \arityof{n}$ enclosed in
  parenthesis next to each nonterminal $n\in\nonterm$, where
  $\rootof{n}$ is overlined and $\rootsof{n}$ are underlined.
  Internal nodes are drawn as black circles and sources are white
  numbered circles.  Binary terminal edges are drawn with solid lines
  and binary nonterminal edges are dashed. Ternary terminal edges are
  filled with grey and ternary nonterminal edges are unfilled.
\end{example}

The first result is that tree-verifiable graph grammars produce
\cmso-definable sets of graphs:

\begin{theoremE}\label{thm:tree-verifiable-grammar-definability}
Let $\langu$ be a tree-verifiable graph language. Then, $\langu$ is
\cmso-definable.
\end{theoremE}
\begin{proofE}
  Let $\grammar = (\nonterm,\rules)$ be a tree-verifiable grammar such that $\lang{\grammar} = \langu$.
  Let $\mathcal{W} \subseteq
  \nonterm$ be the set that partitions $\nonterm$ as required by the
  definition of a  tree-verifiable graph grammar.
  By Lemma~\ref{lemma:singleB}, whose proof also applies to tree-verifiable grammars,
  we can assume that for each pair of non-terminals $u, w \in
  \nonterm$, there is at most one rule $u \rightarrow u \parallel_1
  w^\expon$ of shape \ruleBShape\ in $\rules$.
  Let $\ruleASet, \ruleBSet, \ruleCSet$ be the set of rules of the
  form shape~\ref{it1:tree-verifiable-graph-grammar},
  \ref{it2:tree-verifiable-graph-grammar}, \ref{it3:tree-verifiable-graph-grammar} in
  $\rules$, respectively, where by $n_\ruleASet = \cardof{\ruleASet}$
  and $n_\ruleCSet = \cardof{\ruleCSet}$ we denote the number of rules
  of the respective form. For each of the rules $\ruleA \in \ruleASet,
  \ruleC \in \ruleCSet$ we will use a set variable $\ruleAPred,
  \ruleCPred$.

  For any $u \in \nonterm \setminus \mathcal{W}$, we below state a formula $\psi(u)$ with free first-order variables $y_1,\ldots,y_{\arityof{u}}$ and free monadic variables $\ruleAPred_1 \ldots \ruleAPred_{n_A}, \ruleCPred_1 \ldots \ruleCPred_{n_C}$, which has the following property:
  Given a graph $\graph$ of type $\arityof{u}$, let $\store$ be the valuation defined by $\store(y_i) = \sourceof{\graph}(i)$ for all $i \in [1,\arityof{u}]$.
  Then, $\rmsources{\graph} \Models^\store \exists \ruleAPred_1 \ldots \ruleAPred_{n_A}, \ruleCPred_1 \ldots \ruleCPred_{n_C}. \ \psi(u)$ iff $\graph \in \langof{\grammar}{u}$.
  We can use these formulae $\psi(u)$ to build a \cmso-sentence $\phi$ such that $\langu = \set{\graph \mid \graph \Models \phi}$:
  $$\phi \iffdef \bigvee_{\text{axiom } \rightarrow u \in \grammar } \exists y_1,\ldots,y_{\arityof{u}},\ruleAPred_1 \ldots \ruleAPred_{n_A}, \ruleCPred_1 \ldots \ruleCPred_{n_C}. \ \psi(u).$$

  We next introduce some notation that will be useful in the definition of the formulae $\psi(u)$, in particular, it will allow us to conveniently access the spanning tree that underlies every derivation of a tree-verifiable grammar.
  We consider some rule $\ruleA = w \rightarrow \ksubst{\graph}{e_1}{u_1}{e_k}{u_k} \in \ruleASet$.
  Let $e$ be the single terminal edge of $\graph$.
  We denote by $\rpos(\ruleA) = j \in [1, \lenof{\edgerelof{\graph}(e)}]$ the index such that $\edgerelof{\graph}(e)_j = \sourceof{\graph}(\rootof{w})$, by $\cpos(\ruleA,i) = j \in [1,\lenof{\edgerelof{\graph}(e)}]$ the index such that $\edgerelof{\graph}(e)_j = \edgerelof{\graph}(e_i)_{\rootof{u_i}}$ and by $\cpos(\ruleA) = \{ \cpos(\ruleA,i) \mid i \in [1,k] \}$ the set of these indices (we recall that the definition of a tree-verifiable grammar requires all those vertices to be attached to $e$).
  We also define $\labels(\ruleA) = \labof{\graph}(e)$ as the label of the edge $e$, $\vertedge(\ruleA,i)  = \edgerelof{\graph}(e)_i$ as the $i$-th vertex attached to the edge $e$, and write $\graphofrule(\ruleA) = \graph$ for the graph of rule $\ruleA$.
  We call $\rpos(\ruleA)$ the \emph{root position} of $\ruleA$ and $\cpos(\ruleA)$ the \emph{positions} of the \emph{children} of $\ruleA$.
  We say that a vertex $v$ is \emph{at the root position} of some edge $\edge$, if $\ruleAPred(\edge)$ for some $\ruleA \in \ruleASet$ and $v = \edgerelof{\graph}(\edge)_{\rpos(\ruleA)}$, and we say that $v$ is \emph{at child position} $\cpos(\ruleA,i)$ of $\edge$, if $\ruleAPred(\edge)$ for some $\ruleA \in \ruleASet$ and $v = \edgerelof{\graph}(\edge)_i$.
  Further, for a rule  $\ruleC = u \rightarrow w_1 \parallel_1 \ldots \parallel_1 w_k \in \ruleCSet$ we denote its head by $\head(\ruleC) = u$.

  We now consider some non-terminal $u^\circ \in \nonterm \setminus \mathcal{W}$.
  Let $\nrootsof{u^\circ} = [1,\arityof{u^\circ}] \setminus (\rootsof{u^\circ} \cup \{\rootof{u^\circ}\})$ be the indices that are neither the root nor the future roots of $u^\circ$.
  The formula $\psi(u^\circ)$ is the conjunction of the sentences
  below:~\begin{compactitem}
  \item $\phi_0 \iffdef \bigwedge_{1 \le i < j \le \arityof{u^\circ}} y_i \neq y_j$ states that all sources of the graph are pairwise different.
  \item $\phi_1 \iffdef \forall e \in \edges. \bigvee_{\ruleA \in
    \ruleASet} \ruleAPred(e) \land \bigwedge_{\ruleA_1 \neq \ruleA_2 \in
    \ruleASet} \neg (\ruleAPred_1(e) \land \ruleAPred_2(e))$. This
    formula states that every edge is labelled by exactly
    one rule of shape~\ref{it1:tree-verifiable-graph-grammar}.
  \item $\phi_2 \iffdef \forall v \in  \vertices. \left(
  \bigvee_{\ruleC \in
  \ruleCSet} \ruleCPred(v) \leftrightarrow \bigwedge_{i \in \nrootsof{u^\circ}} y_i \neq v
  \right) \land \bigwedge_{\ruleC_1 \neq \ruleC_2 \in
  \ruleCSet} \neg (\ruleCPred_1(v) \land \ruleCPred_2(v)$.
  %
  %
  This formula states that every vertex is labelled by a rule of shape~\ref{it3:tree-verifiable-graph-grammar} iff this vertex is neither the root nor a future root of $u^\circ$, and that every vertex is labelled by at most one such rule.
\item $\phi_3$ is the conjunction of the formulae $\phi_{3a}$,
  $\phi_{3b}$, $\phi_{3c}$, and $\phi_{3d}$, and states that the graph contains a spanning tree with root $y_{\rootof{u^\circ}}$, where the spanning tree is constructed using only the root and child positions of every edge:
  \begin{compactitem}
  \item The formula
     \begin{multline*}
     \phi_{3a} \iffdef \forall e \in \edges. \bigwedge_{\ruleA \in
     \ruleASet} \ruleAPred(e) \rightarrow \forall x_1 \ldots
     x_{\arityof{\labels(\ruleA)}} \in \vertices.
     \edgrel{\labels(\ruleA)}(e,x_1,\ldots,x_{\arityof{\labels(\ruleA)}}) \rightarrow \\ \bigwedge_{i \neq j \in \cpos(\ruleA) \cup \{\rpos(\ruleA)\} } x_i \neq x_j
    \end{multline*}
     states that for every edge the vertices at its root and child positions are pairwise different.
  \item The formula
    \begin{multline*}
    \phi_{3b} \iffdef  \forall e \in \edges.  \bigwedge_{\ruleA \in
    \ruleASet}  \ruleAPred(e) \rightarrow \forall x_1 \ldots
    x_{\arityof{\labels(\ruleA)}} \in \vertices.
    \edgrel{\labels(\ruleA)}(e,x_1,\ldots,x_{\arityof{\labels(\ruleA)}}) \rightarrow \\
    \bigwedge_{i \in \cpos(\ruleA)} x_i \neq y_{\rootof{u^\circ}}
    \end{multline*}
    states that the root $y_{\rootof{u^\circ}}$ is only incident to an edge $\edge$, if $y_{\rootof{u^\circ}}$ is not at a child position of $\edge$.
  \item The formula \begin{multline*}
    \phi_{3c} \iffdef \forall v \in \vertices. v \neq y_{\rootof{u^\circ}} \rightarrow \exists! e \in \edges.\\
     \bigvee_{\ruleA \in \ruleASet} \left( \ruleAPred(e) \land \exists x_1 \ldots x_{\arityof{\labels(\ruleA)}} \in \vertices.
    \edgrel{a}(e,x_1,\ldots,x_{\arityof{\labels(\ruleA)}}) \land \bigvee_{i \in \cpos(\ruleA)} x_i = v \right )
  \end{multline*}
    states that for every vertex $v \neq y_{\rootof{u^\circ}}$ there is exactly one edge $\edge$ such that $v$ is at a child position of $\edge$.
%
  \item $\phi_{3d}$ states that every vertex, except for the vertices $(y_i)_{i \in \nrootsof{u^\circ}}$, is reachable from $y_\rootof{u^\circ}$ using only the root and child positions of every edge (as this formula is standard we omit it for brevity).
  \end{compactitem}
\item $\phi_4 \iffdef \forall e \in \edges \;.\; \bigwedge_{\ruleA \in
  \ruleASet} \left( \ruleAPred(e) \rightarrow \psi^\ruleA_4(e)
  \right)$ states that every edge $\edge$ labelled by some rule
  $\ruleA = w \rightarrow \ksubst{\graph}{e_1}{u_1}{e_k}{u_k} \in
  \ruleASet$ needs to satisfy the formula
	\[
	\psi^\ruleA_4(e) \iffdef
	\forall v \in \vertices \; . \;
	\bigwedge_{1 \le i \le k}
	\left(
	\atindex_{\cpos(\ruleA,i)}(e,v)
	\rightarrow
	\bigvee_{\ruleC \in \ruleASet, \head(\ruleC) = u_i}
	\ruleCPred(v)
	\right)
	,
	\]
        which states that every vertex attached to $\edge$ at some child position $\cpos(\ruleA,i)$, is labelled by some rule with head $u_i$, where the auxiliary
        formula $\atindex_t(e,v)$ denotes that $v$ is at the $t$-th
        position of edge $\edge$, i.e.,
        \[
        \atindex_t(e,v)
        \iffdef
        \bigvee_{a \in \alphabet}
        \exists x_1, \ldots, x_{t-1}, x_{t + 1}, \ldots, x_{\arityof{a}} \in \vertices.
        \edgrel{a}(e,x_1 \ldots x_{t-1}, \;v,\;
        x_{t + 1}, \ldots, x_{\arityof{a}}).
        \]
      \item $\phi_5 \iffdef \forall v \in \vertices.
        \bigwedge_{\ruleC \in \ruleCSet} \ruleCPred(v) \rightarrow
        \left(\bigwedge_{w \in \mathcal{W}} \psi^C_w(v)\right)$ states
        that each vertex labelled by some rule $\ruleC = u \rightarrow
        w_1 \parallel_1 \ldots \parallel_1 w_k \in \ruleCSet$
        satisfies the formula $\psi^C_w$ for every $w \in
        \mathcal{W}$.  This formula is obtained as follows: Let us
        consider some $w \in \mathcal{W}$.  Let $p$ be the number of
        times $w$ appears among the $w_1 \parallel_1 \ldots
        \parallel_1 w_k$.  Let $\expon$ be the number such that there
        is a rule $u \rightarrow u \parallel_1 w^\expon \in \ruleBSet$
        (we recall that there is at most one rule of this shape), and
        let $\expon = 0$, otherwise.  We now define the formula
	\[
	\psi^C_w(v) \iffdef
	\exists X \subseteq \edges \;.\;
	\cardconstr{X}{q}{p}
	\wedge
	\forall e \;.\;
	e \in X
	\leftrightarrow
	\bigvee_{\ruleA \in \ruleASet, \ruleA \text{ has head } w}
	\left(
	\ruleAPred(e)
	\wedge
	\atindex_{\rpos(\ruleA)}(e,v)
	\right),
	\]
        which states that there are exactly $k\expon+p$ edges $\edge$,
        for some $k\in\nat$, such that $\edge$ is labelled by some
        rule $\ruleA \in \ruleASet$ with head $w$ and such that $v$ is incident to $\edge$ at position $\rpos(\ruleA)$.
\item $\phi_6 \iffdef \forall e \in \edges. \bigwedge_{\ruleA \in
  \ruleASet} \ruleAPred(e) \rightarrow \exists x_1 \ldots
  x_{\arityof{\labels(\ruleA)}} \in \vertices. \ \edgrel{\labels(\ruleA)}(e,x_1,\ldots
  x_{\arityof{\labels(\ruleA)}})$ states that $\ruleAPred(\edge)$ for some rule $\ruleA \in \ruleASet$, then the edge $\edge$ is labelled by $\labels(\ruleA)$.
\item The formula
   \begin{multline*}
  \phi_7 \iffdef \forall e \in \edges. \bigwedge_{\ruleA \in
  \ruleASet} \ruleAPred(e) \rightarrow \forall x_1 \ldots
  x_{\arityof{\labels(\ruleA)}} \in \vertices. \ \edgrel{\labels(\ruleA)}(e,x_1,\ldots
  x_{\arityof{\labels(\ruleA)}}) \rightarrow   \\
  \bigwedge_{i \in [1,\arityof{\labels(\ruleA)}])} \left( \bigvee_{j \in \nrootsof{u^\circ}} x_i = y_j \wedge
  \equalformula_{j,\vertedge(\ruleA,i)}(y_{\rootof{u^\circ}},e) \right)
  \vee
  \left( \bigvee_{\ruleC \in \ruleCSet} \ruleCPred(x_i) \wedge \equalformula_{\rootof{\head(\ruleC)},\vertedge(\ruleA,i)}(x_i,e) \right)
  \end{multline*}
  %
  states for every vertex $x_i$ incident to some edge $\edge$ that this vertex needs to be propagated according to the rules of the grammar.
  This propagation can be implemented by a finite state machine, and we provide an explicit \mso\ formula that implements this finite state machine.
  In case $x_i$ is not part of the spanning tree, $x_i$ needs to be equal to some vertex $y_j$ that is neither a root nor a future root of $u^\circ$.
  Then, we require that the source number $j$ for the non-terminal $u^\circ$ agrees with vertex $\vertedge(\ruleA,i)$ as propagated by the rules of the grammar.
  This is expressed by the formula $\equalformula_{j,\vertedge(\ruleA,i)}(y_{\rootof{u^\circ}},e)$.
  In case $x_i$ is part of the spanning tree, there is some rule $\ruleC \in \ruleCSet$ such that $\ruleCPred(x_i)$ holds.
  Then, we require that source $\rootof{\head(\ruleC)}$ agrees with vertex $\vertedge(\ruleA,i)$ as propagated by the rules of the grammar.
  This is expressed by the formula $\equalformula_{\rootof{\head(\ruleC)},\vertedge(\ruleA,i)}(x_i,e)$.
  We give the details on the construction of the formula $\equalformula_{i,z}(v,e)$ below.
  \item The formula
  \begin{multline*}
  \phi_8 \iffdef \forall v \in \vertices.
        \bigwedge_{\ruleC \in \ruleCSet} \ruleCPred(v) \rightarrow \\
        \bigwedge_{i \in [1,\arityof{\head(\ruleC)}]} \bigvee_{j \in \nrootsof{u^\circ}} \equalformula_{j,i}(y_{\rootof{u^\circ}},v) \vee \exists x. \bigvee_{\ruleC' \in \ruleCSet} ( \ruleCPred'(x) \wedge \equalformula_{\rootof{\head(\ruleC')},i}(x,v) )
  \end{multline*}
      states for every vertex $v$ in the spanning tree, i.e., for every vertex $v$ with $\ruleCPred(v)$ for some $\ruleC \in \ruleCSet$, that every source $i \in [1,\arityof{\head(\ruleC)}]$ needs to be propagated according to the rules of the grammar.
      This propagation can be implemented by a finite state machine, and we provide explicit an \mso\ formula that implements this finite state machine.
      In particular, the $i$-th source of $\head(\ruleC)$ must be equal to a vertex $x$ that is part of the spanning tree or to a vertex $y_j$ that is neither a root nor future root of $u^\circ$.
      For a vertex $x$ that is part of the spanning tree there must be some $\ruleC' \in \ruleCSet$ with $\ruleCPred'(x)$.
      Then, being equal to source $i$ of $\head(\ruleC)$ can be expressed by the formula $\equalformula_{\rootof{\head(\ruleC')},i}(x,v)$.
      For a vertex $y_j$ that is neither a root nor a future root of $u^\circ$, being equal to source $i$ of $\head(\ruleC)$ can be expressed by the formula $\equalformula_{j,i}(y_{\rootof{u^\circ}},v)$.
      We give the details on the construction of the formula $\equalformula_{i_1,i_2}(v_1,v_2)$ below.
\end{compactitem}

%

  The formulae $\equalformula_{i,z}(v,e)$ and $\equalformula_{i_1,i_2}(v_1,v_2)$ are both defined in terms of a formula $\rho$, which we define first.
  The formula $\rho$ uses monadic predicates $K_z$ for $z \in \vertof{\graphofrule(\ruleA)}$ and $\ruleA \in \ruleASet$, and $L_k$ for $k \in [1,\arityof{u}]$ and $u \in \nonterm \setminus \mathcal{W}$, in order to track vertices that are equal according to the propagation by the rules of the grammar.
  $\rho$ also uses the monadic predicates $M$ and $N$ (these two predicates are going to be used as blank symbols, i.e., they do not carry any information).
  $\rho$ is given as the conjunction of the following formulae:
\begin{itemize}
  \item $\rho_1 \iffdef \forall e \in \edges. \bigvee_{P \in \mathcal{P}} P(e) \land \bigwedge_{P \neq Q \in \mathcal{P}} \neg (P(e) \land Q(e))$ for $\mathcal{P} = \{K_z \mid z \in \vertof{\graphofrule(\ruleA)}, \ruleA \in \ruleASet\} \cup \{M\}$    states that the predicates $K_z$ and $M$ partition the edges.
  \item $\rho_2 \iffdef \forall e \in \edges. \bigwedge_{\ruleA \in
  \ruleASet} \ruleAPred(e) \rightarrow \bigvee_{P \in \{K_z \mid z \in \vertof{\graphofrule(\ruleA)}\} \cup \{M\}} P(e)$ states that every edge labelled by some $\ruleAPred$ for some $\ruleA \in \ruleASet$ must be labelled by $M$ or by $K_z$ for some $z \in \vertof{\graphofrule(\ruleA)}$.
  \item $\rho_3 \iffdef \forall v \in \vertices. \left(
      \bigvee_{P \in \mathcal{P}} P(e) \leftrightarrow \bigwedge_{i \in \nrootsof{u^\circ}} y_i \neq v \right) \land \bigwedge_{P \neq Q \in \mathcal{P}} \neg (P(e) \land Q(e))$ for $\mathcal{P} = \{L_i \mid i \in [1,\arityof{\head(\ruleC)}], \ruleC \in \ruleCSet\} \cup \{N\}$ states that the predicates $L_i$ and $N$ partition the vertices, except for the vertices $(y_i)_{i \in \nrootsof{u^\circ}}$, which are not labelled by any predicate.
   \item $\rho_4 \iffdef \forall v \in \vertices. \bigwedge_{\ruleC \in
  \ruleCSet} \ruleCPred(v) \rightarrow \bigvee_{P \in \{L_i \mid i \in [1,\arityof{\head(\ruleC)}]\} \cup \{N\}} P(e)$ states that every vertex labelled by some $\ruleCPred$ for some $\ruleC \in \ruleCSet$ must be labelled by $N$ or by $L_i$ for some $i \in [1,\arityof{\head(\ruleC)}]$.
  \item The formula
    \begin{multline*}
    \rho_5 \iffdef \forall v \in \vertices, e \in \edges. \bigwedge_{\ruleA = w \rightarrow \ksubst{\graph}{e_1}{u_1}{e_k}{u_k} \in \ruleASet} \ruleAPred(e) \rightarrow \forall x_1 \ldots x_{\arityof{\labels(\ruleA)}} \in \vertices. \edgrel{\labels(\ruleA)}(e,x_1,\ldots  x_{\arityof{\labels(\ruleA)}}) \rightarrow \\
    \bigwedge_{i \in [1,\arityof{w}]} L_i(x_{\rpos(\ruleA)}) \leftrightarrow  K_{\sourceof{\graph}(i)}(e) \wedge \bigwedge_{1 \le j \le k, 1 \le i \le \arityof{u_j}}  L_i(x_{\cpos(\ruleA,j)}) \leftrightarrow  K_{\edgerelof{\graph}(e_j)_i}(e)
    \end{multline*}
    states that if edge $\edge$ is labelled by $\ruleA = w \rightarrow \ksubst{\graph}{e_1}{u_1}{e_k}{u_k} \in \ruleASet$,
    then the vertex $v$ at the root position of $\edge$ is labelled by $L_i$ iff $\edge$ is labelled by $K_{\sourceof{\graph}(i)}$, and the vertex $v$ at the position $\cpos(\ruleA,j)$ of $\edge$ is labelled by $L_i$ iff $\edge$ is labelled by $K_{\edgerelof{\graph}(e_j)_i}$.
    This formula is the core of the propogation mechanism and ensures that we track vertices that are equal according to the rules of the grammar.

\item $\rho_6$ states that the vertices and edges labelled by $L_i$ resp. $K_z$ are connected, using only the root and child positions of every edge (as this formula is standard we omit it for brevity).
\end{itemize}
\noindent
We can then set
$$\equalformula_{i,z}(v,e) \iffdef \exists (K_z)_z, (L_i)_i, M, N. \ \rho \wedge L_i(v) \wedge K_z(e)$$
and
$$\equalformula_{i_1,i_2}(v_1,v_2) \iffdef \exists (K_z)_z, (L_i)_i, M, N. \ \rho \wedge L_{i_1}(v_1) \wedge L_{i_2}(v_2).$$

We now consider some $u \in \nonterm \setminus \mathcal{W}$, a graph $\graph$ of type $\arityof{u}$, and a valuation $\store$  with $\store(y_i) = \sourceof{\graph}(i)$ for all $i \in [1,\arityof{u}]$.
We prove that $\rmsources{\graph} \Models^\store \exists \ruleAPred_1 \ldots \ruleAPred_{n_A}, \ruleCPred_1 \ldots \ruleCPred_{n_C}. \ \psi(u)$ iff $\graph \in \langof{\grammar}{u}$.

We first assume $\graph \in \langof{\grammar}{u}$ and show that $\rmsources{\graph} \Models^\store \psi(u)$ for some valuation of $\ruleAPred_1 \ldots \ruleAPred_{n_A}, \ruleCPred_1 \ldots \ruleCPred_{n_C}$.
We proceed by induction on the number of rules by which $\graph$ was derived.
Then, the derivation of $\graph$ starts with zero or more applications of rules $\ruleB \in \ruleBSet$ of form $u \rightarrow u \parallel_1 w^\expon$.
At some point there is an application of a rule $\ruleC = u \rightarrow w_1 \parallel_1 \ldots \parallel_1 w_n \in \ruleCSet$.
Then, indexing by $m$, for each non-terminal $w_m$, the derivation continues with an application of some rule $\ruleA^m = w_m \rightarrow \ksubst{\graph^m}{e_1^m}{u_1^m}{e_k^m}{u_k^m} \in \ruleASet$.
Let $e_m$ be the single terminal edge of the graph $\graph^m$.
Let us now assume that for each $u^m_j$ we have derived a graph $\graph^m_j \in \langof{\grammar}{u^m_j}$.
Then,
we obtain
\begin{align*}
&\graph =
\graph^1[e_1^1/G_1^1,\ldots,e_{n_1}^1/G_{n_1}^1]
\cdots
\parallel_1
\graph^m[e_1^m/G_1^m,\ldots,e_{n_m}^m/G_{n_m}^m]
\parallel_1
\cdots
\\
&\qquad\qquad
\parallel_1
\graph^k[e_1^k/G_1^k,\ldots,e_{n_k}^k/G_{n_k}^k].
\end{align*}
We note that each of the graphs $\graph^m_j$ is derived by less rule applications than $\graph$.
We define the valuation $\store^m_j$ by setting $\store^m_j(y_i^{m,j}) = \sourceof{\graph}(i)$ for all $i \in [1,\arityof{u^m_j}]$.
By the induction assumption, we have that $\rmsources{\graph^m_j} \Models^{\store^m_j} \psi(u^m_j)$ for some sets $\ruleAPred_1^{m,j}, \ldots, \ruleAPred_{n_A}^{m,j},
\ruleCPred_1^{m,j}, \ldots, \ruleCPred_{n_C}^{m,j}$.
We now set $\ruleAPred_t = \bigcup_{m,j} \ruleAPred^{m,j}_t
\cup \bigcup_{m} \{e_m \mid \ruleA^m = \ruleA_t\}$ for all $1 \le t \le n_A$, $\ruleCPred_t = \bigcup_{m,j} \ruleCPred^{m,j}_t$ for all $1 \le t \le n_C$ with $\ruleC_t \neq \ruleC$, and $\ruleCPred_t = \bigcup_{m,j} \ruleCPred^{m,j}_t \cup \{\sourceof{\graph}(\rootof{u})\}$, otherwise.

We now verify that $\rmsources{\graph} \Models^\store \psi(u)$:
\begin{compactitem}
\item $\phi_0$: This formula holds as we require sources to be pairwise distinct.
\item $\phi_1$:
Each edge $e$ is either part of some subgraph $\graph^m_j$, or is equal to some edge $e_m$.
The way we defined the sets $\ruleAPred_t$ still partitions all the edges in $\graph$.
\item $\phi_2$: Each vertex is either a source of $\graph$, an internal node of $\graph^m$, or belongs to $\graph^m_j$.
    We note that for a vertex $v$ of $\graph$ that is neither a source of $\graph$ nor an internal node of $\graph^m$, there is exactly one graph $\graph^m_j$ such that $v \in \vertof{\graph^m_j}$ and we get from $\rmsources{\graph^m_j} \Models^{\store^m_j} \psi(u^m_j)$ that there is exactly one rule $\ruleC \in \ruleCSet$ with $\ruleCPred^{m,j}(v)$.
    Further, for every internal vertex $v \in \internal{\graph^m}$, there is exactly one $u^m_j$ with $v = \store^m_j(y^{m,j}_i)$ for some $i \in \rootsof{u^m_j}$ because of the condition
    $\internal{\graph^m} \cup \{ \sourceof{\graph}(i) \mid i \in \rootsof{w^m} \}= \biguplus_{j=1..n_m} \{\edgerelof{\graph^m}(e^m_j)_i \mid i \in \{\rootof{u^m_j}\} \cup \rootsof{u^m_j}\}$.
    Hence, we get from $\rmsources{\graph^m_j} \Models^{\store^m_j} \psi(u^m_j)$ that there is exactly one rule $\ruleC \in \ruleCSet$ with $\ruleCPred^{m,j}(v)$.
    We now consider a vertex $v = \sourceof{\graph}(i)$ for some $i \in \rootsof{u}$.
    Then, because of the condition $\rootsof{u} = \biguplus_{i=1..k}
    \rootsof{w_i}$ of $\ruleC$ and the conditions  $\internal{\graph^m} \cup \{ \sourceof{\graph}(i) \mid i \in \rootsof{w^m} \}= \biguplus_{j=1..n_m} \{\edgerelof{\graph^m}(e^m_j)_i \mid i \in \{\rootof{u^m_j}\} \cup \rootsof{u^m_j}\}$, there is exactly one $u^m_j$ with $v = \store^m_j(y^{m,j}_{i'})$ for some $i' \in \rootsof{u^m_j}$.
    Hence, we get from $\rmsources{\graph^m_j} \Models^{\store^m_j} \psi(u^m_j)$ that there is exactly one rule $\ruleC \in \ruleCSet$ with $\ruleCPred^{m,j}(v)$.
    Finally, for the root $\sourceof{\graph}(\rootof{u})$ we have set $\ruleCPred(\store(y_{\rootof{u}}))$ for the rule $\ruleC \in \ruleCSet$ that was applied in the beginning of the derivation (after zero or more applications of rules $\ruleB \in \ruleBSet$).
\item $\phi_3$:
    We get from $\rmsources{\graph^m_j} \Models^{\store^m_j} \psi(u^m_j)$ that there is a spanning tree of $\graph^m_j$ with root $\sourceof{\graph}(\rootof{u^m_j})$ that uses only the root and child position of each edge.
    We then observe that adding the edges $e_1,\ldots,e_m,\ldots,e_k$ gives a spanning tree of $\graph$ with root $\sourceof{\graph}(\rootof{u})$ that uses only the root and child position of each edge.
\item $\phi_4$:
    We get from $\rmsources{\graph^m_j} \Models^{\store^m_j} \psi(u^m_j)$ that this property holds for all edges of graph $\graph^m_j$.
    We have that the root $\sourceof{\graph^m_j}(\rootof{u^m_j})$ of every $\graph^m_j$ is labeled by some rule with head $u^m_j$.
    Hence, the property holds for all edges $e_m$.
\item $\phi_5$: We get from $\rmsources{\graph^m_j} \Models^{\store^m_j} \psi(u^m_j)$ that this property holds for all vertices of
  $\graph^m_j$.  We now consider the root $y_{\rootof{u}}$.  We recall that
  the root is labelled by some rule $\ruleC = u \rightarrow w_1
  \parallel_1 \ldots \parallel_1 w_k \in \ruleCSet$.  Let $w \in
  \nonterm \setminus \mathcal{W}$ be some non-terminal.  Let $p$ be
  the number of times $w$ appears among the $w_1 \parallel_1 \ldots
  \parallel_1 w_k$.  Let $\expon$ be the number such that there is a
  rule $u \rightarrow u \parallel_1 w^\expon \in \ruleBSet$ (we recall
  that there is at most one rule of this shape), and let $\expon = 0$,
  otherwise.  It is now easy to verify that there are at least $p$
  edges, which are labelled by rules with head $w$, and the number of
  these edges equals $p$ modulo $q$.  As this holds for every $w \in
  \nonterm \setminus \mathcal{W}$, the property holds for the root
  $y_{\rootof{u}}$.
\item $\phi_6$: Clearly, this property holds for all edges of $\graph^m_j$.
    The property further holds for the edges $e_m$.
\item $\phi_7$:
  We consider some edge $\edge$ that is part of some $\graph^m_j$ labelled by some $\ruleA \in
  \ruleASet$, and some vertex $v = \edgerelof{\graph^m_j}(i)$ for $i \in [1,\lenof{\edgerelof{\graph^m_j}(\edge)}]$.
  We get from $\rmsources{\graph^m_j} \Models^{\store^m_j} \phi_7$ that either we have
  $v = \sourceof{\graph^m_j}(l)$ and $\equalformula_{l,\vertedge(\ruleA,i)}(\sourceof{\graph^m_j}(\rootof{u^m_j}),\edge)$ for some $l\in \nrootsof{u^m_j}$, or there is some $\ruleC \in \ruleCSet$ with $\ruleCPred(v)$ and $\equalformula_{\rootof{\head(\ruleC)},\vertedge(\ruleA,i)}(v,\edge)$.
  In the later case, the property continues to hold for $\graph$ because $\graph^m_j$ is a subgraph of $\graph$.
  Hence, we only need to consider $v = \sourceof{\graph^m_j}(l)$ and $\equalformula_{l, \vertedge(\ruleA,i)}(\sourceof{\graph^m_j}(\rootof{u^m_j}),\edge)$ for some $l\in \nrootsof{u^m_j}$.
  We first consider the case $v \in \internal{\graph^m}$.
  Then, we must have $v = \sourceof{\graph^m_{j'}}(l')$ for some $1\le j' \le n_m$ and $l' \in \{\rootof{u^m_{j'}}\} \cup \rootsof{u^m_{j'}}$ because of the condition $\internal{\graph^m} \cup \{ \sourceof{\graph}(i) \mid i \in \rootsof{w^m} \}= \biguplus_{j=1..n_m} \{\edgerelof{\graph^m}(e^m_j)_i \mid i \in \{\rootof{u^m_j}\} \cup \rootsof{u^m_j}\}$.
  Now, we get from $\rmsources{\graph^m_{j'}} \Models^{\store^m_{j'}} \phi_8$ that there is some vertex $v'$ with $\ruleCPred(v')$ for some $\ruleC \in \ruleCSet$ such that $\equalformula_{\rootof{\head(\ruleC)},l'}(v',\sourceof{\graph^m_{j'}}(\rootof{u^m_{j'}}))$.
  Together with $\equalformula_{l,l'}(\sourceof{\graph^m_j}(\rootof{u^m_j}),\sourceof{\graph^m_{j'}}(\rootof{u^m_{j'}}))$,
  because of $v = \sourceof{\graph^m_j}(l) =\sourceof{\graph^m_{j'}}(l')$,
  we get that $\graph \Models^\store \equalformula_{\rootof{\head(\ruleC)},\vertedge(\ruleA,i)}(v',\edge)$.
  We now assume that $v = \sourceof{\graph}(t)$ for some $1 \le t \le \arityof{u}$.
  In case of $t \in \nrootsof{u}$, there is nothing to show.
  In case of $v = \sourceof{\graph}(\rootof{u})$, the claim follows from the fact that
  $\equalformula_{l,\rootof{u}}(\sourceof{\graph^m_j}(\rootof{u^m_j}),\sourceof{\graph}(\rootof{u}))$.
  Hence, it remains to deal with $t \in \rootsof{u}$.
  We note that because of the conditions $\rootsof{u} = \biguplus_{i=1..k} \rootsof{w_i}$ and $\internal{\graph^m} \cup \{ \sourceof{\graph}(i) \mid i \in \rootsof{w^m} \}= \biguplus_{j=1..n_m} \{\edgerelof{\graph^m}(e^m_j)_i \mid i \in \{\rootof{u^m_j}\} \cup \rootsof{u^m_j}\}$ we must have $v = \sourceof{\graph^{m'}_{j'}}(l')$ for some $1\le m' \le k$, $1\le j' \le n_{m'}$ and $l' \in \{\rootof{u^m_{j'}}\} \cup \rootsof{u^{m'}_{j'}}$.
  We can then conclude similarly as in the case of $v \in \internal{\graph^m}$.
  We finally consider some edge $e_m$.
  We note that the claim follows by arguments similar to the above.

\item $\phi_8$: This property can be proven similarly to property $\phi_7$.
\end{compactitem}

We now assume $\rmsources{\graph} \Models^\store \psi(u)$ for some valuations of $\ruleAPred_1 \ldots \ruleAPred_{n_A}, \ruleCPred_1 \ldots \ruleCPred_{n_C}$.
We will show that $\graph \in \langof{\grammar}{u}$.
We proceed by induction on the height of the spanning tree, whose existence is ensured by subformula $\phi_3$ of $\psi(u)$.
Let $\rootVar = \store(y_{\rootof{u}})$ be the root of the spanning tree, and let $\edge_1, \ldots, \edge_k$ be the edges of $\graph$ such that $r$ is at the root position of each $\edge_i$.
Let $\ruleA^m = w_m \rightarrow    \ksubst{\graph^m}{e_1^m}{u_1^m}{e_{n_m}^m}{u_{n_m}^m} \in \ruleASet$ be the rule such that $\ruleAPred^m(\edge_m)$ and let $\edgerelof{\graph}(\edge_m) = \rootVar^m_1 \ldots \rootVar^m_{\lenof{\edgerelof{\graph}(\edge_m)}}$ be the incident vertices, for $1 \le m \le k$.
We note that $\graph \Models^\store \psi(u)$ implies that $\rootVar = \rootVar^m_{\rpos(u_m)}$ and $\labels(\ruleA^m) = \labof{\graph}(\edge_m)$ for all $1 \le m \le k$.
Moreover, $j = \cpos(\ruleA,i)$, for some $i \in [1,n_m]$, implies that $\ruleCPred(\rootVar^m_j)$, for some rule $\ruleC \in \ruleCSet$ with $\head(\ruleC) = u^m_j$.

For each $j \in \cpos(\ruleA^m)$, we are now going to define a subgraph $\graph^m_j$ of $\graph$:
We first define its sources.
For this, we are going to use that $\rmsources{\graph} \Models^\store \phi_8$.
We instantiate $\phi_8$ with $v = \rootVar^m_j$ and $i \in [1, \arityof{u^m_j}]$, and set $\sourceof{\graph^m_j}(i) = \store(y_l)$ in case of $\rmsources{\graph} \Models^\store  \equalformula_{l,i}(\rootVar,\rootVar^m_j)$ for some $l \in \nrootsof{u}$, or we set $\sourceof{\graph^m_j}(i)$ as the value of $x$ in case of $\rmsources{\graph} \Models^\store \exists x. \bigvee_{\ruleC' \in \ruleCSet} ( \ruleCPred'(x) \wedge \equalformula_{\rootof{\head(\ruleC')},i}(x,\rootVar^m_j)$. We note that these values are unique.
We then consider the subtree rooted in $\rootVar^m_j$ of the spanning tree with root $r$.
Let $V$ resp. $E$ be the vertices that appear in this subtree.
We then set $\vertof{\graph^m_j} = V \cup \{\sourceof{\graph^m_j}(i) \mid i \in [1, \arityof{u^m_j}] \}$, $\edgeof{\graph^m_j} = E$ and $\labof{\graph^m_j} = \domrestrict{\labof{\graph}}{E}$.
We now choose the sets $\ruleAPred_1^{m,j}, \ldots, \ruleAPred_{n_A}^{m,j}, \ruleCPred_1^{m,j}, \ldots, \ruleCPred_{n_C}^{m,j}$ as the restriction of the sets $\ruleAPred_1 \ldots \ruleAPred_{n_A}, \ruleCPred_1 \ldots \ruleCPred_{n_C}$ to the vertices and edges of graph $\graph^m_j$.
It can now be verified that $\rmsources{\graph^m_j} \Models^{\store^m_j} \phi(u^m_j)$ for the valuation $\store^m_j$ defined by $\store^m_j(y_i^{m,j}) = \sourceof{\graph}(i)$ for all $i \in [1,\arityof{u^m_j}]$.
By the induction assumption we then obtain that
$\graph^m_j \in \langof{\grammar}{u^m_j}$.

We now argue for every edge $\edge_m$ and $l \in [1,\lenof{\edgerelof{\graph}(\edge_m)}]$ that $r^m_l = \store(y_i)$ for some $i \in [1,\arityof{u}]$ or that $r^m_l = \sourceof{\graph^m_j}(i)$ for some $i \in [1,\arityof{u^m_j}]$ and $1 \le j \le n_m$ (\#).
For this, we are going to use that $\rmsources{\graph} \Models^\store \phi_7$.
We consider some $l \in [1,\lenof{\edgerelof{\graph}(\edge_m)}]$ with $r^m_l \neq \store(y_i)$ for all $i \in [1,\arityof{u}]$.
Hence, $\vertedge(\ruleA^m,l) \in \internal{\graph^m}$.
Because of $\internal{\graph^m} \cup \{ \sourceof{\graph}(i) \mid i \in \rootsof{w^m} \}= \biguplus_{j=1..n_m} \{\edgerelof{\graph^m}(e^m_j)_i \mid i \in \{\rootof{u^m_j}\} \cup \rootsof{u^m_j}\}$ we have that there is an $j \in [1,n_m]$ with
$\vertedge(\ruleA^m,l) = \edgerelof{\graph^m}(e^m_j)_i$ for $i \in \{\rootof{u^m_j}\} \cup \rootsof{u^m_j}$.
This means that $\rmsources{\graph} \Models^\store \equalformula_{i,\vertedge(\ruleA^m,l)}(r^m_j,\edge_m)$ (+).
Because of $\rmsources{\graph} \Models^\store \phi_7$ we must have $\rmsources{\graph} \Models^\store \equalformula_{\rootof{\head(\ruleC)},\vertedge(\ruleA^m,l)}(r^m_l,\edge_m)$ for the rule $\ruleC \in \ruleCSet$ such that $\ruleCPred(r^m_l)$.
With (+) this establishes $\rmsources{\graph} \Models^\store \equalformula_{\rootof{\head(\ruleC)},i}(r^m_l,r^m_j)$.
Hence, we have $\sourceof{\graph^m_j}(i) = r^m_l$ by the definition of the sources of $\graph^m_j$.
By similar arguments, we establish that $\sourceof{\graph}(i) = r^m_l$ iff $\sourceof{\graph^m}(i) = \vertedge(\ruleA^m,l)$, and that $\sourceof{\graph}(i) = \sourceof{\graph^m_j}(i')$ iff $\sourceof{\graph^m}(i) = \edgerelof{\graph^m}(e^m_j)_{i'}$.
With (\#), we then get
\begin{align*}
&\graph =
\graph^1[e_1^1/G_1^1,\ldots,e_{n_1}^1/G_{n_1}^1]
\cdots
\parallel_1
\graph^m[e_1^m/G_1^m,\ldots,e_{n_m}^m/G_{n_m}^m]
\parallel_1
\cdots
\\
&\qquad\qquad
\parallel_1
\graph^k[e_1^k/G_1^k,\ldots,e_{n_k}^k/G_{n_k}^k] (*).
\end{align*}

Now, let $\ruleC = u \rightarrow w_1 \parallel_1 \ldots \parallel_1 w_n \in \ruleCSet$ be the rule such that $\ruleCPred(r)$.
Let $w \in \nonterm \setminus \mathcal{W}$ be some non-terminal.  Let $p$ be the number of times $w$ appears among the $w_1 \parallel_1 \ldots \parallel_1 w_n$.
Let $\expon$ be the number such that there is a rule $u \rightarrow u \parallel_1 w^\expon \in \ruleBSet$ (we recall that there is at most one rule of this shape), and let $\expon = 0$, otherwise.  Because of $\rmsources{\graph} \Models \psi(u)$ there are at least $p$ edges among the $e_1,\ldots,e_k$, which are labelled by rules with head $w$, and the number of these edges equals $p$ modulo $q$.
As this holds for every $w \in \nonterm \setminus \mathcal{W}$, we can apply zero or more times rules $\ruleB \in \ruleBSet$, followed by a single application of rule $\ruleC = u \rightarrow w_1 \parallel_1 \ldots \parallel_1 w_n \in \ruleCSet$, in order to obtain a partial derivation that can be summarized as $u \rightarrow w_1 \parallel_1 \cdots \parallel_1 w_k$,
where $w_m$ is the head of the rule $\ruleA^m \in \ruleASet$.  Then, we continue the derivation by applying the rule $\ruleA^m$ for every $m \in [1,k]$ in order to obtain the partial derivation
\begin{align*}
&u \rightarrow \ksubst{\graph^1}{e_1^1}{u_1^1}{e_{n_1}^1}{u_{n_1}^1} \parallel_1
\cdots
\parallel_1
\ksubst{\graph^1}{e_1^m}{u_1^m}{e_{n_m}^m}{u_{n_m}^m} \parallel_1
\cdots
\\
&\qquad\qquad
\parallel_1 \ksubst{\graph^1}{e_1^k}{u_1^k}{e_{n_k}^k}{u_{n_k}^k}.
\end{align*}
This partial derivation together with (*) and $\graph^m_j \in \langof{\grammar}{u^m_j}$ then establishes $\graph \in \langof{\grammar}{u}$.
\end{proofE}

The rest of this section is concerned with proving a partial
completeness result. We define a parameter, called \emph{embeddable tree-width}, which is an over-approximation of the standard
tree-width of a graph. We show that the class of \cmso-definable sets
of graphs, for which the embeddable tree-width is bounded, coincides
with the class of tree-verifiable graph languages.

Before defining embeddable tree-width formally, we recall the standard
definitions of \emph{tree decomposition} and \emph{treewidth} of a
c-graph, by considering trees over an alphabet of edge labels that
consists of a single binary element:

\begin{definition}\label{def:tree-decomposition}
  A \emph{tree decomposition} of a c-graph $\graph$ of type $n$ is a pair
  $(\tree,\beta)$, where $\tree$ is a tree, called \emph{backbone}, and $\beta : \vertof{\tree}
  \rightarrow \pow{\vertof{\graph}}$ maps the vertices of $\tree$ into
  sets of vertices from $\graph$, such that:
  \begin{compactenum}
  \item\label{it1:treewidth} for each $e \in \edgeof{\graph}$,
    $\edgerelof{\graph}(e) = \tuple{v_1, \ldots, v_k}$, there exists
    $n \in \vertof{\tree}$ such that $v_1, \ldots, v_k \in \beta(n)$,
  \item\label{it2:treewidth} for each $v \in \vertof{\graph}$
    the set $\set{n \in \vertof{\tree} \mid v \in \beta(n)}$ is
    nonempty and connected in $\tree$.
  \end{compactenum}
  The \emph{width} of the tree decomposition is $\width{\tree,\beta} \isdef \max_{n \in \vertof{\tree}} \cardof{\beta(n)}-1$.
  The \emph{tree-width} of $\graph$ is $\twof{\graph} \isdef \min
  \set{\width{\tree,\beta} \mid (\tree,\beta) \text{ tree decomposition of } \graph}$.
  The tree-width of a graph is the tree-width of some c-graph from it\footnote{Isomorphic c-graphs have equal tree-widths.}.
  A set $\graphs$ of graphs has \emph{bounded tree-width} iff $\set{\twof{\graph} \mid \graph\in\graphs}$ is a finite set.
\end{definition}
We refine the above definition by considering only backbones that are
spanning trees of the graph. The spanning trees considered cover both
the vertices and the edges of the graph:

\begin{definition}
\label{def:embeddable-tree-decomposition}
  An \emph{embeddable} tree-decomposition of a c-graph $\graph$ is a
  tuple $(\tree,\beta,\gamma,\delta)$ such that $(\tree,\beta)$ is a
  tree-decomposition of $\graph$, and there is some subset $W
  \subseteq \vertof{\tree}$ of the nodes of $\tree$ such
  that: \begin{compactitem}
  \item the root of $\tree$ does not belong to $W$,
  \item for every node $u \in \vertof{\tree}$ and every child $w \in
    \vertof{\tree}$ of $u$, we have $w \in W$ iff $u \not\in W$, i.e.,
    the membership in $W$ alternates for the nodes on every path of
    $\tree$,
  \end{compactitem}
  Moreover, $\gamma: \vertof{\tree} \setminus W \rightarrow
  \vertof{\graph}$ is a bijective mapping of the nodes from
  $\vertof{\tree} \setminus W $ to the vertices of $\graph$, and
  $\delta: W \rightarrow \edgeof{\graph}$ is a bijective mapping of
  the vertices in $W$ to the edges of $\graph$, satisfying the
  conditions: \begin{compactenum}
  \item\label{it1:embeddable} for all $u \in \vertof{\tree} \setminus
    W$ we have that $\gamma(u) \in \beta(u)$, and
  \item\label{it2:embeddable}
    for all $w \in W$ and $e = \delta(w)$ we have
    \begin{compactitem}
    \item if $w$ is the child of $u$ in $\tree$ then there is some $i
      \in \interv{1}{\arityof{\labof{\graph}(e)}}$ such that
      $\edgerelof{\graph}(e)_i = \gamma(u)$,
    \item for every child $v$ of $w$ in $\tree$ there is a
      distinct $j \in
      \interv{1}{\arityof{\labof{\graph}(e)}}\setminus\set{i}$ such
      that $\edgerelof{\graph}(e)_j = \gamma(v)$,
    \item $\edgerelof{\graph}(e)_i \in \beta(w)$ for all $1 \le i \le \lenof{\edgerelof{\graph}(e)}$.
    \end{compactitem}
  \end{compactenum}
  The \emph{embeddable tree-width} of $\graph$ is $\etwof{\graph}
  \isdef \min \{\width{\tree,\beta} \mid (\tree,\beta,\gamma,\delta)
  \text{ is an embeddable tree decomposition of } \graph\}$. The
  embeddable tree-width of a graph is the embeddable tree-width of
  some c-graph from it.  A set $\graphs$ of graphs has \emph{bounded
  embeddable tree-width} iff $\set{\etwof{\graph} \mid
    \graph\in\graphs}$ is a finite set.
\end{definition}
For example, Fig.~\ref{fig:etd} shows a graph that is generated by the tree-verifiable grammar from Example~\ref{ex:tree-verifiable-tll} and an embeddable tree decomposition of this graph
(this tree decomposition is the tree decomposition that underlies the proof of Proposition~\ref{prop:tree-verifiable-embeddable}).
The nodes from the inverse
image of $\gamma$ (resp. $\delta$) are depicted by filled circles
(resp. rectangles).
The sets in the image of $\beta$ (called \emph{bags}) associated with the nodes of the tree decomposition are stated next to the tree nodes.

\begin{figure}[t!]
  \centerline{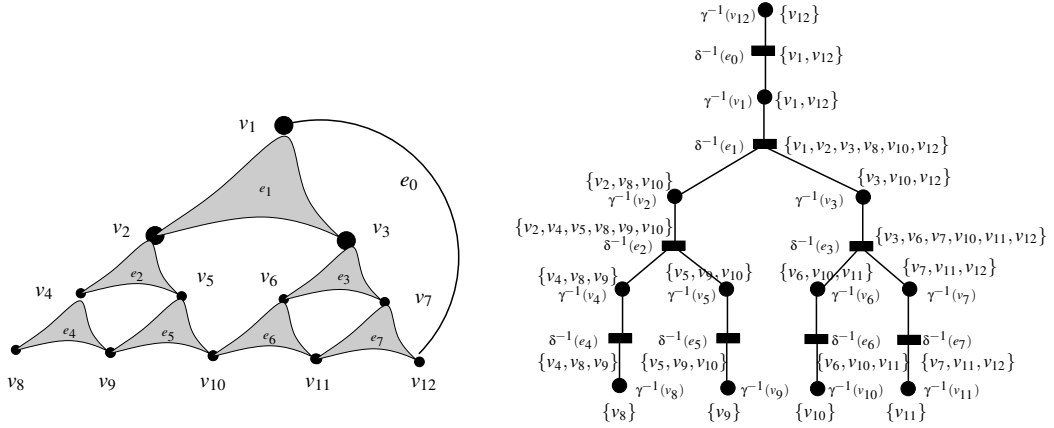}

  \vspace*{-.5\baselineskip}
  \caption{Embeddable Tree Decomposition}
  \label{fig:etd}
  \vspace*{-.5\baselineskip}
\end{figure}

The following variation of
Def.~\ref{def:embeddable-tree-decomposition} will be useful for
working with the notion of an embeddable tree-decomposition in
analyzing graph grammars and for conducting inductive proofs about
graph grammars:

\begin{definition}\label{def:embeddable-tree-decomposition-v2}
  Let $\graph$ be c-graph $\graph$ of type $n$ and let $\iroot \in
  \interv{1}{n}$ and $\iroots \subseteq \interv{1}{n} \setminus
  \set{\iroot}$.  We consider a tuple $(\tree,\beta,\gamma,\delta)$
  and subset $W \subseteq \vertof{\tree}$ such that the conditions of
  Def.~\ref{def:embeddable-tree-decomposition} are satisfied, except
  for the bijectivity requirement on $\gamma$, and the requirement
  that the root of $\tree$ does not belong to $W$.  Let $r$ be the
  root of $\tree$.  We now call $(\tree,\beta,\gamma,\delta)$ a
  \emph{$(\iroot, \iroots)$-embeddable tree-decomposition} of $\graph$
  iff the following hold:~\begin{compactenum}
  \item $\sourceof{\graph}(i) \in \beta(r)$ for all $i \in \interv{1}{n}$,
  \item $\gamma: \vertof{\tree} \setminus W \rightarrow
    \internal{\graph} \cup A$ is a bijective mapping of the vertices
    in $\vertof{\tree} \setminus W$ to the internal vertices of
    $\graph$ plus the set $A = \iroots$ in case of $r \in W$ and $A =
    \{\iroot \} \cup \iroots$ otherwise,
  \item if $r \in W$ then there is some $i \in
    \interv{1}{\arityof{\labof{\graph}(\delta(r))}}$ such that
    $\edgerelof{\graph}(\delta(r))_i = \sourceof{\graph}(\iroot)$ and
    $\gamma(u) \neq \sourceof{\graph}(\iroot)$ for every child $u$ of
    $r$,
  \item else ($r \in \vertof{\tree} \setminus W$) we have that
    $\gamma(r) = \sourceof{\graph}(\iroot)$.
  \end{compactenum}
  We call the tree-decomposition \emph{pointed}, if $r \in W$, and \emph{non-pointed}, otherwise.
\end{definition}

To begin with, we note that tree-verifiable graph grammars are set up
such that the resulting languages always have bounded embeddable
tree-width:

\begin{propositionE}\label{prop:tree-verifiable-embeddable}
  Let $\langu$ be a tree-verifiable set of graphs.
  Then, $\langu$ has bounded embeddable tree-width.
  In more detail, given a tree-verifiable
  graph-grammar $\grammar = (\nonterm,\rules)$, let $k$ be the maximum of $\cardof{\vertof{\graph}}$ for the graphs $\graph$
  appearing in $\rules$ of shape~\ref{it1:tree-verifiable-graph-grammar} and the maximum of $\arity{u}$ for all non-terminals $u \in \nonterm$;
  then we have $\etwof{\graph} \le k$ for all $\graph \in \lang{\grammar}$.
\end{propositionE}
\begin{proofE}
  Let $\grammar = (\nonterm,\rules)$ be some tree-verifiable
  graph-grammar such that $\langu=\lang{\grammar}$, with $\mathcal{W}
  \subseteq \nonterm$ for some subset of the non-terminals.  Let $k$
  be the maximum of $\cardof{\vertof{\graph}}$ for the graphs $\graph$
  appearing in $\rules$ of
  shape~\ref{it1:tree-verifiable-graph-grammar} and the maximum of $\arity{u}$ for all non-terminals $u \in \nonterm$.
  We now prove for every
  $v \in \nonterm$ and $\graph \in \langof{\grammar}{v}$ that there is
  a $(\rootof{v}, \rootsof{v})$-embeddable tree-decomposition with
  width $\le k$.  Further, this tree-decomposition is pointed iff $v
  \in \mathcal{W}$.  The proof proceeds by induction on the length of
  the derivation of $\graph \in \langof{\grammar}{v}$.

  We first consider a nonterminal $u \in \nonterm \setminus
  \mathcal{W}$ and a graph $\graph \in \langof{\grammar}{v}$.  Let $n
  = \arityof{u}$.  We have that the derivation of $\graph \in
  \langof{\grammar}{u}$ starts with several applications of rules of
  shape $u \rightarrow u \parallel_n w^\expon$, followed by one
  application of shape $u \rightarrow w_1 \parallel_n \ldots
  \parallel_n w_k$.  In summary, we have $\graph = \graph^1
  \parallel_n \graph^2 \parallel_n \dots $ for some graphs $\graph^j
  \in \langof{\grammar}{w_j}$ and non-terminals $w_j \in \mathcal{W}$.
  In particular, we can assume that the edges of the graphs $\graph^j$
  are disjoint and that the only common vertices of $\graph^j$ are
  among the vertices $[1,n]$.  By induction assumption we have that
  there is a pointed $(\rootof{w_j}, \rootsof{w_j})$-embeddable
  tree-decomposition $(\tree^j,\beta^j,\gamma^j,\delta^j)$ of
  $\graph^j$ with width $\le k$ and a corresponding subset $W^j
  \subseteq \vertof{\tree^j}$.  We now construct a tree $\tree$ by
  adding a fresh root vertex $r$ that we connect to the roots $r^j$ of
  the trees $\tree^j$, and we set $W = \bigcup_j W^j$, $\beta = \{r
  \mapsto \{ \sourceof{\graph}(i) \mid i \in [1,\arityof{u}] \}\} \cup
  \bigcup_j \beta^j$, $\gamma = \{r \mapsto
  \sourceof{\graph}(\rootof{u})\} \cup \bigcup_j \gamma^j$ and $\delta
  = \bigcup_j \delta^j$.  It is now easy to verify that
  $(\tree,\beta,\gamma,\delta)$ is a non-pointed $(\rootof{u},
  \rootsof{u})$-embeddable tree-decomposition with width $\le k$; we
  note that the bijectivity requirement of $\gamma$ is satisfied
  because of the side-conditions $\rootof{w} = \rootof{u}$ and
  $\rootsof{w} = \emptyset$ of the rules of
  shape~\ref{it2:tree-verifiable-graph-grammar} and the
  side-conditions $\rootof{u} = \rootof{w_i}$ and $\rootsof{u} =
  \biguplus_{i=1..l} \rootsof{u_i}$ of the rules of
  shape~\ref{it3:tree-verifiable-graph-grammar}.

  We now consider a nonterminal $w \in \mathcal{W}$ and a graph
  $\graph \in \langof{\grammar}{w}$. We have that there is an
  application of some rule $w \rightarrow
  \ksubst{\graph^\circ}{e_1}{u_1}{e_k}{u_k}$, for some $u_1,\ldots,u_k
  \in \nonterm \setminus \mathcal{W}$.  Let $e$ be the single terminal
  edge of $\graph^\circ$.  We have that there are some graphs
  $\graph^j \in \langof{\grammar}{u_j}$ such that $\graph =
  \graph^\circ[e_1/\graph^1,\ldots, e_k/\graph^k]$.  In particular, we
  can assume that the edges of the graphs $\graph^j$ are disjoint and
  that the only common vertices of $\graph^j$ are among the
  $\vertof{\graph^\circ}$.  By induction assumption we have that there
  is a non-pointed $(\rootof{u_j}, \rootsof{u_j})$-embeddable
  tree-decomposition $(\tree^j,\beta^j,\gamma^j,\delta^j)$ of
  $\graph^j$ with width $\le k$ and a corresponding subset $W^j
  \subseteq \vertof{\tree^j}$.  We now construct a tree $\tree$ by
  adding a fresh root vertex $r$ that we connect to the roots $r^j$ of
  the trees $\tree^j$, and we set $W = \{r\} \cup \bigcup_j W^j$,
  $\beta = \{r \mapsto \vertof{\graph^\circ} \} \cup \bigcup_j \beta^j$, $\gamma = \bigcup_j
  \gamma^j$ and $\delta = \{r \mapsto e\} \bigcup_j \delta^j$.  It is
  now easy to verify that $(\tree,\beta,\gamma,\delta)$ is a
  non-pointed $(\rootof{u}, \rootsof{u})$-embeddable
  tree-decomposition with width $\le k$; we note that the bijectivity
  requirement of $\gamma$ is satisfied because of the side-condition
  $\internal{\graph^\circ} \cup \{ \sourceof{\graph^\circ}(i) \mid i
  \in \rootsof{w} \} = \biguplus_{i=1..k} (\{\rootof{u_i}\} \uplus
  \{\edgerelof{\graph^\circ}(u_i)_j \mid j \in \rootsof{u_i} \} )$ of
  the rules of shape~\ref{it3:tree-verifiable-graph-grammar}.

  We finally consider some axiom $\rightarrow u$ for some $u \in
  \nonterm \setminus \mathcal{W}$ with $[1,\arityof{u}] =
  \{\rootof{u}\} \cup \rootsof{u}$ (*) and some graph $\graph \in
  \langof{\grammar}{u}$.  By the above we have that there is some
  $(\rootof{u}, \rootsof{u})$-embeddable tree-decomposition
  $(\tree,\beta,\gamma,\delta)$ with width $\le k$ for $\graph$.
  Because of (*) we have that $\gamma: \vertof{\tree} \setminus W
  \rightarrow \vertof{\graph}$ is a bijective mapping of the vertices
  in $\vertof{\tree} \setminus W$ to the vertices of $\graph$.  Hence,
  $(\tree,\beta,\gamma,\delta)$ is an embeddable tree-decomposition
  $(\tree,\beta,\gamma,\delta)$ and we have established
  $\etwof{\graph} \le k$.
\end{proofE}

We now prove the existence of a tree-verifiable grammar that generates
all graphs with embeddable treewidth $\le n$ for some given $n \in
\nat$. While the existence of such a grammar is an interesting result
on its own, we will build on this result in our partial completeness
proof (Theorem~\ref{thm:tree-verifiable-grammar-completeness}), where
we pair the non-terminals of this grammar with the equivalence classes
of some congruence relation.

\begin{propositionE}\label{prop:generic-embeddable-grammar}
  For each finite alphabet $\alphabet$ of edge labels and each $n \in
  \nat$, there exists a tree-verifiable grammar $\grammar$, such that
  $\lang{\grammar}=\set{\graph \in \graphsof{\alphabet}{} \mid \graph \text{ has type } 0 \text{ and } \etwof{\graph} \leq n}$.
\end{propositionE}
\begin{proofE}
  We consider the set of non-terminals $$\nonterm =
  \{u^{\iroot,\iroots,\type},w^{\iroot,\iroots,\type} \mid \type \in
     [0,n], \iroot \in [1,\type] \text{ and } \iroots \subseteq
     [1,\type] \setminus \{\iroot\} \},$$ with
     $\arityof{x^{\iroot,\iroots,\type}} = \type$ for all $x \in
     \{u,w\}$, $\type \in [0,n]$, $\iroot \in [1,\type]$ and $\iroots
     \subseteq [1,\type] \setminus \{\iroot\}$.  Then, we set
     $\mathcal{W} = \{w^{\iroot,\iroots,\type} \mid \type \in [0,n], i
     \in [1,\type] \text{ and } \iroots \subseteq [1,\type] \}$, as
     well as $\rootof{x^{\iroot,\iroots,\type}} = \iroot$ and
     $\rootsof{x^{\iroot,\iroots,\type}} = \iroots$ for all $x \in
     \{u,w\}$, $\type \in [0,n]$, $\iroot \in [1,\type]$ and $\iroots
     \subseteq [1,\type] \setminus \{\iroot\}$.  We define the
     tree-verifiable graph grammar $\grammar = (\nonterm,\rules)$ by
     the following rules:~\begin{compactenum}[A.]
     \item\label{it1:generic-embeddable-grammar} $w \rightarrow
       \ksubst{\graph}{e_1}{u_1}{e_k}{u_k}$, for some
       $w\in\mathcal{W}$ and $u_1,\ldots,u_k \in \nonterm \setminus
       \mathcal{W}$, where $\graph$ has exactly one terminal edge~$e$,
       $\cardof{\vertof{\graph}} \le n$, vertex $\sourceof{\graph}(\rootof{w})$ is incident to $e$,
       vertex $\edgerelof{\graph}(e_i)_{\rootof{u_i}}$ is incident to
       $e$ for every non-terminal $u_i$, and we have:
       \begin{align*}
         \internal{\graph} \cup \{ \sourceof{\graph}(i) \mid i \in
         \rootsof{w} \} = \biguplus_{i=1..k}
         (\{\edgerelof{\graph}(e_i)_j \mid j \in \{\rootof{u_i}\} \cup
         \rootsof{u_i})
       \end{align*}
     \item\label{it2:generic-embeddable-grammar} $u \rightarrow u
       \parallel_\type w$, where $u \in \nonterm\setminus\mathcal{W}$,
       $w \in \mathcal{W}$ with $\arityof{u}=\arityof{w}=\type$, such
       that $\rootof{w} = \rootof{u}$ and $\rootsof{w} = \emptyset$,
     \item\label{it3:generic-embeddable-grammar} $u \rightarrow w_1
       \parallel_\type \ldots \parallel_\type w_k$, where $u \in
       \nonterm\setminus\mathcal{W}$, $w_1,\ldots,w_k\in\mathcal{W}$
       with $\arityof{u}=\arityof{w_i}=\type$, such that $\rootof{u} =
       \rootof{w_i}$, $\rootsof{u} = \biguplus_{i=1..k} \rootsof{u_i}$
       and $\rootsof{u_i} \neq \emptyset$ for all $i \in [2,k]$.
    \end{compactenum}
  We further add axioms of the shape $\rightarrow
  u^{\iroot,\iroots,\type}$ for all $\type \in [0,n]$, $\iroot \in
  [1,\type]$ and $\iroots = [1,\type] \setminus \{\iroot\}$.

  First, we prove that $\rules$ is finite. There are only finitely
  many rules of shape~\ref{it1:generic-embeddable-grammar} because of $\cardof{\vertof{\graph}} \le n$, there are only finitely many non-terminals $\nonterm$ and finitely many different edge labels $a \in \alphabet$ for $e$, and we have $k
  < \arityof{a} = \arityof{\labof{\graph}(e)}$ because of the
  requirement that the roots of all non-terminals must be pairwise different.
  There clearly are only finitely many rules of
  shape~\ref{it2:generic-embeddable-grammar} because there are only
  finitely many non-terminals $\nonterm$.  There are only finitely
  many rules of shape~\ref{it3:generic-embeddable-grammar} because
  there are only finitely many non-terminals $\nonterm$ and the
  constraint $\rootsof{u} = \biguplus_{i=1..k} \rootsof{u_i}$ requires
  the partitioning of $\rootsof{u}$ into sets $\rootsof{u_i}$, which
  need to be non-empty for all $i \in [2,k]$, and there are only
  finitely many ways to partition a finite set into non-empty subsets.

  We now prove that $\lang{\grammar}$ consists of all graphs over $\alphabet$ with embeddable tree-width $\le n$.
  We first observe that $\etwof{\graph} \le n$ for all $\graph \in \lang{\grammar}$, by Proposition~\ref{prop:tree-verifiable-embeddable}.
  We now show that $\grammar$ can indeed generate all graphs with embeddable tree-width $\le n$.
  In order to do so, we prove the following more general statement:
  Let
  $(\tree,\beta,\gamma,\delta)$ be a $(\iroot, \iroots)$-embeddable
  tree-decomposition, with width $\le n$, of graph $\graph$ with type
  $\type$.  Then, $\graph \in
  \langof{\grammar}{w^{\iroot,\iroots,\type}}$ in case the
  tree-decomposition is pointed, and $\graph \in
  \langof{\grammar}{u^{\iroot,\iroots,\type}}$ otherwise.  The proof
  proceeds by induction on the height of the tree-decomposition
  $(\tree,\beta,\gamma,\delta)$.  Let $r$ be the root of $\tree$.  For
  every child $r^j$ of $r$ we consider the induced tuple
  $(\tree^j,\beta^j,\gamma^j,\delta^j)$, where $\tree^j$ is the
  subtree of $\tree$ rooted in $r_j$, and $\beta^j,\gamma^j,\delta^j$
  are obtained by restricting the functions $\beta,\gamma,\delta$ to
  the vertices of $\tree^j$.

  We first assume that the tree-decomposition is pointed.  Then, we
  have $\delta(r) = e \in \edgeof{\graph}$ such that for every child
  $r^j$ of $r$ in $\tree$ that there is some $i \in
  [1,\arityof{\labof{\graph}(e)}]$ with $\edgerelof{\graph}(e)_i =
  \gamma(r)$.  Let $k$ be the number of children $r_j$ of $r$ (we note
  that $k < \arityof{\labof{\graph}(e)}$ because of the requirement
  that children are attached to pairwise different vertices).  Let $V
  = \beta(r) \cup \bigcup_{j \in [1,k]} \beta(r^j)$ be the vertices
  associated to the bags of $r$ and the $r_j$ in the
  tree-decomposition.  Let $\type^j = \cardof{\beta(r^j)}$.  Let
  $v_1^j,\ldots,v_{\type^j}^j$ be some fixed enumeration of the
  vertices of the bag $\beta(r^j)$.  We consider the graph $\graph^j$
  that is obtained by restricting $\graph$ to the images of $\gamma^j$
  and $\delta^j$ and whose sources are defined by setting
  $\sourceof{\graph^j}(i) = v_i^j$ for all $i \in [1,\type^j]$.  Let
  $\iroot^j \in [1,\type^j]$ be the index such that $v_i^j =
  \gamma(r_j)$ and let $\iroots^j = \{i \in [1,\type^j] \mid v_i^j \in
  \img{\gamma^j} \} \setminus \{ \iroot^j\}$ be the subset of the
  indices whose associated vertices appear in the image of $\gamma^j$.
  We now recognize that $(\tree^j,\beta^j,\gamma^j,\delta^j)$ is a
  non-pointed $(\iroot^j,\iroots^j)$-embeddable tree-decomposition of
  $\graph^j$.  By induction assumption we have that $\graph^j \in
  \langof{\grammar}{u^{\iroot^j,\iroots^j,\type^j}}$.  We now define
  the graph $\graph^\circ$ to consist of the vertices $V_\graph$, the same
  sources as $\graph$ (i.e., we set $\sourceof{\graph^\circ}(i) =
  \sourceof{\graph}(i)$ for all $i \in [1,\type]$), the single terminal
  edge $e$, and some non-terminal edges $e_1,\ldots,e_k$, with
  $\edgerelof{\graph^\circ}(e_j) = \left\langle v_1^j, \ldots,
  v_{\cardof{\beta(r^j)}}^j \right\rangle$, labelled by the non-terminals
  $u^{\iroot^1,\iroots^1,\type^1},\ldots,u^{\iroot^k,\iroots^k,\type^k}$.
  We note that we have $\graph =
  \graph^\circ[e_1/\graph^1,\ldots,e_k/\graph^k]$.  We further note
  that we have $\internal{\graph^\circ} \cup \{ \sourceof{\graph}(i)
  \mid i \in \iroots \} = \biguplus_{j=1..k}
  (\{\edgerelof{\graph}(e_j)_i \mid i \in \{\iroot^j\} \cup
  \iroots^j),$ because of $V = \beta(r) \cup \bigcup_{j \in [1,k]}
  \beta(r^j)$, $\beta(r) = \{ \sourceof{\graph}(i) \mid i \in [1,\type]\}$,
  $\gamma = \bigcup_{j \in [1,k]} \gamma^j$ and $\gamma:
  \vertof{\tree} \setminus W \rightarrow \internal{\graph} \cup \{
  \sourceof{\graph}(i) \mid i \in \iroots) \}$ is a bijective
  function.  Hence, we can infer $\graph \in
  \langof{\grammar}{w^{\iroot,\iroots,\type}}$ by an application of a
  rule of shape~\ref{it1:generic-embeddable-grammar}.

  We now assume that the tree-decomposition is non-pointed.  We
  consider the graph $\graph^j$ that is obtained by restricting
  $\graph$ to the images of $\gamma^j$ and $\delta^j$ and that has the
  same sources as $\graph$, i.e., we set $\sourceof{\graph^j}(i) =
  \sourceof{\graph}(i)$ for all $i \in [1,\type]$.  We note that
  $\{\sourceof{\graph^j}(i) \mid i \in [1,\type] \} = \beta^j(r^j)$
  because we have that $\beta^j(r^j) = \beta(r)$ and
  $\{\sourceof{\graph}(i) \mid i \in [1,\type] \} = \beta(r)$ from the
  assumption that $(\tree,\beta,\gamma,\delta)$ is a $(\iroot,
  \iroots)$-embeddable tree-decomposition of $\graph$.  We now set
  $\iroots^j = \{i \in [1,\type] \mid \sourceof{\graph^j}(i) \in
  \img{\gamma^j} \}$ as the subset of the sources that appear in the
  image of $\gamma^j$.  We now recognize that
  $(\tree^j,\beta^j,\gamma^j,\delta^j)$ is a pointed
  $(\iroot,\iroots^j)$-embeddable tree-decomposition of $\graph^j$.
  By induction assumption we have that $\graph^j \in
  \langof{\grammar}{w^{\iroot,\iroots^j,\type}}$.  We note that we
  have $\iroots = \biguplus_j \iroots^j$ because $\gamma:
  \vertof{\tree} \setminus W \rightarrow \internal{\graph} \cup \{
  \sourceof{\graph}(i) \mid i \in (\{\iroot \} \cup \iroots) \}$ is a
  bijective function and we have the following partitioning $\gamma =
  \{r \mapsto \sourceof{\graph}(\iroot) \} \cup \bigcup_j \gamma^j$.
  Hence, we can infer $\graph \in
  \langof{\grammar}{u^{\iroot,\iroots,\type}}$ by applying rules of
  shape~\ref{it2:generic-embeddable-grammar} as often as needed
  followed by a single application of a rule of
  shape~\ref{it3:generic-embeddable-grammar}.

  We finally consider some graph $H$ of type $0$ for which there is an embeddable tree-decomposition
  $(\tree,\beta,\gamma,\delta)$ of width $\le n$. Let $\type =
  \cardof{\beta(r)}$ for the root $r$ of $\tree$. Note that $\type \le
  n$.  Let $v_1,\ldots,v_\type$ be some fixed enumeration of
  $\beta(r)$.  We now define the $\graph$ of type $\type$ as the
  extension of $H$ with the sources $[1,\type]$ by setting
  $\sourceof{\graph}(i) = v_i$ for all $i \in [1,\type]$.  We now
  choose $\iroot \in [1,\type]$ such that $\gamma(r) = v_\iroot$.  We
  further set $\iroots = [1,\type] \setminus \{\iroot\}$.  We now
  recognize that $(\tree,\beta,\gamma,\delta)$ is a $(\iroot,
  \iroots)$-embeddable tree-decomposition of graph $\graph$.  By the
  above, we get $\graph \in
  \langof{\grammar}{u^{\iroot,\iroots,\type}}$.  Clearly,
  $\rmsources{\graph} = H$.
  Thus, $\graph \in \lang{\grammar}$.
  Because this holds for all graphs $H$, we get that $\lang{\grammar}$
  contains all graphs over alphabet $\alphabet$ with embeddable
  tree-width $\le n$.
\end{proofE}

The following result combines ideas from
Theorem~\ref{thm:regular-tree-grammar-completeness} and
Proposition~\ref{prop:generic-embeddable-grammar}:

\begin{theoremE}\label{thm:tree-verifiable-grammar-completeness}
Let $\langu$ a \cmso-definable graph language with bounded embeddable
tree-width.  Then, $\langu$ is a tree-verifiable graph-language.
\end{theoremE}
\begin{proofE}
Let $t = \max \{ \etwof{\graph} \mid \graph \in \langu\}$.
By Theorem~\ref{thm:rec-equals-cmso-def-trees}, $\langu$ is recognizable.
Hence, there is a locally-finite congruence relation $\cgr$ that recognizes $\langu$.
Let $A_\type$ be the finite set of equivalence classes of graphs of type $\type$.
Let $A = \bigcup_{m \in [0,t]} A_\type$ be the union of the equivalence classes of graphs of type $\type \le t$.
We now observe that $\cgr$ induces an operation
$\parallel_\type$ on every $A_\type$.
In particular, we note that $A_\type$ and $\parallel_\type$ form a finite commutative monoid.
Further, every graph operation $(\graph, e_1, \ldots, e_k)$ with $\arityof{e_i} \le t$ induces a (typed) $k$-ary operation on the set $A$,
which we denote by $b = \graph[e_1/a_1,\ldots,e_k/a_k]$ for some $b,a_1,\ldots,a_k \in A$.
Let $n = \cardof{A}$ be the cardinality of $A$.
Because $A$ is finite, we have for every $a \in A$ that there are some indices $0 \le d < n$ and $1 \le \expon \le n$ such that $a^d = a^{d+\expon}$ (*), where $a^\expon$ denotes the $\expon$-fold composition of $a$ with itself wrt the operation $\parallel_\type$ for the type $\type$ of $a$.
We set $\stem(a) = d$ and $\lasso(a) = \expon$ for the least such $d$ and $\expon$ and note that they unique.

We now define a set of non-terminals by setting
$$\nonterm = \{u^{\iroot,\iroots,\type,a},w^{\iroot,\iroots,\type,a} \mid \type \in [0,t], a\in A_\type, \iroot \in [1,\type] \text{ and } \iroots \subseteq [1,\type] \setminus \{\iroot\} \},$$
with $\arityof{x^{\iroot,\iroots,\type,a}} = \type$ for all $x \in \{u,w\}$, $\type \in [0,t]$, $a\in A_\type$, $\iroot \in [1,\type]$ and $\iroots \subseteq [1,\type] \setminus \{\iroot\}$.
Then, we set $\classof{x^{\iroot,\iroots,\type,a}} = a$,
$\rootof{x^{\iroot,\iroots,\type,a}} = \iroot$ and $\rootsof{x^{\iroot,\iroots,\type,a}} = \iroots$ for all $x \in \{u,w\}$, $\type \in [0,t]$, $a\in A_\type$, $\iroot \in [1,\type]$ and $\iroots \subseteq [1,\type] \setminus \{\iroot\}$,
where $\classSymb: \nonterm \rightarrow A$ is a function that we use to denote the equivalence class associated to every non-terminal.
We define then a tree-verifiable graph grammar $\grammar = (\nonterm,\rules)$ by the following rules:
  \begin{compactenum}[A.]
  \item\label{it1:cmso-embeddable-grammar}
    $w \rightarrow \ksubst{\graph}{e_1}{u_1}{e_k}{u_k}$, for some $w\in\mathcal{W}$ and $u_1,\ldots,u_k \in \nonterm \setminus \mathcal{W}$, where $\graph$ has exactly one terminal edge~$e$, vertex $\sourceof{\graph}(\rootof{w})$ is incident to $e$,
    vertex $\edgerelof{\graph}(e_i)_{\rootof{u_i}}$ is incident to $e$
    for every non-terminal $u_i$, and we have $\classof{w} = \graph[e_1/\classof{u_1},\ldots,e_k/\classof{u_k}]$ and
    \begin{center}
     $\internal{\graph} \cup \{ \sourceof{\graph}(i) \mid i \in \rootsof{w} \}= \biguplus_{i=1..k} (\{\edgerelof{\graph}(e_i)_j \mid j \in \{\rootof{u_i}\} \cup \rootsof{u_i}),$
    \end{center}
  \item\label{it2:cmso-embeddable-grammar} $u \rightarrow u
    \parallel_\type w^\expon$, for some $1 \le \expon \le n$, where $u \in \nonterm\setminus\mathcal{W}$, $w \in \mathcal{W}$ with $\arityof{u}=\arityof{w}=\type$,
    such that $\classof{u} = \classof{u} \parallel_\type \classof{w}^\expon$, $\rootof{w} = \rootof{u}$ and $\rootsof{w} = \emptyset$,
  \item\label{it3:cmso-embeddable-grammar} $u \rightarrow w_1^{p_1}
    \parallel_\type \ldots \parallel_\type w_k^{p_k} \parallel_\type v_1 \parallel_\type \cdots \parallel_\type v_l$, for some $0 \le p_i < 2n$, where $u \in \nonterm\setminus\mathcal{W}$, $w_1,\ldots,w_k,v_1,\ldots,v_l \in\mathcal{W}$  with $\arityof{u}=\arityof{w_i}=\arityof{v_i}=\type$, such that $\classof{u} = \classof{w_1}^{p_1} \parallel_\type \cdots \parallel_\type \classof{w_k}^{p_k} \parallel_\type \classof{v_1} \parallel_\type \cdots \parallel_\type \classof{v_l}$, $\rootof{u} = \rootof{w_i}=\rootof{v_i}$,
    $\rootsof{w_i} = \emptyset$ for all $i \in [1,k]$,
    $\rootsof{v_i} \neq \emptyset$ for all $i \in [1,l]$, and
    $\rootsof{u} = \biguplus_{i=1..l} \rootsof{v_i}$.
    \end{compactenum}
   We further add axioms of the shape $\rightarrow u^{\iroot,\iroots,\type,a}$ for all  $\type \in [0,t]$, $\iroot \in [1,\type]$, $\iroots = [1,\type] \setminus \{\iroot\}$ and $a \subseteq \langu$.

We now prove that $\langu = \lang{\grammar}$:
In order to do so, we prove the following more general statement:
Let $(\tree,\beta,\gamma,\delta)$ be a $(\iroot, \iroots)$-embeddable tree-decomposition of graph $\graph \in a$, with $a \in A$, and let $\type = \cardof{\beta(r)}$ for the root $r$ of $\tree$.
Then, $\graph \in \langof{\grammar}{w^{\iroot,\iroots,\type,a}}$ in case the tree-decomposition is pointed, and $\graph \in \langof{\grammar}{u^{\iroot,\iroots,\type,a}}$ otherwise.

We first verify for all non-terminals $x^{\iroot,\iroots,\type,a} \in \nonterm$ with $x \in \{u,w\}$ that $\graph \in \langof{\grammar}{x^{\iroot,\iroots,\type,a}}$ implies that $\graph \in a$;
this by done by a straightforward induction on the length of the derivation.
In particular, this implies that $\langu \supseteq \lang{\grammar}$.

We now consider some graph $\graph \in a$ with type $\type$ and show that $\graph \in \langof{\grammar}{u^{\iroot,\iroots,\type,a}}$ in case there is a non-pointed $(\iroot, \iroots)$-embeddable tree-decomposition $(\tree,\beta,\gamma,\delta)$ of $\graph$, and $\graph \in \langof{\grammar}{w^{\iroot,\iroots,\type,a}}$ in case there is a pointed $(\iroot, \iroots)$-embeddable tree-decomposition $(\tree,\beta,\gamma,\delta)$ of $\graph$.
The proof proceeds by induction on the height of the tree-decomposition $(\tree,\beta,\gamma,\delta)$.
Let $r$ be the root of $\tree$.
For every child $r^j$ of $r$ we consider the induced tuple $(\tree^j,\beta^j,\gamma^j,\delta^j)$, where $\tree^j$ is the subtree of $\tree$ rooted in $r^j$, and $\beta^j,\gamma^j,\delta^j$ are obtained by restricting the functions $\beta,\gamma,\delta$ to the vertices of $\tree^j$.

We first assume that the tree-decomposition is pointed.
Then, we have $\delta(r) = e \in \edgeof{\graph}$ such that for every child $r^j$ of $r$ in $\tree$ that there is some $i \in [1,\arityof{\labof{\graph}(e)}]$ with $\edgerelof{\graph}(e)_i = \gamma(r)$.
Let $k$ be the number of children $r^j$ of $r$
(we note that $k < \arityof{\labof{\graph}(e)}$ because of the requirement that children are attached to pairwise different vertices).
As in the proof of Proposition~\ref{prop:generic-embeddable-grammar},
we now define the tuple $(\iroot^j,\iroots^j)$, the graphs $\graph^j$ and the graph $\graph^\circ$.
We recall that we have $\graph = \graph^\circ[e_1/\graph^1,\ldots,e_k/\graph^k]$ and that $(\tree^j,\beta^j,\gamma^j,\delta^j)$ is a non-pointed $(\iroot^j,\iroots^j)$-embeddable tree-decomposition of $\graph^j$.
Let $a^j \in A_\type$ be the equivalence class such that $\graph^j \in a^j$.
We note that we have $a = \graph^\circ[e_1/a^1,\ldots,e_k/a^k]$.
Further, by induction assumption we have that $\graph^j \in \langof{\grammar}{u^{\iroot^j,\iroots^j,\type^j,a^j}}$.
As in the proof of Proposition~\ref{prop:generic-embeddable-grammar}, we have $\internal{\graph^\circ} \cup \{ \sourceof{\graph}(i) \mid i \in  \iroots \} = \biguplus_{j=1..k} (\{\edgerelof{\graph}(e_j)_i \mid i \in \{\iroot^j\} \cup \iroots^j)$.
Hence, we can infer $\graph \in \langof{\grammar}{w^{\iroot,\iroots,\type,a}}$ by an application of a rule of shape~\ref{it1:cmso-embeddable-grammar}.

We now assume that the tree-decomposition is non-pointed.
As in the proof of Proposition~\ref{prop:generic-embeddable-grammar},
we now define the sets $\iroots^j$ and the graphs $\graph^j$.
We recall that $(\tree^j,\beta^j,\gamma^j,\delta^j)$ is a pointed $(\iroot,\iroots^j)$-embeddable tree-decomposition of $\graph^j$.
We note that we have $\graph = \graph^1 \parallel_\type \graph^2 \parallel_\type \cdots$.
Let $a^j \in A_\type$ be the equivalence class such that $\graph^j \in a^j$.
Hence, we have $a = a^1 \parallel_\type a^2 \parallel_\type \cdots$. (The previous superscripts were enumerating indices, we now switch to using superscripts like exponents).
Now, because $\parallel_\type$ is associative and commutative, we can represent this product as
$a = a_1^{l_1} \parallel_\type \cdots \parallel_\type a_n^{l_n}$ for some enumeration $a_1, \ldots, a_n$ of the equivalence classes of $A_\type$ and $l_i \in\nat$.
By (*), we have $a = a_1^{p_1} \parallel_\type \cdots \parallel_\type a_n^{p_n}$ for some $0 \le p_i < 2n$ such that $l_i = \lasso(a_i)k_i+p_i$ for some $k_i \in\nat$, and $a = a \parallel_\type a_i^{\lasso(a_i)}$ for every $a_i$ with $l_i \ge 2n$.
By induction assumption we have that $\graph^j \in \langof{\grammar}{w^{\iroot,\iroots^j,\type,a^j}}$.
As in the proof of Proposition~\ref{prop:generic-embeddable-grammar}, we have $\iroots = \biguplus_j \iroots^j$ because $\gamma: \vertof{\tree} \setminus W \rightarrow \internal{\graph} \cup \{ \sourceof{\graph}(i) \mid i \in  (\{\iroot \} \cup \iroots) \}$.
Hence, we can infer $\graph \in \langof{\grammar}{u^{\iroot,\iroots,\type,a}}$ by applying rules of shape~\ref{it2:cmso-embeddable-grammar} as often as needed followed by a single application of a rule of shape~\ref{it3:cmso-embeddable-grammar}.

We finally consider some graph $H \in \langu$ for which there is an embeddable tree-decomposition $(\tree,\beta,\gamma,\delta)$ of width $\le t$.
Let $a\in A$ be the equivalence class such that $\graph \in a$.
Note that $a \subseteq \langu$.
Let $\type = \cardof{\beta(r)}$ for the root $r$ of $\tree$.
Note that $\type \le t$.
Let $v_1,\ldots,v_\type$ be some fixed enumeration of $\beta(r)$.
We now define the $\graph$ of type $\type$ as the extension of $H$ with the sources $[1,\type]$ by setting $\sourceof{\graph}(i) = v_i$ for all $i \in [1,\type]$.
We now choose $\iroot \in [1,\type]$ such that $\gamma(r) = v_\iroot$.
We further set $\iroots = [1,\type] \setminus \{\iroot\}$.
We now recognize that $(\tree,\beta,\gamma,\delta)$ is a $(\iroot, \iroots)$-embeddable tree-decomposition of graph $\graph$.
By the above, we get $\graph \in \langof{\grammar}{u^{\iroot,\iroots,\type,a}}$.
Clearly, $\rmsources{\graph} = H$.
Because this holds for all graphs $H \in \langu$, we get that $\langu = \lang{\grammar}$.
\end{proofE}

We are ready to state the main result of this paper:
\begin{corollary}\label{cor:tree-verifiable-grammar-completeness}
Let $\langu$ a graph language of type $0$.
Then, the following are equivalent:
\begin{compactitem}
\item $\langu$ has bounded embeddable tree-width and is \cmso-definable.
\item $\langu$ is tree-verifiable.
\end{compactitem}
\end{corollary}

\section{Comparison with Regular Graph Grammars}
\label{sec:simulation}
In this section we compare our tree-verifiable grammar to the regular
graph-grammars of Courcelle~\cite{journals/tcs/Courcelle91a}.  We will
establish that the tree-verifiable grammars strictly generalize
regular graph grammars.  We will first demonstrate that for every
regular graph grammar there is a tree-verifiable grammar that
generates the same graph language.  
We will then show that cycles are tree-verifiable but not regular, see Example \ref{ex:cycle}. We begin by recalling the \emph{regular graph grammars} of
Courcelle~\cite{journals/tcs/Courcelle91a}, under a slightly different
notation:

\begin{definition}[\cite{journals/tcs/Courcelle91a}]
\label{def:regular-graph-grammar}
  A graph operation $(\graph, e_1,\ldots,e_k)$ is \emph{regular} iff
  the following hold:
  \begin{compactenum}
  \item\label{it1:regular-graph-operation} $\graph$ has at least one
    edge, either \begin{inparaenum}[(a)]
  \item\label{it11:regular-graph-operation} a single terminal edge
    attached to sources only, or
  \item\label{it12:regular-graph-operation} each of its edges is
    attached to an internal vertex, and
  \end{inparaenum}
  \item\label{it2:regular-graph-operation} between any two vertices of
    $\graph$ there is a path that traverses only internal vertices
    (excepting the endpoints) and terminal edges.
  \end{compactenum}
  A graph grammar $\grammar = (\nonterm,\rules)$ is \emph{regular} iff
  there exists a set $\mathcal{W} \subseteq \nonterm$ of nonterminals
  and the rules in $\rules$ are of one of the following forms, either:
  \begin{compactenum}[A.]
  \item\label{it1:regular-graph-grammar} $w \rightarrow
    \ksubst{\graph}{e_1}{u_1}{e_k}{u_k}$, for some nonterminals
    $w\in\mathcal{W}$, $u_1,\ldots,u_k \in \nonterm \setminus
    \mathcal{W}$ and some regular graph operation
    $(\graph,e_1,\ldots,e_k)$,
  \item\label{it2:regular-graph-grammar} $u \rightarrow u \parallel_n
    w^\expon$, for some $\expon \in \nat$, where $u \in
    \nonterm\setminus\mathcal{W}$ and $w \in \mathcal{W}$, such that
    $\arityof{u}=\arityof{w}=n$,
  \item\label{it3:regular-graph-grammar} $u \rightarrow w_1
    \parallel_n \cdots \parallel_n w_k$, $u \in
    \nonterm\setminus\mathcal{W}$ and $w_1,\ldots,w_k\in\mathcal{W}$,
    such that $\arityof{u}=\arityof{w_i}=n$.
  \end{compactenum}
  We further require that axioms are of the shape $\rightarrow u$ for some $u \in \nonterm \setminus \mathcal{W}$,
  and that $\arityof{u} > 1$ implies that $k \ge 1$ for all rules $u \rightarrow w_1 \parallel_n \cdots \parallel_n w_k$ with head $u$.
\end{definition}

We note that the restriction that $\rightarrow u$ and $\arityof{u} >
1$ imply $k \ge 1$ for all rules $u \rightarrow w_1 \parallel_n \cdots
\parallel_n w_k$ with head $u$ is not part of the definition by
Courcelle~\cite{journals/tcs/Courcelle91a}.  This only excludes a
constant number of graphs, namely the graphs of $n$ isolated vertices,
for which there is an axiom $u \rightarrow \singlegraph_n$, with $n = \arityof{u} >
1$, and a rule $\rightarrow u$.  We make this assumption in order to
facilitate the comparison with tree-verifiable graph grammars, which
always generate connected graphs.  We note, however, that we could
alternatively have weakened the definition of tree-verifiable graph
grammars in order to allow for the generation of the graphs $\singlegraph_n$.

Courcelle established the \cmso-definability of regular graph
grammars, see~\cite[Theorems 4.8 and 5.10]{journals/tcs/Courcelle91a}.
We now show that tree-verifiable graph grammars can simulate regular
graph grammars (see Theorem~\ref{thm:simulation} below).  We note that
the simulation result together with
Theorem~\ref{thm:tree-verifiable-grammar-definability} also
establishes the \cmso-definability of regular graph grammars. For the
purposes of the upcoming proofs, we introduce the following simplified
form of regular graph grammars:

\begin{lemmaE}\label{lem:regular-single-root}
  For every regular graph grammar $\grammar$ there is a regular graph
  grammar $\grammar'$ with $\lang{\grammar} = \lang{\grammar'}$ and
  $\arityof{u} = 1$ for all axioms $\rightarrow u$ of $\grammar'$.
\end{lemmaE}
\begin{proofE}
  W.l.o.g. we can assume that there is at most one axiom $\rightarrow
  u^\circ$ with $\arityof{u^\circ} = 1$ and $u^\circ \rightarrow \singlegraph$, and that this is the only rule
  with head $u^\circ$: In case, there is an axiom $\rightarrow u$ with
  $\arityof{u} = 1$ and a rule $u \rightarrow \singlegraph$, we add a new
  non-terminal $u^\circ$ with $\arityof{u^\circ} = 1$, the axiom
  $\rightarrow u^\circ$ and the rule $u^\circ \rightarrow \singlegraph$.  
  Then, we  do the following for all axioms $\rightarrow u$ with $\arityof{u} =
  1$ for which there is a rule $u \rightarrow \singlegraph$: We create a copy
  $u^\star$ of $u$ with $\arityof{u^\star} = 1$, we add the axiom
  $\rightarrow u^\star$, we delete the axiom $\rightarrow u$, and we
  add a rule $u^\star \rightarrow w_1 \parallel_n \cdots \parallel_n
  w_k$ for every rule $u \rightarrow w_1 \parallel_n \cdots
  \parallel_n w_k$ with $k \ge 1$.  Further, we add the rules $u^\star
  \rightarrow u^\star \parallel_n w^\expon$ and $u^\star \rightarrow
  w^\expon$ for every rule $u \rightarrow u \parallel_n w^\expon$.  It
  is clear that these transformations do not change the language
  generated by the grammar.

  We now consider some axiom $\rightarrow u$ with $u \neq u^\circ$.
  Then, we introduce a fresh non-terminal $u'$ with $\arityof{u'} =
  1$, add the axiom $\rightarrow u'$, and delete the axiom
  $\rightarrow u$.  Further, for every pair of rules $u \rightarrow
  w_1 \parallel_n \cdots \parallel_n w_k$, where we recall that $k \ge
  1$ by assumption, and $w_1 \rightarrow
  \ksubst{\graph}{e_1}{u_1}{e_\ell}{u_\ell}$, we add the rule $u'
  \rightarrow w_1'$, for a fresh non-terminal $w_1'$ with
  $\arityof{w_1'} = 1$, and the rules $w_1' \rightarrow
  \ksubstthree{\graph'}{e_1}{u_1}{e_\ell}{u_\ell}{f}{u^-}$ and $u^-
  \rightarrow w_2 \parallel_n \cdots \parallel_n w_k$, for a fresh
  non-terminal $u^-$, where $\graph'$ is obtained from $\graph$ by
  adding a new nonterminal edge $f$, with $\edgerelof{\graph}(f)_i =
  \sourceof{\graph}(i)$ for all $1 \le i \le n$; additionally, we add
  rules $u^- \rightarrow u^- \parallel_n w^\expon$ for every rule $u
  \rightarrow u \parallel_n w^\expon$.  It is easy to verify that
  these transformations do not change the language generated by the
  grammar.
\end{proofE}

\begin{theoremE}\label{thm:simulation}
  For every regular graph grammar $\grammar$ there is a
  tree-verifiable graph grammar $\grammar'$ such that $\lang{\grammar}
  = \lang{\grammar'}$.
\end{theoremE}
\begin{proofE}
  Let $\grammar = (\nonterm,\rules)$ be some regular graph grammar,
  and let $\mathcal{W} \subseteq \nonterm$ be the associated set of
  non-terminals. By Lemma~\ref{lem:regular-single-root} we can assume
  that $\arityof{u} = 1$ for all axioms $\rightarrow u$ of $\grammar$.
  Let $\ruleASet$ be the set of rules $w \rightarrow
  \ksubst{\graph}{e_1}{u_1}{e_k}{u_k} \in \rules$ of
  shape~\ref{it1:regular-graph-grammar} such that the graph operation
  $(\graph, e_1,\ldots,e_k)$ satisfies
  property~\ref{it12:regular-graph-operation}.

  The idea of this construction is to use the internal vertices, whose
  existence is guaranteed by
  property~\ref{it12:regular-graph-operation} as the roots of the
  non-terminals, in more detail, we will consider every position as a
  potential root vertex.  Hence, we consider the set of non-terminals
  $\nonterm^\circ = \{ (u,i) \mid u \in \nonterm, i \in
  [1,\arityof{u}] \}$, where we pair every nonterminal $u \in
  \nonterm$ with a root position $i \in [1,\arityof{u}]$.  We set
  $\rootof{u,i} = i$ and $\rootsof{u,i} = \emptyset$ for all $u \in
  \nonterm, i \in [1,\arityof{u}]$.  For every rule $\ruleA = w
  \rightarrow \ksubst{\graph}{e_1}{u_1}{e_k}{u_k} \in \ruleASet$ and
  $i \in [1,\arityof{w}]$, we will below introduce another set of
  non-terminals $\nonterm_{(\ruleA,i)}$; these sets of non-terminals
  are local in the sense that they are only needed to simulate the
  respective graph operation with regard to some root $i$.  The
  overall set of non-terminals $\nonterm'$ is then the union of
  $\nonterm^\circ$ and the $\nonterm_{(\ruleA,i)}$.  We now define the
  grammar $\grammar' = (\nonterm',\rules')$, whose rules are given as
  follows: For every rule $w \rightarrow \graph$ of
  shape~\ref{it11:regular-graph-operation}, i.e., $\graph$ consists of
  a single edge whose vertices are sources, we add the rules $(w,i)
  \rightarrow \graph$ for all $i \in [1,\arityof{w}]$.  For every rule
  $u \rightarrow u \parallel_n w^\expon$ resp. $u \rightarrow w_1
  \parallel_n \ldots \parallel_n w_k$, we add the rules $(u,i)
  \rightarrow (u,i) \parallel_n (w,i)^\expon$ resp. $(u,i) \rightarrow
  (w_1,i) \parallel_n \cdots \parallel_n (w_k,i)$ for all $i \in [1,\arityof{u}]$.
  Finally, for every rule $\ruleA = w \rightarrow
  \ksubst{\graph}{e_1}{u_1}{e_k}{u_k} \in \ruleASet$ and root position
  $i \in [1,\arityof{w}]$, we add a set of rules $\rules_{(\ruleA,i)}$
  with the following property: By successively applying rules from
  $\rules_{(\ruleA,i)}$, we can obtain the derived rule $(w,i)
  \rightarrow \ksubst{\graph}{e_1}{u_1,j_1}{e_k}{u_k,j_k}$ and
  this is the only derived rule with head $(w,i)$,
  for some where $1 \le j_i \le \arityof{u_i}$, such that the
  right-hand side of the derived rule does not contain any
  non-terminals in $\nonterm_{(\ruleA,i)}$ (*).  We finally add axioms
  $\rightarrow (u,1)$ for every axiom $\rightarrow u$ of $\grammar$.
  With property (*) it is straightforward to prove that
  $\lang{\grammar} = \lang{\grammar'}$ because every derivation of
  $\grammar'$ induces a derivation of $\grammar$ by deleting the root
  positions from the derivation, and every derivation of $\grammar$
  can be extended to a derivation of $\grammar'$ by adding suitable
  root positions to the derivation.

  Let us consider some rule $\ruleA = w \rightarrow
  \ksubst{\graph}{e_1}{u_1}{e_k}{u_k} \in \ruleASet$ and some root
  position $i \in [1,\arityof{w}]$.  We will below define a set of
  non-terminals and rules that simulate the graph operation $(\graph,
  e_1,\ldots,e_k)$, considering $i$ as the root of the non-terminal
  $w$.  Let $e_0 \in \edgeof{\graph}$ be some terminal edge to which
  the vertex $\sourceof{\graph}(i)$ of $\graph$ is incident (note that
  such an edge must exist because $\graph$ has at least one edge, this
  edge must contain an internal vertex $v$, and there is a terminal
  and internal path between any two vertices of $\graph$, in
  particular such a path exists between $\sourceof{\graph}(i)$ and
  $v$, and we can take the first edge of this path).  We now execute a
  graph search in $\graph$, maintaining a worklist and a done list of
  vertices and edges that have yet to be processed resp. already have
  been processed.  We only store internal vertices, but terminal as
  well as non-terminal edges.  At the beginning this worklist contains
  only $e_0$ and the done list is empty.  Assume we are processing
  some edge $e \in \edgeof{\graph}$ from the worklist, then we add
  every internal vertex $v$ that is incident to $e$ to the worklist in
  case the vertex is not yet in the done list; we further store $e$ as
  the parent of $v$.  Assume we are processing some node $v$ from the
  worklist, then we add every terminal edge $e$ to which $v$ is
  incident to the worklist in case the edge is not yet in the done
  list; we store $v$ as the parent of $e$; in case $e$ is a
  non-terminal edge that is not yet in the done list, we directly add
  it to the done list, and also store $v$ as the parent of $e$.  We
  observe that the graph search reaches every internal vertex and
  every edge of $\graph$ because, by the definition of a regular graph
  operation, there is a path between any two vertices of $\graph$ that
  traverses only internal vertices (excepting the endpoints) and
  terminal edges, and every non-terminal edge contains an internal
  vertex.

  We are now ready to state the non-terminals and rules that simulate the graph operation $(\graph, e_1,\ldots,e_k)$.
  The idea of this construction is to use the graph search to break down the graph operation into the steps of the graph search;
  these steps will be in one-to-one correspondence with rule shapes as required by a tree-verifiable graph grammar.
  Intuitively, the graph search defines a tree structure that can be used in order to define the tree shape that underlines the tree-verifiable graph grammars.
  Let $n = \cardof{\vertof{\graph}}$ be the number of vertices of $\graph$.
  We define the set of non-terminals
  $$\nonterm_{(\ruleA,i)} = \{ e^{(\ruleA,i)} \mid e \neq e_0 \in \edgeof{\graph} \text{ is a terminal edge}, i\in [1,\arityof{w}]\} \cup \{ v^{(\ruleA,i)} \mid v \in \internal{\graph}, i\in [1,\arityof{w}]\}.$$
  We set $\arityof{u} = n$ for all $u \in \nonterm_{(\ruleA,i)}$.
  We now fix some bijective function $\position: \vertof{\graph} \rightarrow [1,n]$ and we write  $\sigma$ for the sequence of length $n$ over the set of vertices $\vertof{\graph}$ such that $\position(\sigma_i) = i$ for all $i \in [1,n]$.
  We set $\rootof{e^{(\ruleA,i)}} = \position(v)$ for every edge $e \neq e_0$ where $v$ is the parent of $e$.
  We set $\rootof{v^{(\ruleA,i)}} = \position(v)$ for every internal vertex $v \in \internal{\graph}$.
  We now define the ancestor relation to be the transitive closure of the parent relation.
  We set $\rootsSymb(e^{(\ruleA,i)})$ to be the set of indices $\position(w)$ of all vertices $w \in \internal{\graph}$ such that edge $e$ is an ancestor of $w$,
  and we set $\rootsSymb(v^{(\ruleA,i)})$ to be the set of indices $\position(w)$ of all vertices $w \in \internal{\graph}$ such that vertex $v$ is an ancestor of $w$.
  We are now ready to state the set of rules $\rules_{(\ruleA,i)}$:
  For every terminal edge $e$ of $\graph$,
  we add the rule $w_e \rightarrow \ksubstfour{H}{e_1}{u_1,j_1}{e_\ell}{u_\ell,j_\ell}{f_1}{v_1^{(\ruleA,i)}}{f_k}{v_k^{(\ruleA,i)}}$ with
  $w_e = e^{(\ruleA,i)}$ for $e \neq e_0$, and $w_e = (w,i)$ for $e = e_0$,  where the graph $H$ of type $n$ consists of the vertices $\vertof{\graph}$, with $\sourceof{H}(\position(v)) = v$ for all $v \in \vertof{\graph}$, the single terminal edge $e$ with $\labof{H}(e) = \labof{\graph}(e)$ and $\edgerelof{H}(e) = \edgerelof{\graph}(e)$, the non-terminal edges $e_i$ with $\labof{H}(e_i) = \labof{\graph}(e_i) = u_i$ and $\edgerelof{H}(e_i) = \edgerelof{\graph}(e_i)$ whose parent $z_i$ is an internal vertex incident to $e$ and we have $\edgerelof{\graph}(e_i)_{j_i} = z_i$, and the (new) non-terminal edges $f_1,\ldots,f_k$, with $\edgerelof{H}(f_1) = \cdots = \edgerelof{H}(f_k) = \sigma$,   labelled by the non-terminals $v_1^{(\ruleA,i)},\ldots,v_k^{(\ruleA,i)}$, where the $v_1,\ldots,v_k$ are the internal vertices for which $e$ is their parent.
  For every $v \in \internal{\graph}$, we add the rule $v^{(\ruleA,i)} \rightarrow e_1^{(\ruleA,i)} \parallel_n \cdots \parallel_n e_k^{(\ruleA,i)}$, where $e_1,\ldots,e_k$ are the terminal edges of $\graph$ for which $v$ is the parent.
  We now state the main property of the rule set $\rules_{(\ruleA,i)}$:
  By successively applying rules from $\rules_{(\ruleA,i)}$, we can obtain the derived rule $(w,i) \rightarrow \ksubst{\graph}{e_1}{u_1,j_1}{e_k}{u_k,j_k}$ and this is the only derived rule with head $(w,i)$ such that the right-hand side of the derived rule does not contain any non-terminals in $\nonterm_{(\ruleA,i)}$ (*),
  where we have $\edgerelof{\graph}(e_i)_{j_i} = z_i$ for the parent $z_i$ of $e_i$.
  This can be easily established by following the steps of the graph search and applying the corresponding rules.
\end{proofE}

Next, we prove that tree-verifiable grammars strictly subsume regular
graph grammars. Given a graph $\graph$, we recall that a set of
vertices $C \subseteq \vertof{G}$ is connected in $\graph$ iff between
any two vertices in $C$ there exists a path in $\graph$ that traverses
only vertices from $C$. If $\vertof{G}$ is connected in $\graph$, we
say that $\graph$ is \emph{connected}. A graph is \emph{disconnected}
iff it is not connected. A \emph{cut} $C \subseteq \vertof{G}$ of a
connected graph $\graph$ is a set of vertices, such that the graph
obtained by removing from $\graph$ the vertices in $C$ and the edges
incident to some vertex from $C$ is disconnected.

\begin{propositionE}
  Let $\grammar$ be a regular graph grammar. Then, there is a constant
  $n \in \nat$ such that every graph $\graph \in \lang{\grammar}$
  with $\cardof{\vertof{\graph}} \geq n$ has a cut
  $C\subseteq\vertof{\graph}$ that is connected in $\graph$.
\end{propositionE}
\begin{proofE}
Let $m = \max \{ \cardof{\vertof{\graph}} \mid \text{ graph } \graph \text{ appears in a rule of shape~\ref{it1:regular-graph-grammar} of } \grammar\}$ and $n = 3m$.
We now observe that every graph $H$ with more than $n$ vertices contains at least 3 instances of an application of a rule of shape~\ref{it12:regular-graph-operation} such that either (1) the three rules are linearly ordered by the descendent relation in the derivation tree, or (2) there is one application of a rule of shape~\ref{it12:regular-graph-operation} such that the two other applications are (incomparable) descendants of the first application, or (3) all three applications are (incomparable) descendants of the root of the derivation tree and for each of these application there is no application of a rule of shape~\ref{it12:regular-graph-operation} that is strictly between the root and the respective application with regard to the descendent order.
We now observe for (1) that the vertex set $\vertof{\graph}$ for the graph $\graph$ that appears in the middle rule application forms a connected cut, for (2) the vertex set $\vertof{\graph}$ for the graph $\graph$ that appears in top-most rule application forms a connected cut, and for (3) any of the vertex sets $\vertof{\graph}$ for a graph $\graph$ that appears in one of the rule applications forms a connected cut (note that each of the cuts is connected by property~\ref{it2:regular-graph-operation} and that it generates indeed at least two components because each of the other two rule applications contains internal nodes that are now disconnected).
\end{proofE}



The following example shows a graph language consisting of simple
cycles of arbitrary size. These graphs do not have connected cuts,
because by removing any sequence of one or more consecutive vertices
in a cycle we always obtain a connected acyclic path.

\begin{example}\label{ex:cycle}
  We define cycles by the tree-verifiable grammar from
  Fig. \ref{fig:cycle}, with the set of non-terminals
  $\nonterm=\set{u_{\mathit{axiom}},w_{\mathit{axiom}},u,w}$ and
  distinguished non-terminals
  $\mathcal{W}=\set{w_{\mathit{axiom}},w}$, with arities, roots
  and future roots, as follows:~\begin{compactitem}
  \item $\arityof{u_{\mathit{axiom}}}=\arityof{w_{\mathit{axiom}}}=1$,
    $\rootof{u_{\mathit{axiom}}}=\rootof{w_{\mathit{axiom}}}=1$,
    $\rootsof{u_{\mathit{axiom}}}=\rootsof{w_{\mathit{axiom}}}=\emptyset$,
  \item $\arityof{u}=\arityof{w}=2$, $\rootof{u}=\rootof{w}=1$,
    $\rootsof{u}=\rootsof{w}=\emptyset$.
  \end{compactitem}
\end{example}

\begin{figure}[t!]
  \centerline{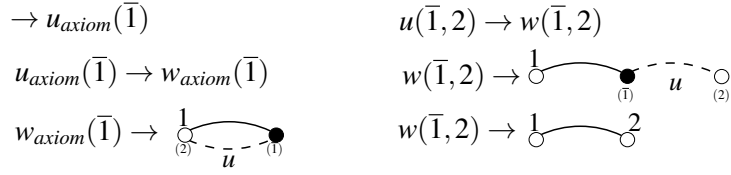}

  \vspace*{-.5\baselineskip}
  \caption{Tree-Verifiable Grammar for Cycles}
  \label{fig:cycle}
  \vspace*{-.5\baselineskip}
\end{figure}

\section{Conclusions}
We introduced the notion of tree-verifiable grammars, a class of
hyperedge-replacement grammars that produce \cmso-definable
languages. Since \cmso-definable context-free graph languages come
with a decidable inclusion problem, the class of tree-verifiable graph
languages is a useful tool for reasoning about and verifying systems
that can be modeled using graphs, such as, e.g., distributed
networks. We characterize tree-verifiable languages by showing their
equivalence with the class of \cmso-definable graph languages that
have bounded embeddable tree-width, a novel parameter, which
strengthens the standard graph-theoretic tree-width. Finally, we
show that the class of tree-verifiable graph languages strictly
subsumes the languages produced by the regular graph grammars
introduced by Courcelle, providing a more expressive class of languages at the intersection of the hyperedge-replacement context-free and the \cmso-definable sets of graphs.




\bibliographystyle{plain}
\bibliography{refs}

\appendix

\section{Proofs}
\label{app:proofs}
\printProofs

\end{document}